\documentclass[11pt,a4paper]{article}
\usepackage{jcappub}

\title{Self-Interacting Dark Matter in Cosmology: accurate numerical implementation and observational constraints}
\author[a,b]{Rafael Yunis}
\author[a,c,d]{Carlos R. Arg\"uelles}
\author[c,d]{Claudia G. Sc\'occola}
\author[e,f]{Diana L\'opez Nacir}
\author[g,h]{Gast\'on Giordano}

\affiliation[a]{ICRANet, Piazza della Repubblica 10, I--65122 Pescara, Italy}
\affiliation[b]{Physics Department, La Sapienza University of Rome, P.le Aldo Moro 5, I--00185 Rome, Italy}
\affiliation[c]{Facultad de Ciencias Astron\'omicas y Geof\'isicas, Universidad Nacional de La Plata, Paseo del Bosque, B1900FWA LaPlata, Argentina}
\affiliation[d]{CONICET (Consejo Nacional de Investigaciones Cient{\'\i}ficas y T\'ecnicas), Argentina}
\affiliation[e]{Departamento de F\'isica  Juan Jos\'e Giambiagi, FCEyN UBA}
\affiliation[f]{IFIBA CONICET-UBA, Facultad de Ciencias Exactas y Naturales, Ciudad Universitaria, Pabell\'on I, 1428 Buenos Aires, Argentina}
\affiliation[g]{Instituto de F\'isica La Plata IFLP}
\affiliation[h]{Departamento de F\'isica FCE-UNLP C.C. 67, 1900 La Plata, Argentina}
\emailAdd{yunis121@gmail.com}

\usepackage{graphicx}
\usepackage{caption}
\usepackage{subcaption}

\usepackage{mathtools}
\usepackage{amsmath}
\usepackage{amsmath}
\usepackage{amssymb}
\usepackage{verbatim}
\usepackage{mathrsfs}

\usepackage{aas_macros}

\numberwithin{equation}{section}

\usepackage{booktabs}
\usepackage{hhline}

\usepackage{lscape}

\usepackage[utf8]{inputenc}	% Required for inputting international characters
\usepackage[T1]{fontenc} 	% Required for outputting international characters
\usepackage{lmodern}		% nicer font
\usepackage{microtype}		% improv. of white-space
\usepackage{changepage}
\usepackage{multirow}

\usepackage{cleveref}		%For cref
\usepackage[all]{nowidow}	%??
\usepackage{pdfpages}		%??

\usepackage{ulem}   %Claudia: lo agrego para poder tachar texto

\DeclareTextFontCommand{\emph}{\boldmath\itshape}

\abstract{{
This paper presents a systematic and accurate treatment of the evolution of cosmological perturbations in self-interacting dark matter models, for particles which decoupled from the primordial plasma while relativistic. 
We provide a numerical implementation of the Boltzmann hierarchies developed in a previous paper [JCAP, 09 (2020) 041] in a publicly available Boltzmann code and show how it can be applied to realistic DM candidates such as sterile neutrinos either under resonant or non-resonant production mechanisms, and for different field mediators. 
At difference with traditional fluid approximations - also known as a $c_{\rm eff}-c_{\rm vis}$ parametrizations- our approach follows the evolution of phase-space perturbations under elastic DM interactions for a wide range of interaction models, including the effects of late kinetic decoupling.
Finally, we analyze the imprints left by different self interacting models on linear structure formation, which can be constrained using Lyman-$\alpha$ forest and satellite counts. We find new lower bounds on the particle mass that are less restrictive than previous constraints.
}}

\begin{document}
\thispdfpagelabel{Title}
\maketitle
\flushbottom

%%%%%%%%%%%%%%%%%%%%%%%%%%%%%%%%%%%%%%%%%%%%%%%%%%%%%
%%%%%%%%%%%%%%%%%%%%%%%%%%%%%%%%%%%%%%%%%%%%%%%%%%%%%

\sloppy
\clearpage
\section{Introduction}
\label{sec:Intro}

%Introduction, models, etc.

%generalidades de DM, un par de oraciones nomas

%hablar un toque de CDM y LambdaCDM, mencionar los modelos de DM
%tratar de darle un enfoque mas linear structure-y
%hablar de formacion de estructura
%mencionar tensiones en short scales y modelos alternativos:SIWDM
%Hablar posta de los modelos ahora: produccion y esas cosas
%mencionar constraints, bounds
%hablar de papers anteriores

The nature, mass and dynamics of the dark matter (DM) particles are still a mistery that have sparked several extensions to the standard model of particle physics. 
Most of the evidence of its existence comes from astrophysical and cosmological observations, from which we can search for clues about its behaviour \cite{PLANCK2018,Vogelsberger2014}.
So far, the most likely explanation for the dark matter phenomenology has some sort of microscopic origin \cite{BertoneNature,Bull:2015stt,Milgrom2019}. 
In the standard model of structure formation,  DM haloes form on a ``bottom-up'' hierarchy: small scales collapse first to form virialized objects, later merging and accreting into larger 
ones \cite{Hut1984,Bahcall:1999xn,PLANCK2018}.  
This paradigm is in remarkable agreement with observations at large and small scales, such as the distribution of large-scale structure \cite{Troster2019,DAmico2019}, CMB anisotropies \cite{PLANCK2018} and the internal structure of DM haloes \cite{Vogelsberger2014}. 
However, recent observations at small scales have become challenging to be described within the paradigm \cite{Bullock2017}. 
Such observations include
a number of satellite galaxies in the Milky Way that is far smaller than predicted (the ``missing satellite problem'')\cite{Griffen2016,Janesh_2019,TheDESCollaboration2015a}; 
the lack of observation of the most massive haloes, which are predicted to be luminous (the ``too big to fail'' problem)\cite{Boylan-Kolchin2011a}; 
and inconsistencies with the inner structure of dSph galaxies (the ``core-cusp'' problem) \cite{Oh2015,Navarro1997,Bullock2017}, among others.

While these tensions by no means rule out the concordance model, as they can be alleviated within $\Lambda$CDM itself, they have raised an increasing amount of interest in alternative DM models which may reduce these tensions in a more natural way. 
Some of the earliest approaches to these alternative models imply modifying the DM fluid's properties themselves. The earliest approaches include a DM mixture of CDM and a ``hot'' version of DM; low mass particles with a velocity dispersion such as to erase small scale structure \cite{Klypin:1992sf,Dolgov:2002wy}. 
Related to this approach, recently there has been growing interest in the field of interacting dark matter: while some of these models propose interactions between CDM and standard model particles, others assume a dark mediator field with low mass, interacting with CDM and contributing to the energy density budget of the universe. 
These types of models are widely known as ``Self Interacting Dark Matter'' (SIDM): such self interactions, apart from modifying the cosmological evolution  of these particles, affect significantly the clustering of matter in N-body simulations, as collisions can flatten the inner regions of most galaxies in CDM, reconciling these values with observation and alleviating the core-cusp problem \cite{Tulin2017,Zavala_2013,2016MNRAS.460.1399V}. 
Some particle physics models for SIDM may involve mirror world DM (see \cite{Khlopov2013,Khlopov2021} and references therein), light or heavy mediators through a broken $U(1)$ symmetry \cite{Tulin2012,Aarssen2012,Spergel1999a}, non Abelian interactions \cite{Buen_Abad_2015} or ``dark atoms'' \cite{Cyr-Racine2013}, to name a few.

Other promising approaches include ``Warm Dark Matter''(WDM): these models slightly reduce the preferred mass ranges of DM enough to include a moderate amount of initial velocity dispersion and free streaming, sufficient to erase some of the smallest scale structure \cite{Bode2000,Lovell2012,Boyarsky2018}. 
If this free streaming length today is smaller than the size of galaxy clusters, it can alleviate the missing satellites and too-big-to-fail problems \cite{Lovell2016b,2012MNRAS.424..684S,Schneider2016}.
Traditionally, these particles were also expected to generate DM cores in haloes and solve the core-cusp tension, however in recent years it has been shown that the WDM particle mass required for this to happen in N-body simulations is inconsistent with phase-space constraints \cite{Maccio2012,Shao2013}. 
Nevertheless, different approaches to DM halo structure may indeed relax these tensions, in particular the ones considering halo formation from a maximum entropy production principle, which can account for the expected DM distribution and halo-sizes in dSphs, while being in agreement with phase-space bounds \cite{Arguelles2019,2021arXiv211106199A}. 

A promising extension to the standard model of particle physics that naturally leads to
these types of WDM models can be found in $\nu$MSM (neutrino minimal standard model), which includes intermediate mass sterile neutrinos \cite{Boyarsky2018,Adhikari2016}. 
For such model, the preferred production mechanism involves active-sterile oscillations in the neutrino fluid, around the time of the quark-hadron transition \cite{ShiFuller,LaineShaposhnikov,Venumadhav2015}. 
It is possible to strongly constrain these models using a set of cosmological and astrophysical data. Indeed, a combination of limits from DM production; constraints from the predicted X-Ray radiative decay of sterile neutrinos; and structure formation bounds such as MW subhalo counts and Lyman-$\alpha$ forest severely reduced the allowed region in the parameter space of the models \cite{Adhikari2016,Perez2017,Cherry2017,Schneider2016,Ng2019a,Enzi2020a}. 
In fact, this last set of observations have, in recent years, almost completely ruled out this particular realization of WDM \cite{Schneider2016,Ng2019a,Enzi2020a}.

These tensions occurring in traditional WDM cosmologies with sterile neutrinos, together with the non-detection of CDM-type candidates (e.g. either within direct or $\gamma$-ray indirect detection experiments), strongly motivate to consider DM candidates that behave somewhat differently, both in their astrophysical impact on structure formation and in their detection abilities. Indeed, we will show in this work that extensions to the $\nu$MSM model which combine both WDM and SIDM approaches -labelled as ``Self Interacting Warm Dark Matter'' (SI-WDM)- can simultaneously address many of the small scale tensions we have mentioned, while considerably relaxing former constraints on the traditional $\nu$MSM model (as explicited in \cref{sec:Obs} and in \cref{fig:Obs_Results_SInuMSM}).%}

It was realized already in previous studies \cite{Boyarsky2018,Johns:2019cwc,deGouvea2019,Yunis2020a} that including self interactions in WDM sterile neutrino models may contribute to its production in the early universe, while in \cite{Yunis2020a} the $\nu$MSM parameter space and its observational bounds were re-analyzed in the context of indirect detection techniques with the inclusion of self interactions.
%that this extension (i.e. the SI-WDM model) may not only combine the benefits of both models, but also alleviate some of the inherent tensions to $\nu$MSM.
%
%
Moreover, it has been recently suggested that self interactions in WDM may play a significant role in the evolution of cosmological perturbations. 
That is, many studies have shown that strongly self interacting light DM (with particle masses ranging from $\mathcal{O}(1-100)$ keV) will remain in kinetic equilibrium until late times: this significantly modifies the linear evolution of cosmological perturbations, with respect to non-interacting models \cite{Huo2019,Heimersheim2020,Garny2018,Egana-Ugrinovic2021} %Being the ..[add the reference of the possibility to constraint the characteristic oscillations in the high-k region of the matter power spectrum with observations [Viel paper]]}
%these dudes do a good job
\footnote{While finalizing this manuscript, the work \cite{Egana-Ugrinovic2021} was presented on arXiv which studies WDM together with self interactions including the effects of kinetic coupling until the non-relativistic transition. 
Though they implemented approximations such as instantaneous self-interaction decoupling and a partial-fluid description which are superseded by our more self-consistent approach, both works show relaxed bounds when including the self interaction effects.}.

Following the footsteps of a previous work \cite{Yunis2020b}, where a full theoretical framework was given in order to treat the linear evolution of SI-WDM perturbations, we seek to improve these earlier numerical studies on several aspects and provide a more accurate framework for studying the evolution of SI-WDM, at the same time providing a comprehensive theoretical framework and developing a straightforward numerical application. 
%github
To this end, we develop an extension of the public Boltzmann solver CLASS v2.7.2 \cite{lesgourgues2011cosmic}, and provide a free distribution link at \href{https://github.com/yunis121/siwdm-class}{github.com/yunis121/siwdm-class}. 
%we explain RTA, TCA
We begin with an accurate approach to the Boltzmann collision term in cosmological perturbation theory as in \cite{Yunis2020b}, and consider several SI-WDM candidates, interactions and production mechanisms. 
Then, we properly introduce the necessary approximations for a feasible numerical application, paying special attention to the relaxation time and tight coupling approximations.
%Fluid does not work
Regarding the evolution of cosmological perturbations, many authors choose to compute the evolution of self interacting warm species using a fluid-like approximation (also known as a $c_{\rm eff}-c_{\rm vis}$ parametrization) either during its entire evolution as in \cite{Heimersheim2020,Garny2018}, or up to self-interaction decoupling \cite{Egana-Ugrinovic2021}. 
However, it has been argued in \cite{Oldengott2014,Oldengott2017} and shown in \cite{CLASSIV} that those fluid-like approximations, typical of CDM scenarios, provide inaccurate results for the computation of the power spectrum for self-interacting neutrinos and WDM respectively, and several doubts have been raised \cite{Cyr-Racine2013,Oldengott2014,Sellentin2014} about its physical meaningfulness. 

%We only use perturbations
Thus, we choose here to fully follow the evolution of the perturbations to the background DM distribution function, accounting for the elastic DM interactions self-consistently throughout the evolution of the species and up to today. 
This implies a more rigorous approach to the evolution of SI-WDM perturbations with respect to previous works on the subject (see above), which either use fluid parametrizations or focus solely on the background evolution.
%These models as extensions to WDM
Throughout this work, we present these self-interacting models as minimal extensions to existing WDM candidates (such as the sterile neutrino $\nu$MSM) instead of effective toy models (such as the scalar field DM models \cite{2018PhRvD..98h3517B}). This has the advantage of presenting numerous constraints from various sources to the parameter space, and including our models in a more concrete and better motivated framework that can account for several other processes such as DM production, baryogenesis and neutrino mass. We leave for a future work the possible effects self interactions may have on these predictions. 
%Lyman-alpha
The evolution of DM perturbations typically leave their strongest imprint in observables such as the CMB, Milky Way satellite counts and the Lyman-$\alpha$ forest \cite{Boyarsky2018,VielLesgourgues,Bullock2017}. We consider these last two and compare our results with observations, obtaining the strongest constraints on the mass of the DM particles from the Lyman-$\alpha$ forest.

The paper is organized as follows. In the remaining of this introduction, we will explore and discuss several production mechanisms for WDM, as the distribution function resulting from these processes will have significant consequences on the evolution of perturbations and, consequently, on structure formation observables. 
Afterwards, following from \cite{Yunis2020b}, we will present in \cref{sec:Boltz} the full formalism for SI-WDM in linear theory. We will recall some previous results on the Boltzmann collision terms, explore the relaxation time approximation, discuss about different regimes of DM self decoupling (introducing the non-relativistic self decoupling regime) and provide simple forms for the relevant quantities ideal for numerical implementation. 
This will be fully explored in \cref{sec:CLASS}, where we give details on the development of a SI-WDM linear cosmology solver based on CLASS, together with relevant approximations. 
We also include in that section a numerical application of non-relativistic self decoupling and present some initial results on the power spectra obtained. 
With that in mind, in \cref{sec:Obs} we compare the results from this solver with observations. We will focus particularly on the predicted number of MW satellites and the Lyman-$\alpha$ forest observations. We explore the results of these analysis for various initial conditions, and compare the self interacting alternative to the standard $\nu$MSM. 
Finally, in \cref{sec:Conclusions} we draw out conclusions.

\subsection{WDM Production in the Early Universe}

\label{sec:Intro_Production}

%Primero explicar por que necesitamos hablar de mecanismos de produccion
%Despues thermal
%Despues no resonante
%Despues resonante

Before introducing the assumptions and governing equations of SI-WDM as a whole, it is important to discuss briefly about its production in the early universe. 
This discussion serves a two-fold purpose. Firstly, a relevant model should allow DM to be sufficiently produced in the early universe as to match its abundance today. Secondly, the production mechanism gives the initial condition for the whole structure formation process, which results in additional constraints on the model \cite{Boyarsky2018,Venumadhav2015}. 
An interesting example that is constrained by the latter but not the former is Hot Dark Matter (HDM) \cite{Primack2000}, which can be efficiently produced in equilibrium with the hot plasma but their resulting high velocity dispersion does not reproduce the observed DM halo assembly history \cite{Hut1984}. 
Thus, it is important to consider the possible production mechanisms for these models as a first step in their analysis.

There have been recent advancements in exploring some models of WDM production in the presence of self interactions, with varying degrees of success in describing the DM abundance as observed today \cite{deGouvea2019,Johns:2019cwc}.
We choose to avoid these questions for now, as they exceed the scope of this work, mostly aimed at discerning these effects at the level of linear theory of perturbations. 
Thus, along this work we will simply explore a few relevant WDM production mechanisms, assuming that they are not significantly altered by the presence of self interactions (other than in a few obvious aspects such as thermalization of the initial distribution function (DF), see \cref{sec:CLASS}) and observe the consequences at the level of structure formation. 
We remind the reader that we do not explore whether or not dark matter production is significantly modified by the presence of self interactions, and we aim to address these questions and provide a more in-depth treatment in future works. 

Having clarified this, we will consider a few production mechanisms for these models: namely thermal production and production through active-sterile neutrino mixing (both on the resonant and non-resonant scenarios). 
The first scenario is the most straightforward, DM particles are produced in equilibrium in the hot plasma through interactions with other species and, as their interaction rate falls, they decouple from the hot plasma and evolve independently from the other species. 
This scenario is very similar to WIMP decoupling, however they decouple while relativistic with a relatively high particle mass, generally seen as incompatible with them being produced at the same time as WIMPS or neutrino HDM, when weak force interactions decouple (see e.g. \cite{VielLesgourgues}). Indeed, the abundance today for a particle that decouples while relativistic is approximately \cite{MoBoschWhiteBook}

\begin{equation}
\Omega_{i,0}h^2 \approx 7.64 \, \times \, 10^{-2} \left[ \frac{g_{i,{\rm eff}}}{g_{*,s}(T_{\rm f})} \right] \left(\frac{m_i}{\rm eV} \right) \ ,
\end{equation}

\noindent where $g_{i,{\rm eff}} = g_i$ for bosons, $g_{i,{\rm eff}} = (3/4)g_i$ for fermions, $g_i$ the spin degeneracy of the species and $T_{\rm f}$ its freeze-out temperature. 
This indeed results in overproduction for typical WDM masses of $m \sim \mathcal{O}({\rm keV})$, unless the particles are produced much earlier, when the number of relativistic degrees of freedom $g_*$ is much higher than $\sim 10^3$ \cite{VielLesgourgues}. 
Thus, thermal models are assumed to be produced at extremely high energies on an equilibrium background distribution, with an overall temperature of $(T/T_\nu)^3 = \Omega_{\rm DM}h^2 ({\rm 94 \ eV}/m)$ if we assume these to take up the whole DM budget in the universe \cite{VielLesgourgues}. 

While these thermal models are easily extended to many DM candidates, the other models we will consider are particular for the case of sterile neutrinos. 
Typical sterile neutrinos are heavier, mostly right handed companions of the active neutrinos, with the lightest of these particles acting as WDM. There exists mixing in these models between active and sterile species, measured by the angle $\theta$, which in the models considered here contributes to both their production and decay.
When discussing these sterile neutrino models, we will mostly focus on a particular realization of these, which involves a minimal extension to the standard model of particle physics: $\nu$MSM (see e.g. \cite{Boyarsky2018,Adhikari2016} for comprehensive reviews).

Indeed, DM production in these models results from the quantum oscillations between active and sterile neutrinos in the hot plasma. 
In the cases relevant to sterile neutrino DM, the Boltzmann equation for the quantum-damped, collisionally driven sterile neutrino production is \cite{Venumadhav2015}
\begin{equation}
\begin{split}
\frac{\partial f_{\nu_s}}{\partial t} - H p \frac{\partial f_{\nu_s}}{\partial p} = &\sum_{\nu_\mu+a+...\rightarrow i + ...} \int \frac{d^3p_a}{(2\pi)^3 2E_a} ... \frac{d^3p_i}{(2\pi)^3 2E_i} (2\pi)^4 \delta_D^4(p+p_a+...-p_i-...) \\
&\times \frac{1}{2} \Big[ \left<P_m(\nu_\mu \rightarrow \nu_s)\right> (1-f_{\nu_s}) \sum \left|\mathcal{M}\right|^2_{i+...\rightarrow a+\nu_\mu + ...} f_i ... (1\pm f_a)(1-f_{\nu_\mu})... \\
&-\left<P_m(\nu_s \rightarrow \nu_\mu)\right> f_{\nu_s} (1-f_{\nu_\mu}) \sum \left|\mathcal{M}\right|^2_{\mu_\nu + a + ... \rightarrow i+...} f_a ... (1\pm f_i)... \Big] \ .
\end{split}
\label{eq:Intro_Production_Boltzmann}
\end{equation}
The r.h.s. of this equation sums up all interactions that produce or consume, in this case, a muon neutrino, with the respective Bose enhancement and Pauli blocking terms $(1 \pm f)$. The $P_m$ are the active-sterile oscillation probabilities in matter modified by interactions with the medium, which are parametrized by neutrino self energy and quantum damping. 
The oscillation probability $P_m$ then reads
\begin{equation}
\begin{split}
&\left<P_m(\nu_\mu \leftrightarrow \nu_s;p,t)\right> = \\
&\frac{1}{2} \Delta^2(p) \sin^2 2\theta \left\{\Delta^2(p)\sin^2 2\theta + D^2(p) + \left[ \Delta(p)\cos 2\theta -V^L - V^{\rm th}(p) \right]^2\right\}
\end{split}
\label{eq:Intro_Production_OscProb}
\end{equation}
where $\Delta (p)$ is the vacuum oscillation rate $\Delta (p) \equiv \delta m_{\nu_\mu , \nu_s}/2p$, $\delta m_{\nu_\mu , \nu_s}$ is the mass difference between active and sterile neutrino, the quantity $D(p)$ is the quantum damping rate (half the interaction rate of active neutrinos) and the neutrino self energy is split into the lepton asymmetry potential $V^L$ and the thermal potential $V^{\rm th}$. More information on these potentials, as well as detailed calculations can be found in \cite{Venumadhav2015}. 

Two distinct regimes appear in this case: first, under no initial lepton asymmetry the potential $V^L$ becomes zero. This situation, as originally proposed by \cite{Dodelson1993a}, is known as non-resonant production. In this case, the resulting distribution function of sterile neutrinos roughly follows the one for active ones, but normalized depending on the mixing angle $\theta$. 
Thus, in this last production scenario, only a small window of mixing angles can provide the correct DM abundance today. 

However, in the presence of a nonzero lepton asymmetry the potential $V^L$ enhances the production rate, allowing for smaller mixing angles and, at the level of the distribution function, resulting in a nonequilibrium function which is generally colder \cite{ShiFuller,LaineShaposhnikov}. 
This mechanism, in the presence of a significant lepton asymmetry, is the scenario known as resonant production. The resulting background distribution functions depend on the mixing angle, particle mass and lepton asymmetry values; and their shapes are typically computed numerically (see \cite{Venumadhav2015} for a typical application, together with a public numerical tool). 

It is important to remark here that the form of the Boltzmann equation \eqref{eq:Intro_Production_Boltzmann}, particularly the interaction terms, are the ones responsible for DM production and most relevant in the early radiation dominated era. 
Later in this work, we will use a different form of the Boltzmann equation to calculate the evolution of perturbations (equations \eqref{eq:Boltz_Formal_F_motion} and \eqref{eq:RTA_Hierarchy_Motion}), in which we neglect this production term and only consider elastic DM collisions. 
This is due to the fact that the perturbations typically evolve later in cosmological history. However, the details of DM production still affect the evolution of perturbations as they act as initial conditions for the background distribution function.

\section{Boltzmann Formalism for Self Interactions in WDM}
\label{sec:Boltz}

%Introducir el paper anterior y que hicimos
In a previous paper \cite{Yunis2020b}, we introduced a full framework to study the effect of self interactions in WDM in Cosmological Perturbation Theory and its impact on the linear power spectrum. 
In particular, we considered an extension to \cite{Oldengott2017} where the full collision integrals are calculated for a scalar mediator and effectively massless particles, and extended it for massive particles, maintaining a certain degree of generality in the mediator model. 

%Cual es el framework/scenario en el que laburamos
While the formalism we develop is fairly general, a key task of the present work is to calculate the effects on cosmological observables under a certain scenario, described by the following \emph{hypotheses:}

\begin{itemize}

\item We consider a typical (fermionic) WDM particle of mass $\mathcal{O}(1-50)$ keV, that is produced and also decouples from the plasma while being relativistic, and becomes non-relativistic in the radiation dominated era (see, for example, \cite{Boyarsky2018}). In any case, the formalism itself makes no assumptions on particle mass, so limiting forms (either relativistic or non-relativistic) can provide accurate treatment for light and heavy relics, provided the other assumptions here listed are met.

\item We assume that, after production/decoupling from the plasma, the only form of interaction is an elastic self interaction. We further assume that these effects can be modeled by a tree-level scattering process following Feynmann rules of e.g. \cite{Gluza1991,Denner1992} (Note: the amplitudes presented here apply only for \textit{Majorana} fermions)
\footnote{A notable example where this is not the case is in light mediator models: as the DM particle becomes non-relativistic, scattering increasingly relies on the effects of point-charge potentials. In terms of Feynmann diagrams, this means that events where many mediator particles are exchanged in each process become increasingly relevant \cite{Feng_2010,Buen_Abad_2015}.}.

\item We neglect the cosmological population of mediator particles, and avoid calculating the evolution of the mediator fields themselves. This is a justified assumption on massive mediator scenarios: the initial population of mediators is assumed to have already decayed at $T_\gamma < m_{\rm med}$ (the mediator mass), and further production is kinematically supressed. The case for massless mediators has been briefly discussed in \cite{Yunis2020b} and thoroughly discussed in \cite{Oldengott2014} for light DM.

\item In order to model these self interactions, we work under a specific ansatz for the scattering amplitude:

\begin{equation}
\left| \mathcal{M} \right|^2 = A_t(s)t^2+B_t(s)t+C_t(s) = A_u(t)u^2+B_u(t)u+C_u(t)
\label{eq:Boltz_Formal_Ansatz}
\end{equation}

where $A_t$, $B_t$ and $C_t$ are simple functions of $s$ and $A_u$, $B_u$ and $C_u$ are simple functions of $t$, where $s,t,u$ are the Mandelstam variables, and the two expressions are related by $s+t+u=4m^2$. This ansatz includes various massive mediator cases, as well as a few others, and we explicitly presented results in \cite{Yunis2020b} for three different mediator models: Constant Amplitude, Massive Scalar and Massive Vector Field. 

\end{itemize}

%Introduccion a la seccion
In this section, we will provide a brief summary of the results in \cite{Yunis2020b} and provide some new concepts. 
First, we will summarize the full framework for studying Boltzmann Hierarchies in the scenarios we introduce in that previous work. 
Then, we present the Relaxation Time Approximation used to construct Boltzmann Hierarchies and specialize the results in \cite{Yunis2020b} to the specific models we consider in \cref{sec:Intro}. 
Finally, we will study the evolution of the background distribution function $f_0$ under these models and provide here a self-consistent study on the effects of self interaction decoupling, superseding (most) of the previous results in the literature. 

\subsection{Formal Boltzmann Framework}

\label{sec:Boltz_Formal}

Following the derivations in \cite{Yunis2020b}, the linear Boltzmann hierarchy for SI-WDM is given by:

\begin{equation}
\mbox{\small $
\begin{split}
\dot{F_0} (k, E_q, \tau) =& - \frac{q k}{E_q} F_1 (k, E_q, \tau ) + \frac{\dot{h}}{6} \frac{\partial f_0}{\partial \ln q} \\
& - G_0 a F_0 (k, E_q, \tau ) \Gamma (E_q, \tau) + G_0 a \int dE_l F_0 (k, E_l, \tau ) \mathcal{K}^{(1)}_0 (E_q, E_l, \tau) \\
\dot{F_1} (k, E_q, \tau) =& \frac{q k}{3 E_q} F_0 (k, E_q, \tau ) - \frac{2 q k}{3 E_q} F_2 (k, E_q, \tau ) \\
& - G_0 a F_1 (k, E_q, \tau ) \Gamma (E_q, \tau) + G_0 a \int dE_l F_1 (k, E_l, \tau ) \mathcal{K}^{(1)}_1 (E_q, E_l, \tau) \\
\dot{F_2} (k, E_q, \tau) =& \frac{q k}{5 E_q} \Big[ 2 F_1 (k, E_q, \tau ) - 3 F_3 (k, E_q, \tau ) \Big] - \frac{\partial f_0}{\partial \ln q} \Bigg[ \frac{1}{15} \dot{h} + \frac{2}{5} \dot{\eta} \Bigg]\\
& - G_0 a F_2 (k, E_q, \tau ) \Gamma (E_q, \tau) + G_0 a \int dE_l F_2 (k, E_l, \tau ) \mathcal{K}^{(1)}_2 (E_q, E_l, \tau) \\
\dot{F_l} (k, E_q, \tau) =& \frac{q k}{(2l+1) E_q} \Big[ l F_{(l-1)} (k, E_q, \tau ) - (l+1) F_{(l+1)} (k, E_q, \tau ) \Big] \\
& - G_0 a F_l (k, E_q, \tau ) \Gamma (E_q, \tau) + G_0 a \int dE_l F_l (k, E_l, \tau ) \mathcal{K}^{(1)}_l (E_q, E_l, \tau) \quad , \quad l \geq 3\\
\end{split}
$}
\label{eq:Boltz_Formal_F_motion} 
\end{equation}

\noindent in the synchronous gauge, defined by the line element $ds^2 = a^2(\tau) \left\{ -d\tau^2 + (\delta_{ij} + h_{ij}) dx^i dx^j \right\} $, where the scalar mode of the perturbation $h_{ij}$ can be described in terms of two fields $h(\vec{k}, \tau)$ and $\eta(\vec{k}, \tau)$, the trace and traceless parts of the $h_{ij}$ perturbation in Fourier space, respectively. An overdot denotes derivation with respect to comoving time $\tau$, and moments $\vec{q}=a\vec{p}$ are comoving proper momenta, with $\vec{p}$ the proper momentum measured by an observer in a fixed spatial coordinate, and $E_q = \sqrt{m^2 + q^2/a^2}$, with $m$ the DM mass.

The various kernel moments are defined as:

\begin{equation}
\Gamma (E_q, \tau) = \frac{1}{E_q q} \int dE_l ds f_0 (E_l) \chi(s) \ ,
\label{eq:Boltz_Formal_F_kernelDef_gamma} 
\end{equation}

\begin{equation}
\mathcal{K}^{(1)}_l (E_q, E_l, \tau ) = -\chi_l (E_q, E_l) f_0 (E_q)  + 2 \frac{1}{E_q q} K_l (E_q, E_l, \tau )  \ ,
\label{eq:Boltz_Formal_F_kernelDef_K} 
\end{equation}

\noindent with $G_0 = 1/[4(2\pi)^3]$, $s$ refering to the Mandelstam variable $s \equiv (\mathbf{q}+\mathbf{l})^2$ and, for the purposes of the integration kernels $\Gamma, \mathcal{K}$ the momenta $q$ is \emph{not} comoving: $q \equiv \sqrt{E_q^2-m^2}$. Integration over these terms (and in all definitions) runs over the whole range of the variable when integration limits are not specified. The $l$-th moment of the perturbed DF, $F_l$, is defined as in 

\begin{equation}
\begin{split}
F(|k|, |q|, \cos \epsilon ) &= \sum _{l=0}^{\infty} (-i)^l (2l+1) F_l(|k|, |q|) P_l(\cos \epsilon )\ ,\\
F_l(|k|, |q|) &= \frac{i^l}{2} \int_{-1}^{1} d \cos \epsilon F(|k|, |q|, \cos \epsilon) P_l(\cos \epsilon ) \ ,
\end{split}
\label{eq:Boltz_Formal_F_LegendreQ} 
\end{equation}

\noindent  and $K_l$, $\chi_l$ the Legendre transforms of the $K$, $\chi$ kernel functions are defined as in 

%\begin{equation}
%\frac{i^l}{2} \int_0^{2\pi} \frac{d\psi}{2\pi} \int_{-1}^{1} d\cos \epsilon P_l(\cos \epsilon) \left( \frac{\partial f}{\partial \tau}\right)_{k}^{(1)} = \int dE_l \mathcal{K}_l (E_q, E_l, \tau ) F_l (|k|, |l|, \tau) \ ,
%\label{eq:Boltz_Formal_F_PropEq2} 
%\end{equation}
%
%\noindent with

\begin{equation}
\mathcal{K}_l (E_q, E_l, \tau ) \equiv \int ds \mathcal{K} (E_q, E_l, s, \tau ) P_l (\cos \theta (s) ) \ .
\label{eq:Boltz_Formal_F_PropEq3} 
\end{equation}  

The $K$, $\chi$ functions, the proper kernel functions, encode the behaviour of the particular mediator model and are defined in terms of the coefficients \eqref{eq:Boltz_Formal_Ansatz} as:

\begin{equation}
\chi(s) = \frac{1}{3} A_t (s-4m^2)^2 + \frac{1}{2} B_t (s-4m^2) +C_t \,,
\label{eq:Boltz_Formal_ChiDef}
\end{equation}

\begin{equation}
\mbox{\small $
\begin{split}
K(E_q, E_{q'}, t, \tau) = \Bigg\{ \frac{A_u}{8 |\vec{q} - \vec{q'}|^5} \Bigg\{ & \left< f_0 \right>_2 \Bigg[ 4 t \left(3 (E_q + E_{q'})^2 t - ((E_q - E_{q'})^2 - t) (-4 m^2 + t)\right)\Bigg] \\
	+ & \left< f_0 \right>_1 \Bigg[ 4 t (4 (E_q - E_{q'})^2 (E_q + 3 E_{q'}) m^2  \\
	  & \hphantom{\left< f_0 \right>_1 \Bigg[} - 4 (E_q (E_q - E_{q'}) (E_q + 2 E_{q'}) + (E_q + 3 E_{q'}) m^2) t + (E_q + 3 E_{q'}) t^2) \Bigg]  \\
	+ & \left< f_0 \right>_0 \Bigg[ (48 (E_q - E_{q'})^4 m^4 - 16 (E_q - E_{q'})^2 m^2 (2 E_q^2 - 3 E_q E_{q'} + 6 m^2) t  \\
	  & \hphantom{\left< f_0 \right>_0 \Bigg[} + 8 (E_q^2 (E_q - E_{q'})^2 + (7 E_q^2 - 12 E_q E_{q'} + 3 E_{q'}^2) m^2 + 6 m^4) t^2  \\ 
	  & \hphantom{\left< f_0 \right>_0 \Bigg[} - 4 (2 E_q^2 - 3 E_q E_{q'} + 6 m^2) t^3 + 3 t^4)  \Bigg] \Bigg\}  \\
	  +\frac{B_u}{2 |\vec{q} - \vec{q'}|^3} \Bigg\{ & \left< f_0 \right>_1 \Big[ t (E_q+E_{q'}) \Big]+ \left< f_0 \right>_0 \Big[ 2 (E_q - E_{q'})^2 m^2 + 2 E_q (-E_q + E_{q'}) t - 4 m^2 t + t^2 \Big] \Bigg\}  \\
	  + \frac{C_u}{|\vec{q}-\vec{q'}|} \hphantom{\Bigg\{ } & \left< f_0 \right>_0 \quad \Bigg\}
\end{split}
$}
\label{eq:Boltz_Formal_KDef_intFinal}
\end{equation}

\noindent  with the background distribution moments $\left< f_0 \right>_i$ defined as:

\begin{equation}
\left< f_0 \right> _n (E_q , E_{q'}, t , \tau) = \int_{R_2}^{\infty} d E_{l'} f_0 (E_{l'}, \tau) E_{l'}^{n}\ ,
\label{eq:Boltz_Formal_f0MeanDef}
\end{equation}

\noindent which are functions of $(E_q,E_{q'},t)$ only through $R_{2}$, defined as

\begin{equation*}
R_{2} = \frac{1}{2} \left\{ E_q - E_{q'} - |\vec{q}-\vec{q'}|\sqrt{1 - \frac{4m^2}{t}} \right\} \ .
\end{equation*}

%Motivations, final words
The Boltzmann Hierarchy \eqref{eq:Boltz_Formal_F_motion} provides an accurate, relatively model-independent approach to the collisional Boltzmann hierarchies in the DM cosmological scenario above described. 
Once the interaction model amplitude is specified in equation \eqref{eq:Boltz_Formal_Ansatz}, these Boltzmann hierarchies can be solved coupled to the Einstein equations in order to provide an accurate description of SI-WDM. 
While the expressions are lengthy, most of these model-dependent kernels have forms relatively straightforward to evaluate and all these kernels can be precomputed, once the evolution of $f_0$ is known. 
The evolution of the background distribution $f_0$ due to elastic self interactions is also given in \cite{Yunis2020b}, and we refer the reader to section 3.2 there for a full expression of the collision terms. 

%Introduction for next section: why we need to approximate
Still, in order to provide a better understanding of these collision terms and their broad effects on the cosmological evolution, it is always useful to provide some more straightforward approximation to the collision terms in these hierarchies. 
%What we do
In the rest of this section we will review how to bring this exact operator into a simpler form, provide approximate Boltzmann hierarchies (yet portraying accurately the general behaviour of elastic collisions) and use this formalism to better comprehend the cosmological evolution of the SI-WDM component.
%Fluid Approximation Beware!
This approach of providing approximate forms for the collision terms in the Boltzmann equation comes into contrast with previous works on the evolution of SI-WDM, which choose to avoid calculating perturbations to the distribution function entirely by relying on a fluid approximation of the energy-momentum tensor \cite{Ma1995,Heimersheim2020,Garny2018}. 
However, it has been shown in \cite{Oldengott2014,Oldengott2017} to not accurately portray the evolution of species that decouple while relativistic and methods based on the full Boltzmann hierarchies were favored. 

\subsection{Relaxation Time Approximation}

\label{sec:RTA}

%The theoretical basis for the RTA: equilibrium and the collision terms
The evolution of the collisional DM fluid can be approximated in a simpler way by noting the general behavior of the collisions in the phase space density $f(\vec{k},\vec{q},\tau)$ as a whole. 
Statistical systems evolving through the Boltzmann equation with a significant collision term tend to relax to their local equilibrium distributions due to entropy maximization: these equilibrium forms carry the particularity of nullyfying the Liouville operator, the collision term and it can be shown to satisfy the detailed balance condition $\ln \left[ f_{eq}(\mathbf{q}) \right] + \ln \left[ f_{eq}(\mathbf{l}) \right] = \ln \left[ f_{eq}(\mathbf{q'}) \right] + \ln \left[ f_{eq}(\mathbf{l'}) \right]$ with the 4-momenta $\mathbf{q},\mathbf{l},\mathbf{q'},\mathbf{l'}$ satisfying conservation of total 4-momentum, where the primed momenta correspond to the outgoing particles in an interaction.

%General forms and theorems for equilibrium in an expanding universe
In general, it can be shown \cite{Bernstein_1988} that in a FLRW expanding universe there is no general equilibrium distribution function, i.e., there is no function $f_{eq}$ that satisfies $\mathcal{L}(f_{eq})=0$ at all times under this metric. However, solutions can be found in two limiting cases:

\begin{equation}
f_{eq}(p, t) = 
\begin{cases}
f_{rel}(p, t) \propto e^{-\beta p a(t)}, & \text{if} \ p \gg m \\
f_{nrel}(p, t) \propto e^{-\beta \frac{p^2}{2 m} a^2(t)}, & \text{if} \ p \ll m
\end{cases}
\label{eq:RTA_Equilibrium_General}
\end{equation}

\noindent with $\beta$ a constant, usually interpreted as the inverse of the temperature today and $p$ the physical momentum, and we have suppressed quantum effects in $f_{eq}$ for simplicity, see \cite{Kolb_Turner_1990} for a full expression. We will refer to $f_{rel}$ as the \emph{relativistic} equilibrium DF and to $f_{nrel}$ as the \emph{non-relativistic} equilibrium DF.

%We can use this to reconstruct a basic operator! What we saw in the previous paper 
Taking into consideration this behavior, we can construct a much simpler operator for the collision term that recreates the expected behavior for small departures from equilibrium:

\begin{equation}
\left( \frac{\partial f}{\partial t} \right)^{(1)}_{col} = -\frac{f(\vec{x},\vec{p},\tau) - f_{eq}(\vec{p}, \tau)}{\tau_{rel}(q)}
\label{eq:RTA_Definition_CollisionTerm}
\end{equation}

\noindent where  $\tau_{rel}$ is a quantity known as \emph{relaxation time} and $f_{eq}$ is a local equilibrium distribution function - in the context of cosmology, we will specify this local equilibrium function to be the background DF,  $f_0$, that describes the isotropic evolution.

%What is relaxation time
This approximate form is known as the \emph{Relaxation Time Approximation} (RTA). In general, this operator will qualitatively ``erase'' all perturbations to $f_0$ bringing the system back into equilibrium (even erasing collisional invariants such as number and energy densities). 
The quantity $\tau_{rel}$ governs the time scale in which these perturbations relax back into equilibrium and, in general, the rate of change of the distribution function as a whole. 
This approximate form can be obtained from the full collision term by assuming that the perturbation to the DF is highly localized in momenta or, equivalently, forcing the evolution of the perturbations to be local in momentum space. %Pages marked for this derivation, create a notebook please!

%Discussion about validity
%First, it is best when perturbations are small, so ideal for this case
As we mentioned above, the validity of this approximation  lies on the assumption that the proper distribution is very close to equilibrium, i.e. that the perturbation to equilibrium is very small \cite{Krapivsky_2010}.
{This makes it particularly useful to  describe the  evolution of  the  perturbation   to the background (isotropic) distribution in the linear regime of most of the  species in the universe. This is indeed the approach followed by many authors   in dealing with collisions between these species   \cite{Hannestad00,Oldengott2017,Egana-Ugrinovic2021}.  More importantly, it has been explicitly shown in \cite{Oldengott2017} that this approximation provides accurate results when compared to exact models of relativistic neutrinos (i.e akin to our model), significantly outperforming fluid approximations (also known as $c_{\rm eff}$-$c_{\rm vis}$)\footnote{It is important to notice that the background distribution for the species must take also its equilibrium form. This assumption may break during transitions, when the equilibrium form for these species becomes ill defined: we study a scenario like this one in more detail in the following section \eqref{sec:SID}.}}
%
%This makes it particularly useful in describing cosmological perturbation in the linear regime, when it is assumed that the perturbations to the background (isotropic) distribution of most species is in the linear regime.
%
%This is indeed the approach followed by many authors in dealing with collisions in these species in the first place \cite{Hannestad00,Oldengott2017,Egana-Ugrinovic2021}.
%\color{red}{But more important, it has been explicitly shown in \cite{Oldengott2017} that this approximation provides accurate results when compared to exact models of relativistic neutrinos (i.e akin to our model), significantly outperforming fluid approximations (also known as $c_{\rm eff}$-$c_{\rm vis}$)\footnote{It is important to notice that the background distribution for the species must be also its equilibrium form. This assumption may break during transitions, when the equilibrium form for these species becomes ill defined: we study a scenario like this one in more detail in the following section \eqref{sec:SID}.}.}}
%Important! Background not defined sometime, we discuss this in following sections
%It is important that the background is the equilibrium DF

%The general form for the relaxation time! (provide a slightly clearer result than on previous paper, it's important!)
%About the normalization: I use the normalization factor approach: I do everything as if it were f_0 unnormalized, then I normalize at the end! (see non-relativistic Decoupling.ipynb)
It has been shown in a previous paper \cite{Yunis2020b} that the relaxation time $\tau_{rel}$ can be expressed in terms of the collision kernel $\chi$ as:

\begin{equation}
\tau_{rel}(q)^{-1} = - \frac{\mathcal{D}_2[f_{eq}](q, \tau)}{f_{eq}(q, \tau)} = \frac{g_i^3}{32 (2\pi)^3 E_q q} \int dE_l ds f_{eq}(E_l, \tau) \chi(s)
\label{eq:RTA_TauDefinition_D2}
\end{equation}

\noindent where $f_{eq}$ is now an equilibrium DF, $g_i$ is the number of spin degrees of freedom of the DM particle and, in the last equality, momenta and energies are defined as in \eqref{eq:Boltz_Formal_F_motion}. 
In general this expression is momentum dependent, but it is possible to further simplify this approach by constructing a momentum independent thermal average of the relaxation time $\left< \tau_{rel} (q) \right>_{th} \equiv \tau_{rel} = \int d^3q \tau_{rel}(q) f_{eq}(q, \tau) / \int d^3q f_{eq}(q, \tau)$. Thus, the thermal averaged relaxation time (hereby, just relaxation time) can be expressed in terms of collision kernels as:

\begin{equation}
\tau_{rel}^{-1} = \frac{g_i^3}{32 (2\pi)^3} \frac{\int dE_q \, dE_l \, ds \, f_{eq}(E_q, \tau)  \, f_{eq}(E_l, \tau) \, \chi(s)}{\int dE_q \, q \, E_q \, f_{eq}(E_q, \tau)} \ ,
\label{eq:RTA_TauDefinition_dE}
\end{equation}

\noindent assuming the azimuthal angle $\psi$ between $\vec{q}$ and $\vec{l}$ has already been averaged out (see \cite{Oldengott2014}).

At this stage, one can draw a parallelism between this kernel function $\chi$ and the invariant rate defined in \cite{Cannoni_2014}, as $R \propto \chi$. Also, through the invariant flux $F=\frac{1}{2}\sqrt{s(s-4m^2)}$~\footnote{In general, as seen in \cite{Cannoni_2014}, the general expression for the invariant flux is $F=n_1 n_2 \frac{p_1 . p_2}{E_1 E_2} V_r$ with $n_{1,2}$ the number density and $V_r$ the special relativity relative velocity. In this case, the expression reduces to this form assuming that one particle states are normalized to $2E$ and that incoming and outgoing particles are of the same species.}, we find that $\sigma F = 4 E_q E_l \sigma v_{mol} = \frac{g_i^2}{4} \chi$ with $\sigma$ the total cross section of the interaction and $v_{mol}$ the Moller velocity. This allows us to put this relaxation time in terms of more well-known expressions as:

\begin{equation}
\Gamma \equiv \tau_{rel}^{-1} = n \left< \sigma v_{mol} \right> = n \frac{\int d^3q \, d^3l \, \sigma v_{mol} \, f_{eq}(q, \tau) \, f_{eq}(l, \tau)}{\int d^3q \, d^3l \, f_{eq}(q, \tau) \, f_{eq}(q, \tau)}
\label{eq:RTA_TauDefinition_d3qSigmaV}
\end{equation}

\noindent where $\Gamma$ is the interaction rate, $n$ is the DM number density and, to keep the same notation/normalization as in \cite{GondoloGelmini91,Lee_Weinberg_1977}, we use $\sigma v_{mol}= \frac{g_i^2/4}{4 E_q E_l} \chi $. 
This form allows us to compare with their results and to condense the overall normalization of the distribution function in $n$\footnote{It is important to differentiate this definition for $\sigma v_{mol}$ from the one used in \cite{Yunis2020b}. In that case, the normalization and definition for the cross section matches \cite{DeGroot1980}, and differs from this one on a numerical factor.}.

%How does that look in a Boltzmann Hierarchy?
With this simpler form for the relaxation time, constructing a Boltzmann Hierarchy becomes straightforward: in terms of the perturbations to the cosmological background DF, the approximate form \eqref{eq:RTA_Definition_CollisionTerm} becomes

\begin{equation}
\left( \frac{\partial f}{\partial t}\right)^{(1)}_{col,l} \approx - \frac{F_l(k, q, \tau)}{\tau_{rel}}
\label{eq:RTA_lth_CollisionTerm}
\end{equation}

\noindent where we have used the momentum averaged relaxation time $\tau_{rel}$ and, by making use of this, the full Boltzmann Legendre expansion is relatively simple:

\begin{equation}
\begin{split}
\dot{F_0} (k, E_q, \tau) \simeq& - \frac{q k}{E_q} F_1 (k, E_q, \tau ) + \frac{\dot{h}}{6} \frac{\partial f_0}{\partial \ln q}  \\
\dot{F_1} (k, E_q, \tau) \simeq& \frac{q k}{3 E_q} F_0 (k, E_q, \tau ) - \frac{2 q k}{3 E_q} F_2 (k, E_q, \tau ) \\
\dot{F_2} (k, E_q, \tau) \simeq& \frac{q k}{5 E_q} \Big[ 2 F_1 (k, E_q, \tau ) - 3 F_3 (k, E_q, \tau ) \Big] - \frac{\partial f_0}{\partial \ln q} \Bigg[ \frac{1}{15} \dot{h} + \frac{2}{5} \dot{\eta} \Bigg] - a \frac{F_2(k, E_q, \tau)}{\tau_{rel}} \\
\dot{F_l} (k, E_q, \tau) \simeq& \frac{q k}{(2l+1) E_q} \Big[ l F_{(l-1)} (k, E_q, \tau ) - (l+1) F_{(l+1)} (k, E_q, \tau ) \Big] - a \frac{F_l(k, E_q, \tau)}{\tau_{rel}} \quad , \quad l \geq 3 \ .
\end{split}
\label{eq:RTA_Hierarchy_Motion}
\end{equation}

%Math Symbols

\subsection{Evolution of $f_0$ and Self Interaction Decoupling}

\label{sec:SID}

%Now: What is the background distribution f_0? Introduction to decoupling

The only missing piece of information regarding the calculation of the relaxation time and its application in a Boltzmann Hierarchy, is the equilibrium distribution function $f_{eq}$. 
We will now discuss  its behavior, how it evolves in time and what are the mechanisms that may alter it. In the general case, the background distribution function is governed by the zero order Boltzmann equation

\begin{equation}
\frac{\partial f_0}{\partial t} - H(t) p \frac{\partial f_0}{\partial p} = \left( \frac{\partial f_0}{\partial t} \right)_{col}^{(0)}
\label{eq:SID_Boltz0_pvars}
\end{equation} 

\noindent where $H$ is the Hubble parameter, $\dot{a}/a$, and for clarity, we have expressed this equation in terms of time $t$ and physical (not comoving) momentum $p \propto a^{-1}$. 
In general, this zero order collision term includes all possible interactions that the DM particle suffers: production, annihilation, scattering, etc.

%General concept of decoupling: mechanism, decoupling from hot plasma. Follow from M&W sec 3.3.5

The process by which a particle loses equilibrium with the hot plasma (``decouples'') has been well studied in the past years, and several reviews are dedicated to this problem \cite{MoBoschWhiteBook,Kolb_Turner_1990}. In equation \eqref{eq:SID_Boltz0_pvars} the two timescales involved are the gravitational (Hubble) timescale and the timescales of all involved interactions $\left\{ \Gamma_{\rm prod}, \Gamma_{\rm annh}, \Gamma_{\rm scatt}, ...\right\}$ for production, annihilation, scattering, etc. 
Whenever the expansion rate of the universe (measured by the Hubble expansion $H$) overcomes the total interaction rate $\Gamma_{\rm tot} = \sum_i \Gamma_i$, such a species gets out of equilibrium with the hot plasma and effectively decouples from the evolution of the universe, its evolution governed only by the cosmological redshift of momenta.

While a given species is in equilibrium with the hot plasma, its distribution function is governed by a Fermi-Dirac or Bose-Einstein function:

\begin{equation}
f_{eq}(\vec{p}, \tau) = \frac{1}{\exp\left[\frac{E(p) - \mu(\tau) }{T(\tau)}\right] \pm 1} \ ,
\label{eq:SID_Equilibrium_FD-BE}
\end{equation}

\noindent where $E$ is the particle energy, $T$ is the temperature and $\mu$ is the chemical potential (equal to zero for Majorana particles). At the time of decoupling, the temperature of the species is approximately equal to the photon temperature, and after this time the particles move in geodesics, with their physical momenta redshifting with the expansion of the universe. 
Thus, their distribution functions evolve as:

\begin{equation}
f(\vec{p}, \tau) \equiv f(\vec{p} \frac{a}{a_f},t_f)\ ,
\label{eq:SID_Equilibrium_Freezeout}
\end{equation}

\noindent i.e., the distribution function is ``frozen-in'' at the value it had at decoupling. In the case that the decoupling happens while the particle is still ultra-relativistic ($T \gg m$), this distribution evolves as:

\begin{equation}
f_{rel}(\vec{p}, \tau) = \frac{1}{\exp\left[\frac{p}{T(t)}\right] \pm 1} \ ,
\label{eq:SID_Equilibrium_Rel}
\end{equation}

\noindent where we have neglected the impact of the chemical potential $\mu$, and we interpret the temperature evolving as $T = T_f a_f / a \propto a^{-1}$. In the case that it decouples \emph{from the hot plasma} while non-relativistic, the types of interactions that are involved must be carefully considered.
For such a species, if it is being kept in equilibrium by non-elastic interactions, its distribution function becomes kinematically suppressed, and it can be expressed as

\begin{equation}
f_{nrel}(\vec{p}, \tau) = \exp\left[-\frac{m}{T_f}\right] \exp\left[-\frac{p^2}{2 m T(\tau)}\right] \ ,
\label{eq:SID_Equilibrium_NRel_Supressed}
\end{equation}
where now $T$ evolves as $T = T_f a_f^2 /a^2 \propto a^{-2}$. 
While these distribution functions can be used to calculate abundances, it is often required to  fully solve the dynamics of freeze-out in equation \eqref{eq:SID_Boltz0_pvars} in order to obtain accurate results. Studying properly the freeze-out mechanism is necessary to calculate the expected thermal abundance of a given DM species, and is often a major component of DM production in the early universe.

%What happens in the case of decoupling of self interactions? Conservation of number density, etc
In this work, however, we will focus on a slightly different type of freeze-out mechanism. Throughout this paper, we will assume that number non-conserving interactions have frozen out deep into the relativistic dominated era and well before the epoch relevant to the evolution of perturbations. 
We will indeed assume that the only relevant interaction in this era is the self interaction of WDM, and here we study this special scenario where the species has decoupled from the plasma long ago and inelastic interactions are no longer relevant.
It is important to clarify that these (elastic) self interactions are considered here as a separate process from any interactions the DM component may have with the plasma. All of these interactions between DM and standard model particles (both elastic and inelastic) will be considered to have decoupled at an early time, while the self interactions may decouple at a time relevant to the evolution of perturbations.

Qualitatively, the main effect of this assumption is that, as self interactions conserve number density, total momentum and energy density (referred as \emph{collisional invariants}), these quantities are effectively ``locked'' since the species gets out of equilibrium with the plasma (or is produced, see \cite{Adhikari2016}). 
While elastic self interactions might change the form of the background distribution function and the evolution of perturbations, they \emph{cannot} alter the collisional invariant quantities.
Furthermore, we will assume that the original production mechanism of the SI-WDM has indeed produced the correct abundance of the species as to be the sole DM component, and leave the discussion of how this might be achieved for future discussions.
%No recoupling
Regarding the decoupling of these self interactions, we assume that at the early times these maintain the SI-WDM component in kinetic equilibrium and, at some point, they would decouple and equilibrium would be lost until today. 

%Remark the difference between T_dec and T_SID, present the separate treatment of the two cases
In order to clarify these concepts, we will introduce the following notation: we will label the decoupling temperature \emph{from the hot plasma}, i.e., the typical ``freeze-out'' temperature, with the standard notation $T_f$. 
The other relevant scale we will refer to is the self interaction decoupling temperature $T_{\rm SID}$, the scale at which $\Gamma_{SI}(T_{\rm SID})=H$, with $\Gamma_{SI}$ referring to the interaction rate of self interactions only. 
According to our assumptions, $T_f$ would have happened at some very high temperature (relative to the scales of Cosmological Perturbation Theory, CPT) and after this moment the only relevant form of interactions are self interactions, which are assumed to decouple at some point during the computation of perturbations: 
$T_f \gg T_{\rm early} \gg T_{\rm SID} \gg T_0$, where $T_{\rm early}$ is the earliest computed time in CPT, and $T_0$ is the temperature of the CMB photons today.
We will distinguish between two cases: when $T_{\rm SID} \gg m$ (relativistic self decoupling) and when $T_{\rm SID} \ll m$ (non-relativistic self decoupling).

\subsubsection*{Relativistic Self Decoupling}

%Classic treatment: what happens with the distribution function, free streaming

In the case that the self decoupling happens while the particle is still relativistic, i.e. $T_{\rm SID} \gg m$, the scenario is rather similar to what happens with no self interactions. 
In this case, after the decoupling from the hot plasma at $T_f$ takes place, the DM gets out of equilibrium from  the rest of the species in the universe. However, as it maintains kinetic equilibrium with itself, it keeps its distribution function in equilibrium according to the DF of a relativistic species \eqref{eq:SID_Equilibrium_FD-BE}:
\begin{equation}
f_0 (p, T_f>T(t)>T_{\rm SID}) \equiv f_{\rm 0,R} = \mathcal{C}_{\rm R} \frac{1}{\exp\left[\frac{p a(t)}{T_{\rm 0,R}} \right] \pm 1} = \mathcal{C}_R \frac{1}{\exp\left[\frac{q}{T_{\rm 0,R}} \right] \pm 1} \ ,
\label{eq:SID_SelfCoupledEquilibrium_Rel}
\end{equation}
\noindent where $p$ is the local momentum, $q$ is the comoving momentum $q = a p$, $T_{\rm 0,R}$ is the temperature the species would have today if it maintains a relativistic DF, and we have introduced the normalization factor $\mathcal{C}_{\rm R}$ in order to enforce the proper abundance today\footnote{ $\mathcal{C}_{\rm R} $ is a phenomenological   parameter introduced to account for non-thermal production mechanisms by matching the corresponding collisional invariants as described below.  $\mathcal{C}_{\rm R} =1$  if the production is thermal.}. 
Here, the temperature $T$ is to be interpreted to evolve as $T \propto a^{-1}$, and the whole distribution function can be cast in terms of the temperature today $T_{\rm 0,R}$ by using comoving quantities. 
It is important to note that as this species is evolving accordingly to its own temperature, it is not affected by the entropy release in the hot plasma as other species decouple. 

When the self interactions finally decouple at $T_{\rm SID}$, the specie's temperature is frozen out at this temperature and the evolution of the DF becomes governed by the redshift of momenta according to \eqref{eq:SID_Equilibrium_Rel}. This, incidentally, gives the same form for the background DF as in \eqref{eq:SID_SelfCoupledEquilibrium_Rel}, yet maintaining it frozen in this ultra relativistic form even when $T<m$.

\subsubsection*{Non-Relativistic Self Decoupling}

%What happens now? non-relativistic Distribution Function!
The situation becomes drastically different in the case $T_{\rm SID} \ll m$ and self interactions remain coupled through the non-relativistic transition, $T=m$. 
In this case, the situation is the same as in the relativistic self decoupling as long as $m < T \ll T_f$: the species maintains its own temperature $T_{\rm 0,R}$ but remains effectively decoupled from the rest of the hot plasma. As soon as the temperature approaches the particle's mass, self interactions cannot maintain equilibrium in the distribution function anymore, as the equilibrium distribution itself becomes ill-defined. 

Once the (pseudo, see \cite{Hofmann_2001}) temperature of the species falls well below the particle's mass $m \gg T \gg T_{\rm SID}$, self interactions are able to restore equilibrium into the DM species, but now fall into a \emph{non-relativistic} distribution function, as in \eqref{eq:RTA_Equilibrium_General}, 
\begin{equation}
f_0 (p, m > T(t) > T_{\rm SID}) \equiv f_{\rm 0,NR} = \mathcal{C}_{\rm NR} \exp\left[-\frac{p^2 a^2(t)}{2 m T_{\rm 0,NR}}\right] = \mathcal{C}_{\rm NR} \exp\left[-\frac{q^2}{2 m T_{\rm 0,NR}}\right] \ ,
\label{eq:SID_SelfCoupledEquilibrium_NRel}
\end{equation}
with $\mathcal{C}_{\rm NR}$ some appropriate normalization and $T_{\rm 0,NR}$ its temperature today (generally, different from the temperature $T_{\rm 0,R}$ for the relativistic decoupling case). Given the properties of the self interactions, this distribution function would be considerably different from \eqref{eq:SID_Equilibrium_NRel_Supressed}. 
Since collisional invariance protects the number density, velocity and energy density, this new function should preserve these quantities through the non-relativistic transition, fixing the values for temperature and normalization once the DF at $T_f$ is specified.

After self interaction decoupling, the evolution of this DF should follow from the redshift of momenta according to \eqref{eq:SID_Equilibrium_Freezeout}, with the particles moving along a geodesic. 
This ultimately results in this distribution function being frozen out in this same form even after $T_{\rm SID}$, making the form \eqref{eq:SID_SelfCoupledEquilibrium_NRel} valid even for $T<T_{\rm SID}$ in the case the species is coupled through $T=m$. 

\subsubsection*{Transition Approximations and Contour Conditions}

%Present where do we set the transitions
Now, we turn to the question of how to properly specify the conservation conditions for these collisional invariants. 
We will not consider overall momentum: while it is conserved, we assume traslational invariance for the background distribution function. Instead, we will focus ourselves in other conserved quantities in order to specify the non-relativistic distribution function. 

In the case of number density, it is safe to assume that it is conserved throughout the non-relativistic transition. 
As it evolves due to the cosmic expansion as $n \propto a^{-3}$ regardless of the decoupling scenario, it is enough to impose that the total number density for this new distribution function today \eqref{eq:SID_SelfCoupledEquilibrium_NRel} to be the same as the relativistic version \eqref{eq:SID_SelfCoupledEquilibrium_Rel}, valid only in $T_f \gg T \gg m$ but extrapolated until today. Thus, we have our first contour condition

\begin{equation}
n_{\rm 0,R} = \mathcal{C}_{\rm R} \frac{3}{4} \frac{\zeta (3)}{\pi^2} g m T_{\rm 0,R}^3 = \mathcal{C}_{\rm NR} g m \left( \frac{T_{\rm 0,NR} m}{2 \pi}\right)^{3/2} = n_{\rm 0,NR} \ .
\label{eq:SID_CC_nCondition}
\end{equation}

%Present the two types of contour conditions

For the next condition, we would have to rely on the conservation of the energy density $\rho$. 
Ideally, the evolution of the energy density should be identical in this case to the case where there are no self interactions at all, as the species only interacts with itself through elastic collisions.
Indeed if we demand the condition that the number densities follow the same evolution for the two distribution functions (relativistic and non relativistic), it can be readily shown that this condition is automatically fulfilled. 
In order to determine both temperature and normalization, we would have to make a few assumptions about how particular thermodynamical quantities evolve during the non-relativistic transition itself. Here, we will present three sets of assumptions, which all lead to variations of the same result. 
 
%microncanonical
For the first set of contour conditions, we will take a look at the evolution of the energy density. We make the (rough) approximation of extrapolating the expression for energy density in the ultra relativistic regime (with a relativistic DF, temperature $T_{\rm R}(t)$ and normalization $\mathcal{C}_{\rm R}$) and the energy density in the non-relativistic regime (with a non-relativistic DF, temperature $T_{\rm NR}(t)$ and normalization $\mathcal{C}_{\rm NR}$) into the exact moment of the non-relativistic transition $T=T_{\rm R}=T_{\rm NR}=m$, and impose continuity in the energy density. 
So, we impose the condition $\rho_{\rm R}(T_{\rm R} = m) = \rho_{\rm NR}(T_{\rm NR} = m)$ at the transition temperature, with $\rho_{\rm R}$ the relativistic limit of the energy density and $\rho_{\rm NR}$ the non-relativistic limit. This condition, together with the condition $n_{\rm 0,R} = n_{\rm 0,NR}$ will be our first set of contour conditions, which we will name the \emph{microcanonical} contour conditions.

%canonical
For the next two sets of contour conditions, we will perform similarly, but considering different thermodynamic variables.
First, if we impose continuity in the temperature between the two different distribution functions, we end up with the condition $T_{\rm 0,NR}=T_{\rm 0,R}^2/m$. Together with the condition on number density today, we name this set as the \emph{canonical} contour conditions. 
%isoentropic
In the case of the final set, we take a look at the entropy density, calculated as $s=(\rho+p)/T$. As before, this case constitutes the \emph{isentropic} contour condition. We summarize the results for all the different contour conditions in \cref{tab:Contour_Conditions}.

\begin{table}
\centering
\begin{tabular}{c||c|c} 
\toprule
$f_0 (T<m) = \mathcal{C}_{\rm NR} \exp \left[ - \alpha q^2/ T_{\rm 0,R}^2 \right ]$                                             & $\mathcal{C}_{\rm NR}/\mathcal{C}_{R}$                          & $\alpha$                                                                              \\ 
\hline\hline
\begin{tabular}[c]{@{}c@{}}\textit{Micro-Canonical}\\$n_{\rm 0,R} = n_{\rm 0,NR}$ , $\rho_{\rm R}(T=m)=\rho_{\rm NR}(T=m)$\end{tabular} & $\frac{7}{8} \frac{\pi^2}{30} (2\pi)^{3/2} \sim$ $4.534$        & $\frac{1}{2} \left( \frac{180}{7} \frac{\zeta (3)}{\pi^4}\right)^{3/2} \sim$ $1.075$  \\ 
\hline
\begin{tabular}[c]{@{}c@{}}\textit{Canonical}\\$n_{\rm 0,R} = n_{\rm 0,NR}$ , $T_{\rm R}(T=m)=T_{\rm NR}(T=m)$\end{tabular}             & $\frac{3}{4} \frac{\zeta (3)}{\pi^2} (2\pi)^{3/2} \sim$ $1.439$ & $1$                                                                                   \\ 
\hline
\begin{tabular}[c]{@{}c@{}}\textit{Isoentropic}\\$n_{\rm 0,R} = n_{\rm 0,NR}$ , $s_{\rm R}(T=m)=s_{\rm NR}(T=m)$\end{tabular}           & $\frac{7}{360} \pi^2 (2\pi)^{3/2} \sim$~$ 3.022$                & $\frac{\pi^{8/3}}{18}\left( \frac{10}{7} \zeta (3) \right)^{3/2} \sim$ $0.8201$       \\
\bottomrule
\end{tabular}
\caption{Summary of results for the distribution functions at $T<m$ for the different contour conditions mentioned in section \ref{sec:SID}. These distributions are characterized by the factors $\alpha$ and $\mathcal{C}_{\rm NR} / \mathcal{C}_{\rm R}$}
\label{tab:Contour_Conditions}
\end{table}

%Present the problem of interpolation! Mention following sections
These contour conditions are set in order to avoid the complicated non-equilibrium dynamics involved in the transition to the non-relativistic coupled regime. 
A more accurate treatment can be obtained by fully solving the problem and including the results of a full zero-order Boltzmann solution to the DF \eqref{eq:SID_Boltz0_pvars} considering a full expression for the collision term. 
Even when we choose to avoid this problem altogether by making use of these conditions, the question of what sort of background $f_0$ to use in the regime $T \sim m$ still holds. We have studied a phenomenological approach to this, by interpolating between the two known solutions, and will come back to this question in \cref{sec:CLASS_NR}.

\subsection{Limiting forms for the Relaxation Time}

\label{sec:SID_Limiting}

Finally, let us obtain limiting expressions for the relaxation time $\tau_{rel}$ both in the deep relativistic and non-relativistic regimes. 
In the relativistic case, we can make use of the results in \cite{GondoloGelmini91}, where a compact expression for the thermal average $\left< \sigma v_{mol} \right>$ \eqref{eq:RTA_TauDefinition_d3qSigmaV} is given assuming a background DF $f_0 \propto \exp \left[ -E/T \right]$, where $T$ is the temperature \emph{of the species}. In this case, this quantity reduces to the following form, in terms of the kernel $\chi(s)$:
\begin{equation}
\left< \sigma v_{mol} \right> = \frac{1}{4 m^4 T K_2^2(m/T)}\int_{4m^2}^{\infty} ds \chi (s) \frac{ g_i^2}{4} \sqrt{s-4 m^2} K_1 (\sqrt{s} / T ) \ ,
\label{eq:SID_Limiting_GG}
\end{equation} 
where $K_i$ are the modified Bessel functions, and we have corrected an error in the prefactors in \cite{Yunis2020b} due to differences in the normalization of $\chi (s)$. 
While in \cite{GondoloGelmini91} this expression is always valid, in our case it only applied to the relativistic case.
Indeed, in the non-relativistic limit (in both the relativistic and non-relativistic self decoupling scenarios) the background DF is \emph{not} well approximated by $\exp \left[ -E/T \right]$, which carries the kinematic suppression factor in \eqref{eq:SID_Equilibrium_NRel_Supressed}. For this case, we take the limit of equation \eqref{eq:RTA_TauDefinition_d3qSigmaV} in both the relativistic and non-relativistic self decoupling scenarios.

First, we start by reducing the form of the kernels $\chi$ in the non-relativistic limit. In this case, we take $s \rightarrow m^2(4+v_r^2)$ with $v_r^2 = (\vec{v}_q-\vec{v}_l)^2$ the relative velocity of incoming particles. In this limit, the kernel $\chi$ takes the form:

\begin{equation}
\begin{split}
\chi \left[ s \rightarrow m^2 ( 4 + v_r^2 ) \right]&= \frac{C_t^0 v_r}{2}+\frac{1}{16} v_r^3 \left(4 B_t^0 m^2-C_t^0+8 C_t^1\right) \\
&+\frac{1}{768} v_r^5 \left(128 A_t^0 m^4-24 B_t^0 m^2+192 B_t^1 m^2+9 C_t^0-48 C_t^1+384 C_t^2\right)+\mathcal{O}\left(v_r^7\right) \ ,
\end{split}
\label{eq:SID_Limiting_chiNR}
\end{equation}

\noindent where we have separated $C_t = \sum_i C_t^i v_r^{2i}$ in powers of $v_r^2$, and the same for $B_t$ and $A_t$. Then, we can take the non-relativistic limit in equation \eqref{eq:RTA_TauDefinition_d3qSigmaV} \cite{Lee_Weinberg_1977}, where $v_{mol}$ is now replaced by $v_r$ which reads, in terms of $\chi(s)$:

\begin{equation}
\left< \sigma v_{mol} \right>_{\rm NR} = \frac{\int d^3v_1 \, d^3v_2 \, \sigma v_r \, f_{eq}(v_1) \, f_{eq}(v_2)}{\int d^3v_1 \, d^3v_2 \, f_{eq}(v_1) \, f_{eq}(v_2)} \equiv \frac{\int d^3v_1 \, d^3v_2 \,  \chi \left[ s \rightarrow m^2(4+v_r^2) \right] g_i^2/4   \, f_{eq}(v_1) \, f_{eq}(v_2)}{4 m^2 \int d^3v_1 \, d^3v_2 \, f_{eq}(v_1) \, f_{eq}(v_2)} \ .
\label{eq:SID_Limiting_sigmaVNR_General}
\end{equation}

In the case of the relativistic self decoupling scenario, this would be a distribution matching the one of a relativistic, decoupled particle species. 
If we assume a Maxwell-Boltzmann form (ignoring quantum effects in $f_0$, in this case), the integral \eqref{eq:SID_Limiting_sigmaVNR_General} can be carried out directly

\begin{equation}
\begin{split}
\left< \sigma v_{mol} \right>_{\rm NR}^{\rm R-SID} &= \frac{35}{16 m^2} \Bigg( \frac{C_t^0}{2} \left( \frac{T}{m} \right) + \frac{9}{4} \left( 4 B_t^0 m^2-C_t^0+8 C_t^1 \right) \left( \frac{T}{m} \right)^3 \\
& + \frac{99}{32} \left( 128 A_t^0 m^4-24 B_t^0 m^2+192 B_t^1 m^2+9 C_t^0-48 C_t^1+384 C_t^2\right) \left( \frac{T}{m} \right)^5 \Bigg) \\
& + \mathcal{O}\left(T/m\right)^7
\end{split}
\label{eq:SID_Limiting_sigmaVNR_RDec}
\end{equation} 

In the case of a non-relativistic distribution function, the integral can also be carried out directly: following the results in \cite{Cannoni_2014}, we reach the following expression

\begin{equation}
\begin{split}
\left< \sigma v_{mol} \right>_{\rm NR}^{\rm NR-SID} &= \frac{1}{\sqrt{\pi}m^2}\Bigg( \frac{C_t^0}{2}\left(\frac{T}{m}\right)^{1/2} + \frac{1}{2} \left(4 B_t^0 m^2-C_t^0+8 C_t^1\right) \left( \frac{T}{m} \right)^{3/2}\\
&+\frac{1}{8} \left(128 A_t^0 m^4-24 B_t^0 m^2+192 B_t^1 m^2+9 C_t^0-48 C_t^1+384 C_t^2\right) \left( \frac{T}{m} \right)^{5/2} \Bigg)\\
&+ \mathcal{O}\left(T/m\right)^{7/2}
\end{split}
\label{eq:SID_Limiting_sigmaVNR_NRDec}
\end{equation}

\begin{figure}
\centering
\includegraphics[width=\textwidth]{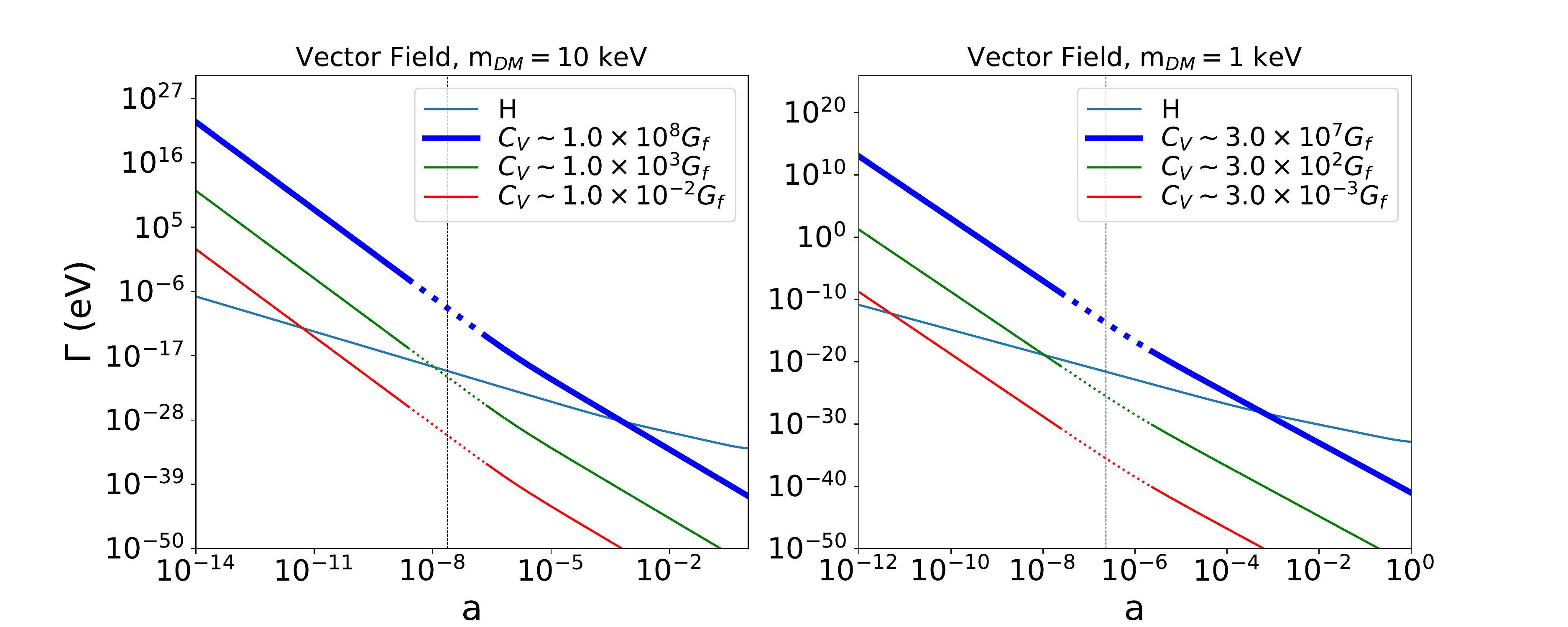}
\caption{\small
Self Interaction rate for a vector field SI-WDM model for two values of the DM particle mass: $1$~keV (\emph{right panel}) and $10$~keV (\emph{left panel}), compared to the Hubble expansion rate $H$ as a function of scale factor, considering the effects of non-relativistic self decoupling. 
Dotted lines show the intermediate regime $0.1 m_{\rm DM} < T < 10 m_{\rm DM}$, where none of the approximate forms described in \cref{sec:SID_Limiting} are used and instead the interpolating form \eqref{eq:SID_Limiting_interp} is used. 
The interaction constant $C_V = g_V^4 m_{V}^{-4} \cos^{-4} \theta_W'$ (with $g_V$ the Lagrangian coupling constant, $m_V$ the mediator mass and $\theta_W'$ the dark sector Weinberg angle) corresponding to the Bullet Cluster upper limit constraints (see \cite{amrr}) are shown in thick lines, and vertical lines label $T_\gamma=m$.
}
\label{fig:SID_RelaxationTime_vs_H}
\end{figure}

We can see the evolution of the relaxation time for, for example, a vector field case in \cref{fig:SID_RelaxationTime_vs_H}. 
There we compare the self interaction rate $\Gamma$ with the Hubble rate $H$ to give a more appropriate idea of the decoupling process: whenever the (self) interaction rate falls below the Hubble rate, the species can be considered as (self) decoupled, thus having lost kinetic equilibrium (with itself). 
For this case, we can see from both the figure and the limiting expression \eqref{eq:SID_Limiting_sigmaVNR_NRDec} a well known behaviour (e.g. \cite{Hannestad00}) of the thermal averaged cross section, behaving as $\left< \sigma v_{mol} \right> \propto T^{1/2}$, but remarkably only when the background distribution function is non relativistic, which is not always the case for a self interacting species.
In any case, in both scenarios the interaction rate behaves as $\Gamma \propto a^{-4}$ with the scale factor.

Both in the deep relativistic as well as in the non-relativistic regimes, the expressions \eqref{eq:SID_Limiting_sigmaVNR_NRDec}, \eqref{eq:SID_Limiting_sigmaVNR_RDec} and \eqref{eq:SID_Limiting_GG} are used along this paper to calculate the relaxation time and the interaction rates for all mediator models. 
However, in the intermediate regime, where none of these approximations are valid, we are left without any straightforward calculation method for those quantities. Thus, we choose to calculate the interaction rate in the intermediate regime as an interpolation between the relativistic and non-relativistic regimes with the form:

\begin{equation}
\Gamma^i \simeq \frac{1}{2} \left( {\rm P}(\Gamma_{\rm NR}^i) + {\rm P}(\Gamma_{\rm R}^i) \right)
\label{eq:SID_Limiting_interp}
\end{equation}

\noindent with $\rm P$ indicating a power-law fit and $\Gamma_{\rm R}^{i}$, $\Gamma_{\rm NR}^{i}$ indicating the relativistic and non-relativistic approximations of the interaction rates for model $i$, respectively (valid for both non-relativistic and relativistic self decoupling modes). 
This approximation provides good results in the case of massive scalar and vector field mediators and, while it is a rough approximation to the true relaxation time, it is a subdominant component of the overall error in our calculations introduced by other approximations.

Now, with all of the ingredients ready (the relaxation time approximation of the Boltzmann Hierarchies, a study on the evolution of the background $f_0$ and a way of computing the relaxation time $\tau_{rel}$), we are set to implement the SI-WDM hierarchies in a Boltzmann solver. 
In the next section, we will see the details regarding the implementation of this species in CLASS, and present the main results for this work.

\section{Numerical Implementation in CLASS}

\label{sec:CLASS}

%We did all RTA in CLASS version 2.7.2, free distribution links?

We implemented the Relaxation Time Approximation \eqref{eq:RTA_Hierarchy_Motion} in the public Boltzmann solver CLASS (version 2.7.2), as a modification to the existing module handling WDM and non cold relics with the objective of assessing the impact of SI-WDM on the power spectrum and on CMB anisotropies
\footnote{We provide a free distribution link to this modified version of CLASS in \href{https://github.com/yunis121/siwdm-class}{github.com/yunis121/siwdm-class}, requiring a pre-existing installation of CLASS v2.7.2 (which we also provide in the repository). This particular version does not precompute the relaxation time within CLASS, but is instead supplied with a relaxation time table as an external file.}.
To this end, we have modified the code appropriately in order to handle the thermal history outlined in \cref{sec:Boltz} for this new species and developed or adapted a few approximations and procedures to provide fast and accurate results, which we describe in this section.

%Momentum discretization does not affect RTA
The CLASS Boltzmann solver introduced a complex method of optimized momentum sampling in order to deal with non cold relics, such that it maximizes the precision in the integrals of the perturbations of the form $q^n F_l(k, E_q, \tau)$ with $n=2,3,4$ and $l=0,1,2$ (see \cite{CLASSIV}, sec. 5 for more details). 
As the approximated Boltzmann hierarchy \eqref{eq:RTA_Hierarchy_Motion} does not couple different momentum bins, this optimization method remains unchanged in this version of SI-WDM implementation, except for a few minor modifications in the non-relativistic self decoupling scenario which we describe in \cref{sec:CLASS_NR}.

%MODULE: Background module:
%			-Decoupling: thermalization + mention NR decouping
Once the relaxation times are already pre-computed, the thermal history for this species is calculated according to what was outlined in \cref{sec:Boltz}. 
If this thermal history is met (the species is self coupled at the earliest computed time, and decouples at some point afterwards), a few procedures are put in order to obtain self consistence. As the program can be initialized under any arbitrary background distribution function, a few steps are taken in order to 
i) thermalize the distribution: i.e. if a non-equilibrium distribution is given, construct an equilibrium distribution that matches its number and energy densities, according to

\begin{align}
n_{eq} = \mathcal{C}_{\rm R} \left(\frac{T_{0}^{eq}}{a}\right)^3 F_2(0) &= 4 \pi \left( \frac{T_{0}^{ncdm}}{a} \right)^3 \int_0^{\infty} dq q^2 f_0(q) = n_{init} \\
\rho_{eq} = \mathcal{C}_{\rm R} \left(\frac{T_{0}^{eq}}{a}\right)^4 F_3(0) &= 4 \pi \left( \frac{T_{0}^{ncdm}}{a} \right)^4 \int_0^{\infty} dq q^2 f_0(q) \epsilon(q,m) = \rho_{init}
\label{eq:CLASS_Intro_nrhoConsistency}
\end{align} 

\noindent where $T_{0}^{eq} \ , \ T_{0}^{ncdm}$ are the equilibrium and initial temperatures of the species today, 
and ii) re-normalize the distribution function in order to match the desired abundance today (a function already implemented in standard CLASS for non cold relics). 
These steps are taken in order to fulfill the assumptions we have taken in \cref{sec:Boltz} and reflect the expected effects of self interactions at early times. 

Once these self consistency steps are taken, the program further checks what decoupling scenario it should implement, whether that may be relativistic or non-relativistic self decoupling. 
As the distribution (in comoving variables) is not modified in the first scenario, the evolution of $f_0$ continues the same way as for standard WDM, but for the non-relativistic decoupling scenario a few extra steps are taken, described in \cref{sec:CLASS_NR} and according to what is discussed in \cref{sec:SID}.

%Present perturbations module! Introduce Tight Coupling
Once the evolution of $f_0$ and the decoupling scenario is specified, the evolution of perturbations $F_i$ in the RTA \eqref{eq:RTA_Hierarchy_Motion} can be implemented, based on the existing implementation of non cold relics in CLASS \cite{CLASSIV}. 
The hierarchy \eqref{eq:RTA_Hierarchy_Motion} is implemented by adding a momentum independent collision term to the existing WDM hierarchies, with a time dependent relaxation time calculated as in \cref{sec:SID_Limiting}. 
As in \cite{CLASSIV,Ma1995}, the infinite hierarchy is closed at a finite $l_{\rm max}$ (typically between 15 and 40 in this application) using a closing assumption for $F_{l_{\rm max}+1}$ \cite{Ma1995}.
Afterwards, the steps regarding the calculation of the angular power spectrum of CMB anisotropies, $C_l$'s, as well as obtaining the linear power spectrum today follow directly from the WDM implementation. 
However, the timescale $\tau_{rel}$ of this new interaction term can overcome the other phenomenological timescales of the system $1/\mathcal{H} = a/\dot{a} $, $\tau_k=\epsilon/qk$, and thus suppressing high-$l$ multipoles while warranting a Tight-Coupling approximation (TCA), as described in \cite{Oldengott2017} and extended here to SI-WDM in \cref{sec:CLASS_TCA}.

\subsection{The Tight-Coupling Approximation}

\label{sec:CLASS_TCA}

%TCA: what it is,  mention timescales
The Tight Coupling Approximation, described in \cite{Oldengott2017}, holds whenever the timescale of self interactions, in this case $\tau_{rel}$ dominates over all of the other timescales of the system. 
For the Boltzmann hierarchies \eqref{eq:RTA_Hierarchy_Motion}, these would be the conformal Hubble rate $1/\mathcal{H}= a/\dot{a}$ related to the metric perturbations and the combination $\tau_k = \epsilon/qk$ related to the streaming of particles to higher $l$ moments. 

In the case TCA holds, the interaction terms have the effect of strongly suppressing the perturbations $l>1$ as the term $C_l[F] = -a \Gamma(\tau) F_l(k, q, \tau)$ dominates. 
Thus, the species becomes essentially a perfect fluid with zero shear stress, and the only non trivial equations for the system become the ones corresponding to the $l=0,1$ moments. 
The proper implementation is especially important in the case of numerical integration, as the strong suppression of high l-moments that takes place in this regime can result in a numerically ``stiff'' system.

\begin{figure}[htb]
\centering
\includegraphics[width=0.8\textwidth]{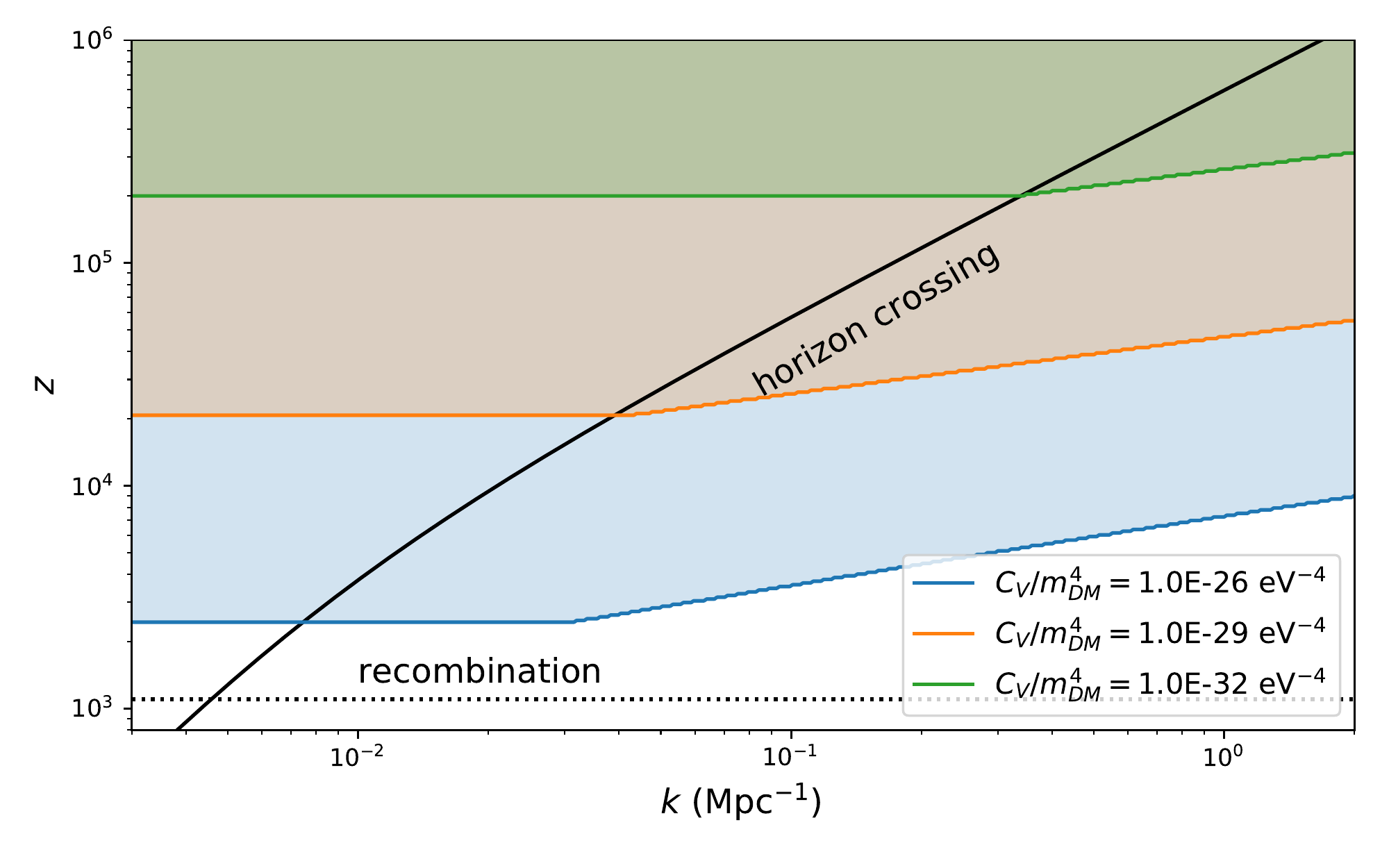}
\caption{\small 
Regions of validity of the Tight-Coupling approximation for different values of the coupling constant, in the case of a vector field mediator for $m_{DM} = 10\,\mathrm{keV}$. 
The black line represents the moment when a mode k undergoes horizon crossing ($k=aH$). For each $C_V$, the TCA is valid within their respective colored regions. 
The colored lines represent the moment when TCA stops being valid, i.e. $\tau_{rel}^{-1} \equiv a \Gamma = \mathrm{max}(aH,(qk/\epsilon)\rvert_{q=1})$. 
We have used, for simplicity, $\Gamma \sim C_V T_{ncdm}^5$ \cite{Oldengott2017}.
}
\label{fig:TC_Validity}
\end{figure}

%PLOTS
While the validity of TCA is in this case with $m \neq 0$ and dependent on the momentum bin considered, for the sake of simplicity we take the approximation as valid as long as it holds for $q \sim 1$. 
We can see in \cref{fig:TC_Validity} the regions of validity of the TCA approximation in the $k,z$ plane. 
As was pointed out in \cite{Oldengott2017}, given a fixed interaction coupling constant $G_{\rm eff}$ the TCA always fails at large wavenumbers, then starts failing later in time for progressively smaller $k$ until it formally fails for all $k$ at the same time. 

\subsubsection*{High l-moments in TCA}

%Describe the higher moment calculation: stiffness and possible precision gain?
%		About precision gain: possible negligible, but allows for 
%		lower tolerances on an inexpensive way

A further refinement on the tight coupling approximation can be achieved by making use of the suppressed moments on $l \geq 2$. If we assume the rate of variation of the interaction term is much higher than the variation of other source terms, the equations for these high $l$-moments in the TCA regime can be approximated as:

\begin{equation}
\begin{split}
\frac{\Psi}{\tau_{rel}} &\sim \frac{qk}{5\epsilon} (2\Psi_1 - 3\Psi_3) - \left( \frac{1}{15} \dot{h} + \frac{2}{5} \dot{\eta} \right) \frac{d \ln f_0}{d \ln q} \ ,\\
\frac{\Psi_{l \geq 3}}{\tau_{rel}} &\sim \frac{qk}{(2l+1)\epsilon} [l\Psi_{l-1} - (l+1)\Psi_{l+1}] \  .
\end{split}
\label{eq:CLASS_TCA_highMoments}
\end{equation}

\noindent where $\Psi_l(k,q,\tau) = F_l(k,q,\tau)/f_0(q,\tau)$. 
By making use of the closing assumption for $\Psi_{l_{max}+1} \approx \Psi_{l_{max}}(2 l_{max} + 1 ) \epsilon / (qk\tau) - \Psi_{l_{max}-1} $ (see \cite{Ma1995}) and only using the value of the first moment in the hierarchy (obtained through the Boltzmann solver), this system can be inexpensively solved by tridiagonal matrix solving algorithms \cite{Higham2002}. 
Thus, a value for $F_2$ (suppressed by a factor of $\tau_{rel}$) can be included back on the evolution equation for $F_1$. This has the advantage of increasing the precision and providing extra numerical stability as the TCA approaches the limit of its validity range, allowing us to be less reliant on the precise region of validity of TCA at a marginal computational cost.

\subsection{Non-Relativistic Decoupling}

\label{sec:CLASS_NR}

%Describe the non-relativistic decoupling module! In case of detection above...
%		-Calculate the interpolator between Non rel and Rel functions
%		-Introduce an extra time dependent factor on background f_0 calculations
%		-Re-Optimize integrators to deal with both functions
%		-Replace calls to f_0 and dlnf0/dlnq in perturbations with new function, uses interpolation  in the middle and precomputed values in the limiting cases
Once the background evolution for $f_0$ (described at the beginning of this section) has been calculated, we are in the position of discerning between the two scenarios described in section \cref{sec:SID}. In the case that non-relativistic self decoupling is detected, a few extra steps are taken in order to implement this scenario in the evolution of the perturbations and background quantities.

%Interpolator
First, we tackle the problem of specifying the new background distribution itself. Once the contour conditions are specified (from the ones we calculated in \cref{sec:SID}), we have both limiting cases for the background distribution function: at $T \gg m$ we have $f_{\rm 0,R} \sim (\exp[q/T] \pm 1)^{-1}$ \eqref{eq:SID_SelfCoupledEquilibrium_Rel} and when $T \ll m$ we have $f_{\rm 0,NR}  \sim [-q^2/2mT]$ \eqref{eq:SID_SelfCoupledEquilibrium_NRel}. 
However, in the intermediate regime, the distribution function still has to be specified: as a first approximation we choose to interpolate between these two regimes using the following function:

\begin{equation}
f_{\rm Sw}(q, T/m) = {\rm Sw} (T/m) f_{\rm 0,NR}(q) + \left(1 - {\rm Sw} (T/m) \right) f_{\rm 0,R} (q) \ ,
\label{eq:CLASS_NR_f0Sw}
\end{equation}

\noindent where the switching function ${\rm Sw}(x)$ is defined as:

\begin{equation}
{\rm Sw}(x) = \frac{1}{2}{\rm erfc}\left[\log (x) \right] \ ,
\label{eq:CLASS_NR_SwDef}
\end{equation}

\noindent and ${\rm erfc(x)}$ is the complementary error function.

%Intercepted calls + precomputing
When the non-relativistic decoupling scenario is specified, all function calls to the background distribution function can be replaced by this new function. This includes both calls to $f_0$ itself and the logarithmic derivative $\partial \ln f_0 / \partial \ln q$ that appears in the Boltzmann hierarchy \eqref{eq:Boltz_Formal_F_motion}. 
In practice, we also precalculate values for $f_0(q_i)$ and $\left[ \partial \ln f_0 / \partial \ln q \right] (q_i)$, where $q_i$ are the momentum samples, in the deep relativistic and non-relativistic regimes and replace the function evaluations for these precomputed values whenever appropriate.

%Evolution of Perturbations through the transition

An interesting question is how to evolve the perturbations through the relativistic to non-relativistic transition itself. 
As it was the case for the background distribution function, the problem of transitioning between the relativistic and non relativistic regimes appears in the perturbations too.
Ideally, a full approach to the Boltzmann hierarchies, complete with moment dependent collision terms such as the ones implemented in \eqref{eq:Boltz_Formal_F_motion} would naturally include 
this transition, and no extra care would be needed.
However, when using an approximate framework such as the RTA, it is important to consider this transition separately. 

One simple approach to this would be to simply assume that the RTA hierarchies \eqref{eq:RTA_Hierarchy_Motion} are valid during the whole evolution, particularly during this transition. In that case, the perturbation moments $F_l(q,k,\tau)$ would evolve without changes through the transition.
However, as the background distribution gets suppressed severely during the transition for high $q$, the ratio $F/f_0 \lvert_{q \gg 1}$ would greatly increase and the perturbations would leave the linear regime. 
This is not representative of the behavior of these relaxation processes, as we expect the interactions that transfer phase space density to lower momenta at this stage to both affect background and perturbation.
This phase space density transfer from high to low momenta cannot be described through the RTA, as it is local in momentum space, making this simple approach a poor choice.
We instead opt for another way of accounting for these effects that does not imply solving the full collision integrals: we set the background-scaled perturbation $\Psi_l = F_l / f_0$ to be constant through the transition, as a straightforward approximation to this complex process. 

%Integration of background quantities
Special care needs to be taken in the case of computing the integrals resulting in the energy-momentum perturbations. These quantities are obtained via integration rules of the form

\begin{equation}
\mathcal{I} =\int_0^{\infty} dq f_0(q) g(q) \approx \sum_{i=1}^n W_i g(q_i)
\label{eq:CLASS_NR_integralRuleDef}
\end{equation}

\noindent where the integrated function can be either $g(q) = \hat{g}(q)$ describing a background quantity or $g(q) = \hat{g}(q) \Psi_i(q, \tau)$ describing an energy-momentum perturbation. 
The weights $W_i$ are optimized via a complex method (see \cite{CLASSIV}) in order to obtain maximum precision in the calculation of these integrals for $\hat{g}(q)=\{q^2,q^3,q^4\}$ (describing energy, momentum, pressure and shear integrals). 
In order to take advantage of the methods already implemented in CLASS, but extend it to this time dependent background DF we choose to re-define the integration rule \eqref{eq:CLASS_NR_integralRuleDef} as:

\begin{equation}
\sum_{i=1}^n W_i g(q_i) \rightarrow \sum_{i=1}^{n'} \left( \frac{f_{\rm Sw} (q_i, T/m)}{f_{\rm 0,R}(q_i)} \right) W_i' g(q_i) \ ,
\label{eq:CLASS_NR_weigthsRedef}
\end{equation}

\noindent where the weights $W_i'$ are now optimized in order to obtain the desired precision in the function

\begin{equation}
t(q) = a_2 q^2 + a_3 q^3 + a_4 q^4 + b_2 \left( \frac{f_{\rm 0,NR}}{f_{\rm 0,R}} \right) q^2 + b_3  \left( \frac{f_{\rm 0,NR}}{f_{\rm 0,R}} \right) q^3 + b_4  \left( \frac{f_{\rm 0,NR}}{f_{\rm 0,R}} \right) q^4
\label{eq:CLASS_NR_NewTestFunction}
\end{equation}

\noindent and the coefficients are chosen such as

\begin{equation*}
a_n \int dq \frac{q^n}{e^q+1}=1 \quad , \quad b_n \int dq q^n e^{-q^2} = 1 \ .
\end{equation*}

As before, we precompute the values for $( f_{\rm Sw} (q_i, T/m) / f_{\rm 0,R}(q_i) )$ in the deep non-relativistic limit and use them appropriately. 

With this machinery in place, it is possible to compute the evolution for SI-WDM under the RTA approximation \eqref{eq:RTA_Hierarchy_Motion} and obtain desired observables such as CMB perturbations and power spectrum, among others. We will further explore the results of this extension to CLASS in \cref{sec:Obs}, in the light of the particle models we describe in \cref{sec:Intro}.

\subsection{Effects on the Matter Power Spectrum}

\label{sec:CLASS_PS}

Having developed this extension to CLASS in order to implement the evolution of SI-WDM, we take a first look in this section to the cosmological results in the linear regime. We show some of the main features that these alternative DM models introduce in the power spectrum, in order to discuss observable effects in section \ref{sec:Obs}.
One of the most noticeable effects is a suppression on the matter power spectrum at high $k$. We compare here the suppression produced by SI-WDM to the one produced by other alternative models, such as  WDM or SIDM models.
Typically, both the tightest constraints and the alleviation of the small scale tensions mentioned in \cref{sec:Intro} have been related to this suppression in the power spectrum for the case of WDM \cite{Bullock2017,Viel2013}, therefore in this work we will focus on it and leave the analysis of other quantities such as the $C_l$'s to future works.

\begin{figure}[htb]
\centering
\includegraphics[width=0.95\textwidth]{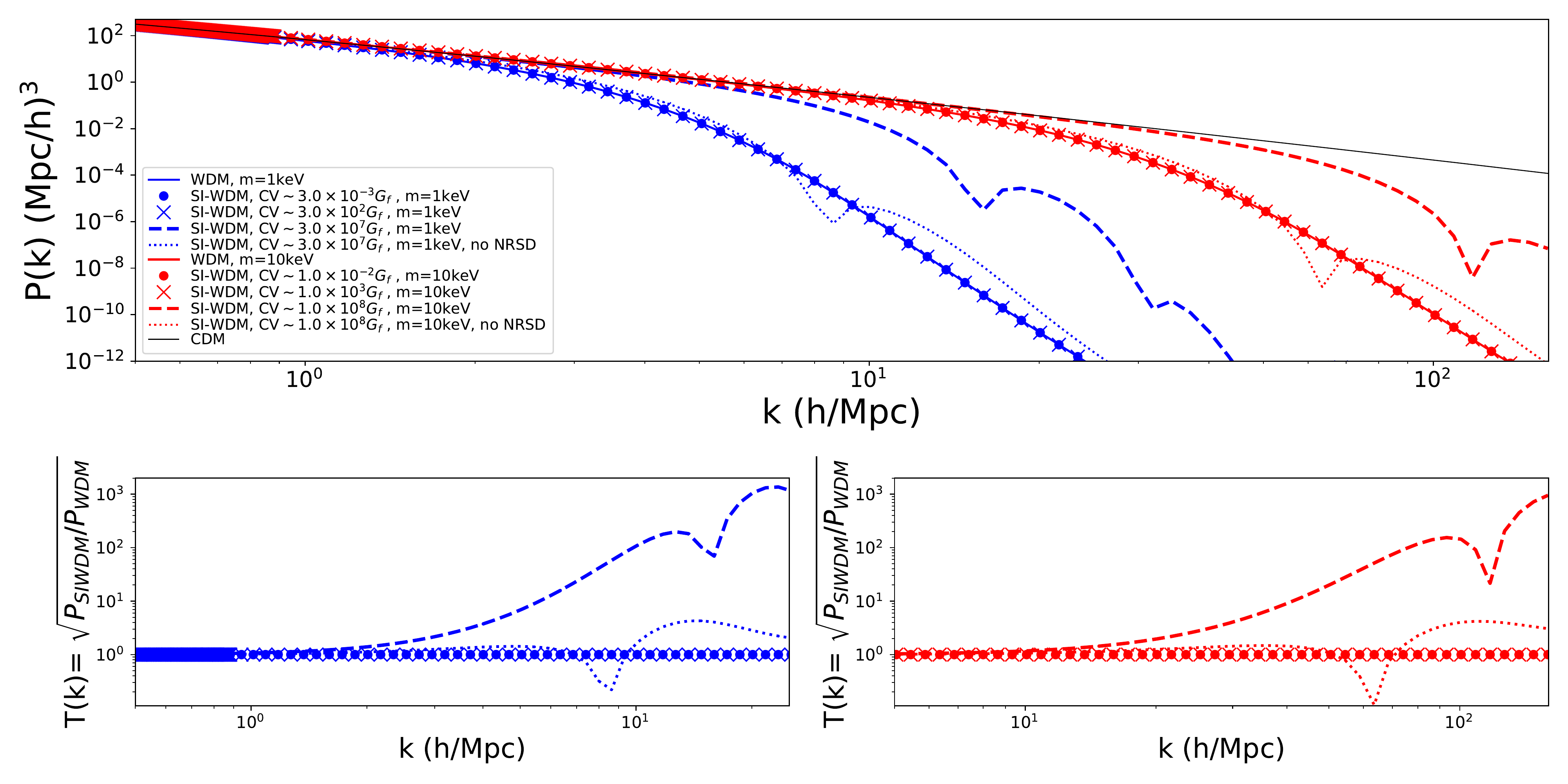}
\caption{\small 
Power Spectrum (\textit{top panel}) and Transfer Functions with respect to standard WDM (\textit{bottom panels}) for a vector field SI-WDM model, simulated using the modification to CLASS described in \cref{sec:CLASS} and considering the effects of non-relativistic self decoupling when appropriate. 
We assume the relaxation time approximation (\ref{eq:RTA_Hierarchy_Motion}), and consider two values of the DM particle mass: $1$ and $10$ keV. Also plotted are the power spectra of a CDM model and of WDM models with DM mass of $1$ and $10\ {\rm keV}$. 
All WDM and SI-WDM models consider a nonresonant production scenario (Dodelson-Widrow mechanism, \cite{Dodelson1993a}) with $T \sim (4/11)^{1/3} T_\gamma$. 
We also show, for comparison, cosmologies where the interaction constant is high enough to be in the non-relativistic self decoupled regime but these effects are ignored, instead using a relativistic DF (dotted lines).
}
\label{fig:PowerSpectrum_Combined}
\end{figure}

%First View
We show a few results of the matter power spectrum $P(k)$ for different masses and interaction constants in figure \ref{fig:PowerSpectrum_Combined} , for the case of a massive Vector Field self interaction. 
%
%In particular, \cref{fig:PowerSpectrum_NRSD} shows the full effect of the non-relativistic self interaction decoupling mentioned in \cref{sec:CLASS_NR}, while in \cref{fig:PowerSpectrum_noNRSD} this phenomenon is ignored altogether. 
%
This very same plot, this time for the case of a Massive Scalar mediator, can be found in figure \ref{fig:PowerSpectrum_MS_Combined_Appx} in  \cref{sec:Appendix_AllPlots}. 

%\begin{figure}[htb]
%\centering
%\includegraphics[width=0.95\textwidth]{plots/power-spectra/Plots_No_NR_SI/SIWDM_VF_Transfer_1and10keV_separateTransfers.pdf}
%\caption{\small 
%Power Spectrum (\textit{top panel}) and Transfer Functions with respect to standard WDM (\textit{bottom panels}) for a vector field SI-WDM model, simulated using the modification to CLASS described in \cref{sec:CLASS}. In this case, the effects of non-relativistic self decoupling are ignored and we assume a relativistic DF \eqref{eq:SID_Equilibrium_FD-BE} holds until today.
%
%We assume the relaxation time approximation (\ref{eq:RTA_Hierarchy_Motion}), and consider two values of the DM particle mass: $1$ and $10$ keV. Also plotted are the power spectra of a CDM model and of WDM models with DM mass of $1$ and $10\ {\rm keV}$. 
%
%All WDM and SI-WDM models consider a nonresonant production scenario (Dodelson-Widrow mechanism, \cite{Dodelson1993a}) with $T \sim (4/11)^{1/3} T_\gamma$.
%}
%\label{fig:PowerSpectrum_noNRSD}
%\end{figure}

%\begin{figure}[htb]
%\centering
%\includegraphics[width=0.95\textwidth]{plots/power-spectra/SIWDM_VF_Transfer_1and10keV_separateTransfers.pdf}
%\caption{\small 
%Simulated Power Spectrum and Transfer Functions with respect to standard WDM for several SI-WDM models under the same assumptions as in \cref{fig:PowerSpectrum_noNRSD}, but considering the effects of non-relativistic self decoupling when appropriate, as described in \cref{sec:SID} and \cref{sec:CLASS_NR}.
%}
%\label{fig:PowerSpectrum_NRSD}
%\end{figure}

%General Description of model parameters
In these figures, we compare the power spectrum with their most similar counterpart, standard WDM, as well as a reference CDM model. We have chosen two different particle masses of $1$ and $10$ keV, the first one to provide a parallelism with \cite{Hannestad00}, and the second one to show a model with predictions that agree better with current observations \cite{Perez2017,Ng2019a}. 
We have assumed a simple background model: a Fermi-Dirac background with a temperature $T_{0} = (4/11)^{1/3}T_\gamma$ identical to SM neutrinos, renormalized to give the correct DM abundance (non-resonant sterile neutrino production \cite{Dodelson1993a}). 
Over these, we show the results of several SI-WDM cosmologies, for the mediator models described in \cite{Yunis2020b} and in \cref{sec:Intro}. We show three examples for each, ranging from a very weak interaction to one matching the upper limits for the scattering amplitude of DM self interactions coming from Bullet Cluster constraints ($\sigma / m_{\rm DM}  \leq 0.1 \, \mathrm{ cm}^2/g$, see \cite{amrr}). We show also the transfer functions for the self interacting models with respect to their WDM counterparts, defined as $T_i(k)=(P_i(k)/P_{\rm WDM}(k))^{1/2}$.

We parametrize these mediator models via their effective interaction constants, which are defined according to the expressions in section 3.3 in \cite{Yunis2020b}. 
In particular, we consider in figure \ref{fig:PowerSpectrum_Combined} a Massive Vector Field interaction, where $\bar{C_V} = \left( g_V m_{V}^{-1} \cos^{-1} \theta_W' \right)^4$, where $g_V$ is the coupling constant appearing in the Lagrangian interaction term, $m_V$ the mediator mass and $\theta_W'$ is the (dark sector) Weinberg angle, in units of the Fermi constant $G_F$.
The other interaction model we consider a Massive Scalar mediator interaction in figure \ref{fig:PowerSpectrum_MS_Combined_Appx}, is also defined according to \cite{Yunis2020b} as $C_m = (g_m/m_\Phi)^4$, with $g_m$ the coupling constant of the interaction and $m_\Phi$ the mediator mass.

%Features: Similar suppression on several scales to WDM
The first obvious conclusion we can draw is that, for low enough interaction constants the results are, predictably, indistinguishable from the ones of WDM.
This is evident for a large range of values of the interaction constants, and is indeed the case if interactions are completely irrelevant during the matter dominated era when metric perturbations start to grow. Even for constants when this is not the case, the ``base'' WDM model can give us an idea of the general behaviour of SI-WDM: a suppression in the power spectrum, typical of WDM models, is also seen in SI-WDM models. This suppression is identical for low enough interaction constant and, as can be seen from these figures, it would remain almost identical even for higher constants were it not for the effect of non-relativistic Self Decoupling.

%Features: non-relativistic Self Decoupling
This brings us to a second feature of these power spectra: the effects of non-relativistic Self Decoupling. For the models shown in \cref{fig:PowerSpectrum_Combined}, the highest interaction constants would correspond to a situation of non-relativistic Self Decoupling as explained in \cref{sec:SID} and \cref{sec:CLASS_NR}. 
We can see that both of these high-constant spectra are the ones that show significant departures from its corresponding WDM equivalent but only a small oscillation at high $k$ is present if we don't consider these effects.
If we consider the effects described in \cref{sec:CLASS_NR}, then the suppression scale moves to a significantly higher $k$ and a completely different pattern of oscillation appears. This is remarkable, as a high interaction constant might make a SI-WDM model ``colder'' than its WDM counterpart, i.e. as if it corresponds to a much higher particle mass: we will explore this further in \cref{sec:Obs}. 
Also, regarding the various contour conditions that were considered in \cref{sec:SID}, we have found that all of them produce virtually indistinguishable results in the power spectrum. 

%Features: Oscillation, reference to Appendix

The final evident feature in these spectra are the oscillations themselves. Indeed, these oscillations are not necessarily exclusive to SI-WDM models, and we can see them in several other self interaction models \cite{2016MNRAS.460.1399V} as well as a few approximation schemes to WDM itself \cite{CLASSIV}. 
In fact, this is common to most fluid or pseudo-fluid approximations to the full Boltzmann hierarchy and we will explore quantitatively how, for the case of SI-WDM, these oscillations come as a consequence of the tight-coupling regime in \cref{sec:Appx_Osc}.

%Features: Independence of contour conditions

Finally, we can discuss about the effects of the contour conditions and the particularities of the non-relativistic transition itself. We have tested all the contour conditions for the transition that figure in  \cref{tab:Contour_Conditions}, and several variants of the transition approximation \eqref{eq:CLASS_NR_f0Sw} and obtained nearly indistinguishable results to what is shown in \cref{fig:PowerSpectrum_Combined}.

\section{Comparison with Observables}

\label{sec:Obs}

%Introduction: objectives for this section
Having discussed about the underlying particle models for SI-WDM in \cref{sec:Intro}, developed theoretical knowledge about their cosmological evolution in \cref{sec:Boltz}, and incorporated this theory into a numerical tool in \cref{sec:CLASS}, it is time to use this framework to put bounds in the parameter space of these models. 
As we mentioned before, in this work we have focused on obtaining and constraining the simulated matter power spectrum, which is likely the quantity that is more sensitive to the evolution of small scale structure. Thus, in this section we will use some observables of the small-scale structure in order to identify the simulated spectra that are allowed by cosmology and thus restrict the available parameter space for SI-WDM.

%About small scale structure formation
However, comparing the outcome of simulated \emph{linear} cosmology with the structure observed today may not be a straightforward task to begin with. In fact, the smallest gravitationally bound structures today have typically collapsed long ago, a highly nonlinear phenomenon that is not well captured by linear cosmology and is traditionally modeled via N-body numerical simulations \cite{Navarro1997,Maccio2012,Fitts2019}.
%Its relation to power spectrum: getting nonlinear stuff from LPS
However, certain predictions about the collapsed nonlinear structure are possible by exploiting the statistical properties of the cosmic density field. In particular, the extended Press-Schechter (PS) formalism allows us to obtain predictions about, for example, the estimated mass functions of haloes and subhaloes \cite{Schneider2016}. These predictions are directly related to the \emph{linear} power spectrum extrapolated until today, thus we can obtain constraints on it by using these theoretical predictions and comparing them against direct observation of structure such as, in this case, the number of subhaloes in the Milky Way.

%Method parametrization: talk about Murgia 17
In \cite{Murgia2017} it is shown that the linear power spectrum can be parametrized for various particle physics models via a phenomenological transfer function defined as 

\begin{equation}
T(k) = \left[1+(\alpha k)^\beta\right]^\gamma \ ,
\label{eq:Obs_TkFittingFormula}
\end{equation}

\noindent where $\{\alpha , \beta ,\gamma \}$ are fitting coefficients to the transfer function $T(k)$. Such a \textit{transfer function based} approach was used in that work to derive predictions about both Milky Way satellite galaxy counts and Lyman-$\alpha$ forest data. 

In this work, however, we use exclusively the simulated data from our modified version of CLASS in order to put constraints on these observables.
Still, we observe that our numerical results can be parametrized using this phenomenological function and, moreover, that the coefficients $\{\alpha,\beta,\gamma\}$ can be put in terms of the physical parameters $\{m,C_i\}$.
We remind the reader that, while this phenomenological fitting function does not capture some richer behaviour, such as the oscillatory patterns, it describes well the power spectrum damping and is accurate when used to calculate the observables relevant for this section, such as subhalo counts and Lyman-$\alpha$ forest measurements.

Thus, we fit this phenomenological transfer function \eqref{eq:Obs_TkFittingFormula} to the simulated data and we find formulas for the fitting coefficients $\{\alpha , \beta ,\gamma \}$ in terms of the model parameters that fit well our results in \cref{sec:Appendix_Table}. 
We remark that these fitting functions, while providing relatively accurate approximations for the transfer function, can incur in significant errors in the absolute values of the parameters $\{\alpha,\beta,\gamma\}$, along with the parametrization of the transfer function itself ignoring more complex effects such as power spectrum  oscillations as described in \cref{sec:Appx_Osc}.

\begin{figure}[htb]
\centering
\includegraphics[width=0.9\textwidth]{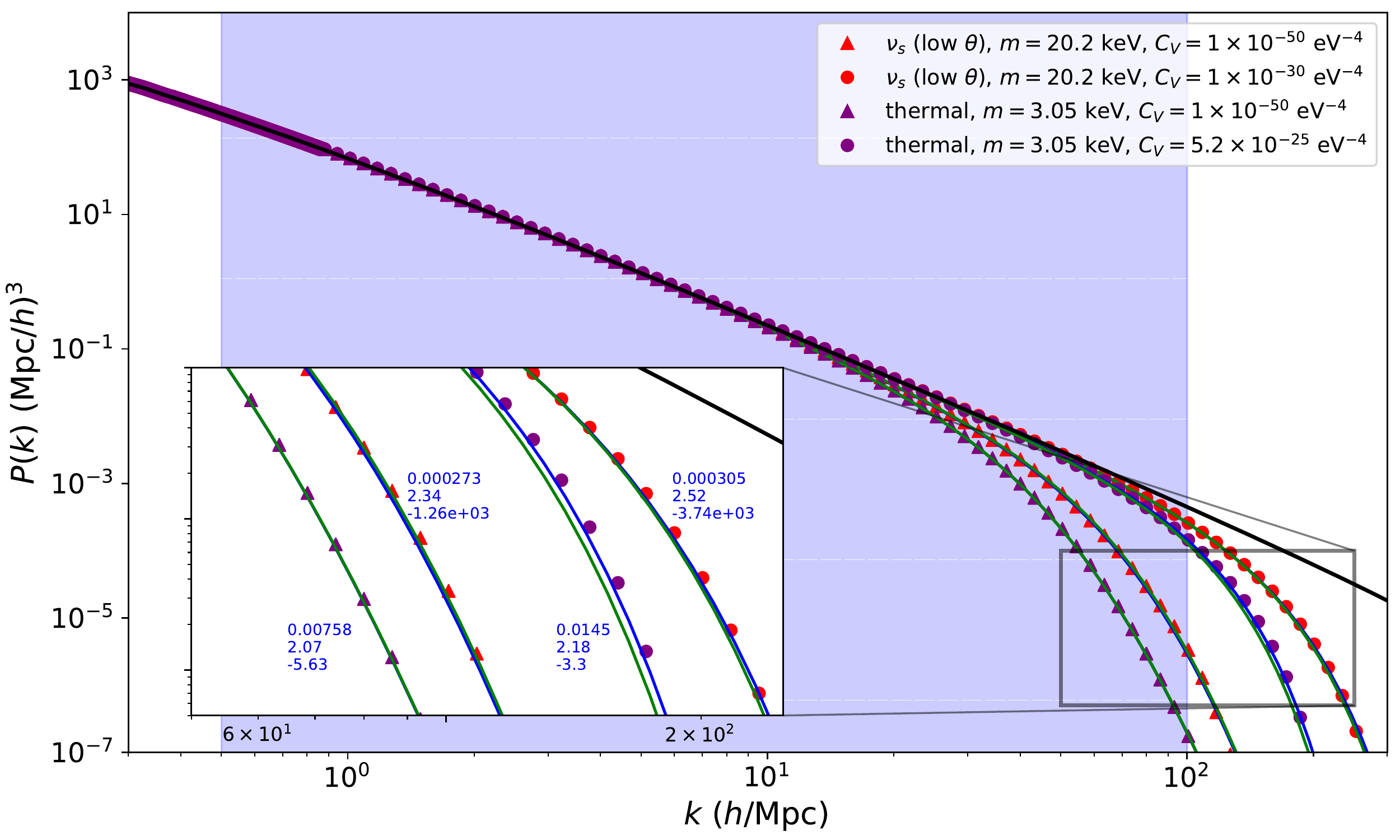}
\caption{\small 
Power spectra obtained through fitting functions, compared to the simulated data. 
We considered four different simulations, for the cases of thermal production (shown in red) and low $\theta$ sterile neutrino WDM (shown in purple, see \cref{fig:Obs_Plots_sterileDMParameterSpace}) and the cases of relativistic (triangles) and nonrelativistic (circles) self interaction decoupling (note that $\theta$ is the value of the sterile neutrino mixing angle with the SM neutrinos).
On top of these, we plot the power spectra obtained from the best fit of the transfer function \cref{eq:Obs_TkFittingFormula} and its corresponding values $\{\alpha,\beta,\gamma\}$, in blue. 
We show as well the power spectra corresponding to the values of $\{\alpha,\beta,\gamma\}$ obtained from the fitting formulae shown in \cref{sec:Appendix_Table}, in green. 
As a reference we also show a CDM power spectrum (black), and the shaded region corresponds to the relevant values of $k$ for the Lyman-$\alpha$ analysis outlined in \cref{sec:Obs_deltaA}. 
}
\label{fig:Obs_FitExamples}
\end{figure}

We show a few examples of simulated power spectra in \cref{fig:Obs_FitExamples}. For these examples, we also show the best fit spectra using the parametrization \ref{eq:Obs_TkFittingFormula} for the transfer function in terms of $\{\alpha,\beta,\gamma\}$. We also compare these with the resulting parameters from the phenomenological fitting formulae outlined in \cref{sec:Appendix_Table}, in terms of $\{m,C_i\}$ and the relevant DM interaction model and thermal history scenario. 
This opens the possibility of comparing these models with future observational constraints in a model-independent way, as this parametrization allows us to compare our predictions of small scale suppression with any similar model. 
Still, it may be possible to draw a comparison using phenomenological  ``fluid'' parameters, such as sound speed or viscosity \cite{Garny2018,2016MNRAS.460.1399V}, and also certain features other than the power spectrum suppression can be compared with similar models, such as the oscillation pattern \cite{Munoz2020}. 

\subsection{Milky Way Satellite Counts}

\label{sec:Obs_Nsub}

%Method summary
We provide here short summaries of the methods used in \cite{Murgia2017} to compare the linear power spectrum with both the observations of Milky Way satellite counts and the Lyman-$\alpha$ forest data. 
In general, the $\Lambda$CDM model predicts a number of satellite galaxies of the Milky Way that is too large, as many as $\sim 1000$, compared to the observed $\sim 50$ \cite{Bullock2017}. 
In order to get an estimate number of observed satellites of the Milky Way, we add the number of classical MW satellites to the number of ultra-faint ones observed by SDSS, accounting for the limited sky coverage of the survey and a 10\% sampling variance, arriving to a conservative number of observed MW satellites of $N_{sat}=57$.

%Method Description
We compare that with the estimated number of subhaloes predicted by the PS formalism. According to \cite{Murgia2017,Schneider2016}, the subhalo mass function, for a given host halo of mass  $M_{halo}$, is given by:

\begin{equation}
\frac{d N}{d M_{sub}} = \frac{1}{44.5} \frac{1}{6 \pi^2} \frac{M_{halo}}{M_{sub}^2} \frac{P(1/R_{sub})}{R_{sub}^3 \sqrt{2 \pi (S_{sub}-S_{halo})}} \ ,
\label{eq:Obs_Nsub_dNdMsub}
\end{equation}

\noindent with $M_{sub}$, $S_{sub}$ the mass and variance of a given subhalo and $M_{halo}$, $S_{halo}$ the ones for the main host halo, defined as:

\begin{equation}
S_{i} = \frac{1}{2\pi^2} \int_0^{1/R_i} dk k^2 P(k) \quad , \quad M_{i} = \frac{4\pi}{3} \Omega_m \rho_c (c R_i)^3 \quad , \quad c = 2.5 
\label{eq:Obs_Nsub_SiRi_definition}
\end{equation}

\noindent with $\Omega_m$ the matter density parameter, $\rho_c$ the critical density of the universe (today) and $P(k)$ the linear power spectrum for a given cosmological model, extrapolated to $z=0$. 
We can obtain a prediction for the total number of subhaloes for the MW using an assumption for it mass of $M_{\rm MW} \sim 1.2 \times 10^{12} M_{\odot}/h$ \cite{Karukes2019} and integrating expression \eqref{eq:Obs_Nsub_dNdMsub} from a minimum subhalo mass of $10^{8}M_\odot / h$ up to $M_{\rm MW}$. 
We obtain a constraint by requiring that the predicted number of subhaloes must be equal or greater than the conservative bound for the observed MW satellites, and report the results for various SI-WDM cosmologies in \cref{sec:Obs_Plots}.

\subsection{Lyman-$\alpha$ Constraints}

\label{sec:Obs_deltaA}

%Method Summary
Another relevant probe of the small scale matter distribution is the so called Lyman-$\alpha$ forest: at a quick glance, the Lyman-$\alpha$ absorption spectrum produced by the intergalactic medium on the spectra of distant quasars. This absorption spectra is produced by the inhomogeneous distribution of neutral hydrogen and has been widely used as an accurate probe of the matter power distribution on small and intermediate scales, namely $0.5\ h/ \mathrm{Mpc} \leq k \leq 100\ h / \mathrm{Mpc}$ \cite{Murgia2017,Viel2013,Baur:2015jsy,Irsic2017}. 
While computing accurate bounds would imply a full statistical analysis, we here instead follow the methods of \cite{Murgia2017,Schneider2016} in order to investigate the plausibility of small deviations to already well-explored non-cold DM cosmologies, namely the thermal 3.5 keV constraints obtained though a full analysis of Lyman-$\alpha$ forest data \cite{Irsic2017}.

%Method Description
We use the criterion set in \cite{Murgia2017} to characterize the deviation of a model from a $\Lambda$CDM cosmology. At a given scale $k$, we quantify this by the ratio of the 1-dimensional power spectra

\begin{equation}
r(k) = \frac{P_{1D}(k)}{P_{1D}^{\Lambda CDM}(k)}
\label{eq:Obs_deltaA_rk_definition}
\end{equation}

\noindent which is itself defined as the following integral of the 3-dimensional power spectrum

\begin{equation}
P_{1D}(k) = \frac{1}{2\pi} \int_k^\infty dk' k' P(k')
\label{eq:Obs_deltaA_P1D_definition}
\end{equation}

This quantity allows us to obtain a rough estimate of how suppressed is a given power spectrum with respect to $\Lambda$CDM. 
We thus can adopt the following criterion in order to determine whether or not a particular cosmology is allowed by this analysis: if a given power spectrum is ``more suppressed'' than the reference 3.5 keV thermal WDM model, then this model would be excluded. We can quantify this suppression at all scales $k$ by introducing the following estimator 

\begin{equation}
\delta A = \frac{A_{\Lambda CDM} - A}{A_{\Lambda CDM}}
\label{eq:Obs_deltaA_deltaA_definition}
\end{equation}

\noindent where $A$ is the integral of the ratio $r(k)$ in the scales probed by the Lyman-$\alpha$ forest observations: $0.5 \mathrm{Mpc}/h \leq \lambda \leq 100 \mathrm{Mpc}/h$ for the MIKE/HIRES+XQ-100 combined dataset used in \cite{Murgia2017,Irsic2017}.

%The actual bounds
Thus, by using the 3.5 keV thermal WDM power spectrum, the authors in \cite{Murgia2017} obtain an estimated suppression value of $\delta A = 0.38$. 
%(which is approximately confirmed by our calculations)
This serves as the most extreme model that can be allowed according to this analysis: any model that shows a higher value of this suppression estimator $\delta A$ is considered to be excluded by the Lyman-$\alpha$ forest data at at least 95\% C.L. 
These authors also include a tighter constraint of $\delta A = 0.21$ for a fixed thermal history, which we will include in our analysis for completeness.

%Discussion about validity?
It is important, however, to discuss about the validity of this method. A full analysis of the Lyman-$\alpha$ forest data requires a careful statistical analysis and a complex set of hydrodynamical simulations in order to compute the most convenient statistics for this set of observations, namely the flux power spectrum $P_F(k)$ \cite{Schneider2016,Viel2013,Viel2007}. 
However, in \cite{Murgia2017,Schneider2016} it is argued that as the relation between flux and linear 3-dimensional matter power spectra can be accurately modeled as $P_F(k) = b^2(k) P_{3D}(k)$, with $b^2(k)$ a bias factor. This bias factor differs little between $\Lambda$CDM and modified DM models such as SI-WDM, which are ``fairly close'' to the reference case, and thus equation \eqref{eq:Obs_deltaA_rk_definition} can be reasonably applied to the flux power spectra as well. 
We remind the reader that this is an approximate method and a proper, robust bound should be obtained via a full hydrodynamical simulation as in, for example, \cite{Viel2013}.

\subsection{Parameter Space Constraints}

\subsubsection{Exploring the SI-WDM parameter space}

\label{sec:Obs_Plots}

%Introductory notes to the plots

\begin{figure}[t!]
\centering
\includegraphics[width=\textwidth]{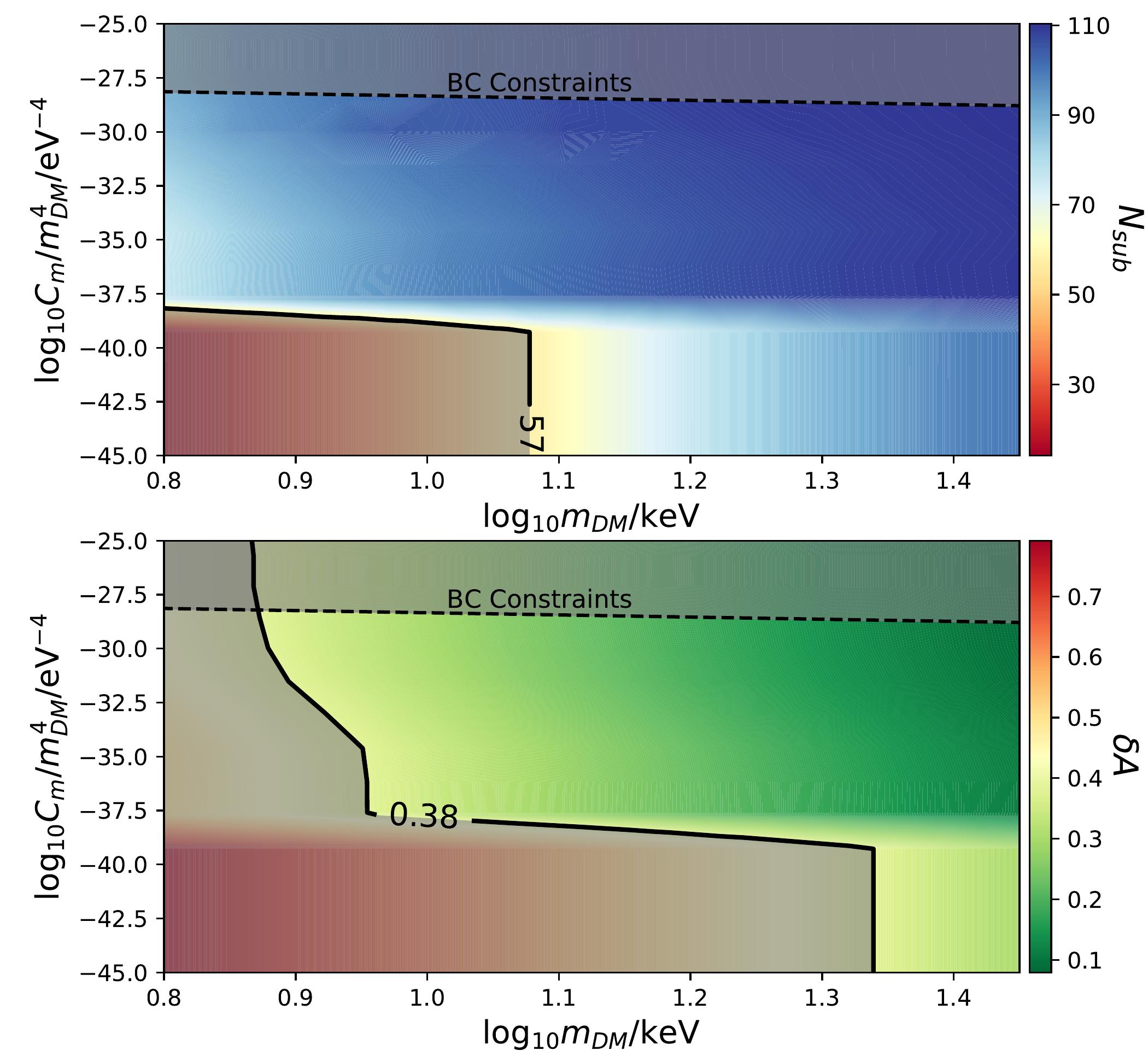}
\caption{\small 
Parameter space constraints for SI-WDM cosmologies according to the analysis outlined in \cref{sec:Obs}, for the case of a Vector Field mediator model and resonant sterile neutrino production (see figure \ref{fig:Obs_Plots_sterileDMParameterSpace} for the High Asymmetry case, and \cref{sec:Appendix_AllPlots} for examples with other mediators and production mechanisms).
For this model, we plot in colors the results of the predicted total number of subhaloes for the MW in the top panel, and the results of relative suppression with respect to $\Lambda$CDM regarding Lyman-$\alpha$ observations in the bottom panel, as a function of DM mass and effective coupling constant.
The results shown are the product of an interpolating function over $\sim$400 data points, each corresponding to an individual SIWDM simulated cosmology. 
In grey we mark the regions of parameter space ruled out by these analyses: for the upper panels the lower shaded regions mark $N_{sub} \leq 57$ and for the lower panels they mark the conservative constraint $\delta A \geq 0.38$ \cite{Murgia2017}. 
The upper grey shaded regions in all panels mark regions excluded by Bullet Cluster constraints \cite{amrr}: we take the approximate bound $\sigma / m \leq 0.1 \mathrm{ cm}^2/g$, with $\sigma \sim C_m^2 / m^2 $ the self interaction cross section today.
}
\label{fig:Obs_Results_Grid}
\end{figure}

In this section, we present the results of the Milky Way satellite count and Lyman-$\alpha$ analysis for a suite of SI-WDM cosmological parameters. 
We analyze the resulting power spectra for different mediator models and initial conditions for the background distribution function, each of them with approximately $\sim 400$ simulated spectra. For each combination of mediator model and initial conditions, we covered a uniform grid in the SI-WDM parameters (particle mass and interaction constant), and performed the analysis we outline in \cref{sec:Obs_Nsub}, \cref{sec:Obs_deltaA} for each of them.
We plot the results for a vector field mediator under a high asymmetry, resonantly produced sterile neutrino model in \cref{fig:Obs_Results_Grid}, and we show the results for the other mediator-initial condition combinations in \cref{sec:Appendix_AllPlots}. 

For the specific case of resonant sterile neutrino production, we considered two different scenarios depending on the values of the sterile neutrino mixing angle with the SM neutrinos ($\theta^2$). 
As in this scenario sterile neutrinos are produced both by the non-resonant \cite{Dodelson1993a} and resonant \cite{ShiFuller,Venumadhav2015} mechanisms, and the latter depends on the amount of initial lepton asymmetry in the universe, one can arrive to the proper DM abundance for a multitude of SM mixing angles by varying this asymmetry value. 
Therefore, in this case we have considered two cases: one with high $\theta$ (low asymmetry) and one with low $\theta$ (high asymmetry), sampled from the slices in the sterile neutrino parameter space for $\nu$MSM we show in \cref{fig:Obs_Plots_sterileDMParameterSpace}. For each mass value sampled along these lines, we have obtained the initial distribution functions at $T=10$ MeV by using the \texttt{sterile-dm} routine in \cite{Venumadhav2015}. 

%Discussion 1: No impact on low constants + results comparable to other studies?
We can see many features in \cref{fig:Obs_Results_Grid}, and most of them are shared across all production mechanisms and interaction models, albeit with small differences in their constraints. 
First of all, we can see that for small enough interaction constants (depending on the interaction model itself), the results for each initial conditions are comparable to the ones for standard WDM. We find similar results to the analysis on \cite{Murgia2017,Schneider2016} in the lower end of interaction constants, and indeed during the whole parameter space corresponding to the relativistic self decoupling scenario. 
%For some reason ~30\% diffrenece in MWsat respect to Murgia only?
Following the same arguments as in the discussion following \cref{fig:PowerSpectrum_Combined}, we can attribute this to the fact that these perturbations do not grow during the radiation dominated era, thus the effects of self interactions deep into the particle's relativistic regime (which is itself deep into the radiation dominated era) is indeed negligible, and the results are comparable to WDM cosmology.

%Discussion 1.5: How tight Lyman-a containts are
Consistently with the findings of \cite{Murgia2017,Schneider2016,Irsic2017}, we find that the results for Lyman-$\alpha$ forest analysis are indeed much tighter than the ones from MW satellite counts for all models, and in particular looking at \cref{fig:Obs_Results_Grid} is   possible to exclude almost all of the available parameter space for resonantly produced $\nu$MSM sterile neutrinos using these observations, at least in the non interacting case. We see, however, that the situation changes drastically if we instead include strong enough self interactions.

%Discussion 2: The clear gap at NRSD: a big deal! Makes the spectra colder
A clear transition in both $N_{sub}$ and $\delta A$ can be seen whenever the models transition to the non-relativistic self decoupling regime. A sharp change takes place for this regime, as the spectra become immediately much colder (as we have discussed after \cref{fig:PowerSpectrum_Combined}). 
This basically results in spectra which are much less suppressed than in the relativistic decoupling regime when compared to $\Lambda$CDM, with a much higher suppression scale in $k$ space. 
This can be thought of as a ``shift to the left'' in mass space: models that originally correspond to a mass that is highly suppressed now result in similar predictions as models with a much higher mass through the introduction of self interactions. Thus, spectra corresponding to WDM become ``colder'' through the effect of self interactions.

%Discussion 3: Disallowed masses become allowed thanks to NRSD
For the models themselves, these colder spectra have two main consequences: first it re-allows models previously excluded by observations. As these models now become much more similar to a $\Lambda$CDM cosmology, the predictions arising from the linear power spectrum become increasingly more consistent with observation. 
At the same time WDM models that are allowed by observations become, through the inclusion of self interactions in this regime, not able to alleviate small scale structure tensions to $\Lambda$CDM through the linear power spectrum alone. 
Still, we remind the reader that some of these tensions can be alleviated via the inclusion of self interactions in the later, non-linear stages of structure formation \cite{Bullock2017,Fitts2019}.

\subsubsection{The $\nu$MSM Parameter Space in the Presence of Self Interactions}

From the previous section we have concluded that, even in a resonant production scenario relevant to $\nu$MSM, sufficiently strong self interactions may significantly relax the constraints on MW satellites and Lyman-$\alpha$. 
Indeed, we have seen that, for a given particle mass and mixing angle $\theta$, the differences between low interaction constant under relativistic self decoupling and high interaction with non-relativistic self decoupling are particularly marked, and a number of models that are excluded by the analysis in the first scenario are admitted in the second. 

Thus, we can go a step further and analyze the full parameter space of $\nu$MSM, discerning exactly how much of the $\{\theta,m\}$ parameter space can be readmitted under the effects of self interactions. 
This new model would be comprised of three parameters now: particle mass $m$, mixing angle $\theta$ and the interaction coupling constant $C_i$. For this analysis, we are interested in which section of the $\{\theta,m\}$ parameter space can be readmitted by using \emph{any} interaction model (quantified by $C_i$). 
As we have seen from the past section, for fixed $\{\theta,m\}$ a higher interaction constant always results in equal or colder spectra. Thus if for these fixed parameters the model with the highest allowed interaction constant is allowed by a given observation, we can conclude that there is indeed a finite range of interacting models for which that particular value of $\{\theta,m\}$ is allowed. 

\begin{figure}[htbp]
\centering
\includegraphics[width=0.9\textwidth]{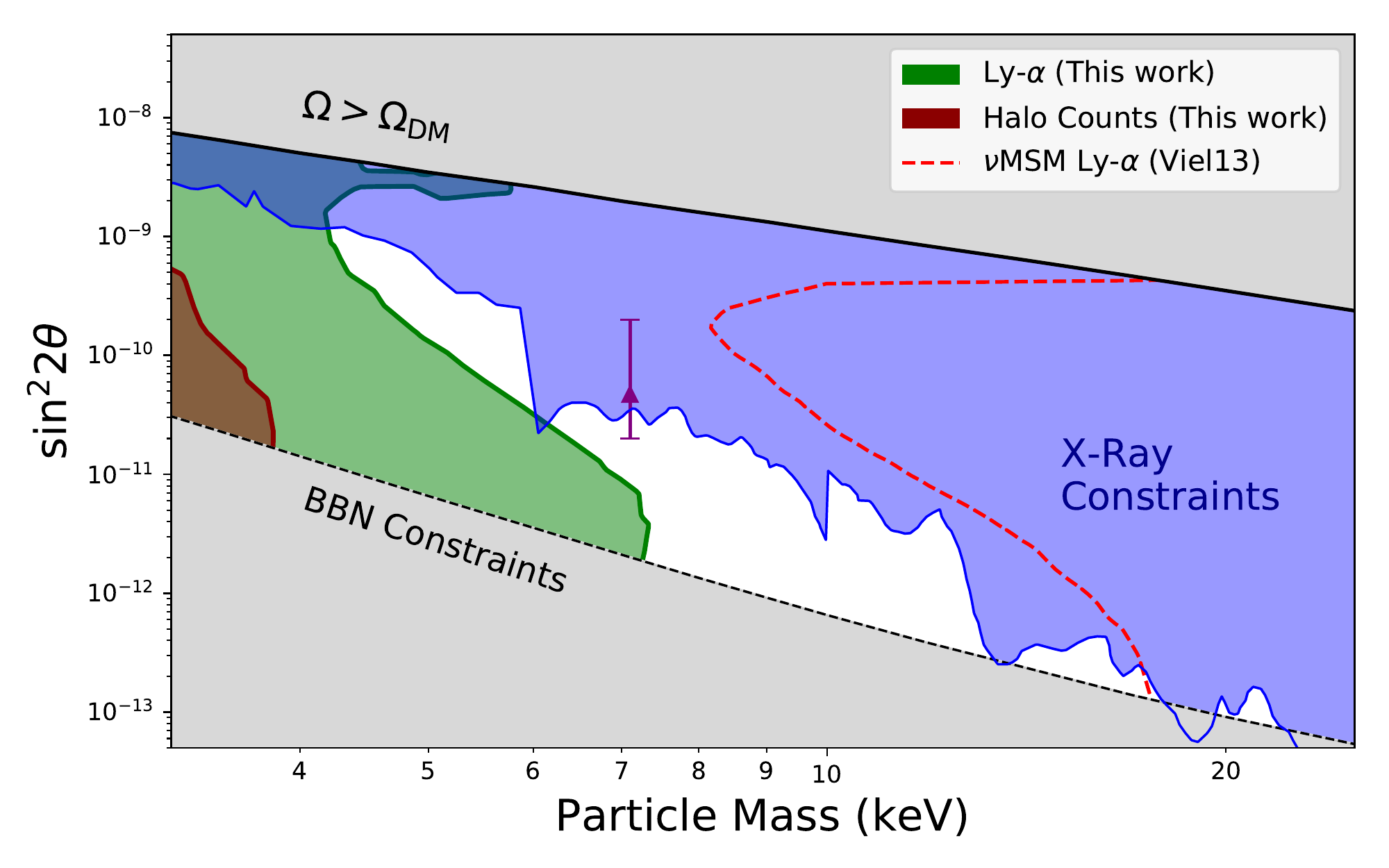}
\caption{\small 
Parameter space constraints for $\nu$MSM, where MW satellite halo counts and Lyman-$\alpha$ forest bounds are analyzed under a self interacting model as outlined above. 
For each point $(\theta,m)$ in the parameter space we consider a self interacting model under a vector field mediator, with its interaction constant given by $\sigma / m  \sim 0.144 C_v^2 / m^3 = 0.1 \mathrm{ cm}^2/g$, the upper limit given by Bullet Cluster constraints \cite{amrr}. A particular point in the parameter space is considered to be admitted by any given analysis if it is so for this maximal model. 
For comparison, we plot the Lyman-alpha bounds for the non interacting case for a comparable analysis, according to the results in \cite{Viel2013}. 
We also plot other bounds to the $\nu$MSM parameter space for informative purposes, namely X-Ray indirect detection bounds \cite{Schneider2016,Cherry2017,Ng2019a} (in blue) and sterile neutrino production bounds \cite{Boyarsky2018,Venumadhav2015} (in grey), again reminding the reader that these bounds do not include the effects of self interactions in sterile neutrino production. 
Also for informative purposes, we plot the sterile neutrino model compatible with a tentative 3.5 keV DM signal, subject of debate in recent years, as a purple triangle \cite{Bulbul2014,JeltemaProfumo,Cappelluti2018}.
}
\label{fig:Obs_Results_SInuMSM}
\end{figure}

So, we calculated these constraints in the $\nu$MSM parameter space and show these results in figure \ref{fig:Obs_Results_SInuMSM}. 
In this analysis, for each point in parameter space, we consider that particular model to be admitted by each observable (MW satellites and Lyman-$\alpha$) if they are allowed for the highest value of the interaction constant consistent with Bullet Cluster constraints (our current upper limit on the interaction strength). 

On top of these bounds, we can see the other sources of constraints for this particular model, namely production bounds (based on upper limits to the amount of lepton asymmetry consistent with Big Bang Nucleosynthesis) and X-Ray constraints from observation of various galactic and extragalactic objects. 
For comparison, we show also the Lyman-$\alpha$ constraint for standard $\nu$MSM as was calculated in \cite{Schneider2016}. We see a striking difference with the non interacting case, with the latter expanding considerably the available parameter space with respect to the former, which nearly excludes the whole available space for these particular WDM models.  

\section{Conclusions}

\label{sec:Conclusions}

Following on the initial theory of self interactions in WDM linear theory set in \cite{Yunis2020b}, we aimed in this paper to provide a relatively complete practical application of this framework. 
This included the relevant approximations that can be used to simplify the treatment, a numerical application in CLASS and comparison of the available parameter space of this self interacting models with its WDM counterparts. 

Thus, in section \ref{sec:Boltz} we started introducing the Boltzmann formalism required to implement these SI-WDM models. After stating the particular scenario of thermal history considered, we recalled the results obtained in \cite{Yunis2020b} and expanded upon them. 
In that section, we provided an in-depth explanation of the relaxation time approximation and the resulting Boltzmann hierarchies, introduced the situation of non-relativistic self decoupling and considered an approximate way of treating the relativistic to non-relativistic transition for both background and perturbations. 
We considered as well limiting forms for the relaxation time in all of these scenarios. 

Then, having discussed thoroughly about these approximations to the full linear evolution of perturbations in SI-WDM, we implemented them in the public Boltzmann solver CLASS. 
We made this modification publicly available at \href{https://github.com/yunis121/siwdm-class}{github.com/yunis121/siwdm-class}, and discussed its implementation in section \ref{sec:CLASS}. 
We first considered a few further approximations in various regimes that are necessary to the numerical implementation, such as the tight-coupling regime, as well as several improvements to the built-in integrator and the background DF modules for the case of non-relativistic self decoupling. 
Afterwards, we took a first look at the resulting power spectra and discussed both the effects of self interactions and non-relativistic self decoupling in the resulting distributions. 

Having obtained these spectra, in section \ref{sec:Obs} we compared these results with observation. From comparing these spectra with observables from structure formation such as the number of Milky Way subhaloes and the Lyman-$\alpha$ forest measurements, we can obtain a very interesting resulting parameter space for several underlying (self interacting extensions of) WDM models, for various mediator models. 

In this work, we obtained several interesting advancements in the area of SI-WDM, together with conclusions relevant to the field of WDM. 
Not only we show a readily applicable version of the results in \cite{Yunis2020b}, a proper discussion of the evolution of these species in the case of the Relaxation Time Approximation and an introduction to the scenario of non-relativistic self decoupling; but also we provide a full numerical application, which implies a very relevant leap forward in the development of SI-WDM models. 
This can pose these models as a very interesting alternative and extension to typical WDM models and providing a careful analysis of linear perturbations can both help encase past works which included such WDM extensions in a more complete cosmological framework, as well as motivate further study in the topic. 
Particularly in the case of self interacting $\nu$MSM, we have seen in section \ref{sec:Obs} that such an approach can severely relax the structure formation constraints, and possibly readmit a large portion of its parameter space, as was suggested in \cite{Yunis2020a}. 

It is important to know, however, that what is presented here is an approximate view of a general class of SI-WDM models, and proper consideration of any particular realization of such models would indeed require further analysis, from their particle production in the early universe, to studies on their clustering as well as observational constraints. 
Particularly, it would be very interesting if the characteristic signatures of SIWDM in the matter power spectrum could allow us to distinguish observationally between WDM and its self interacting counterpart, as was done for SIDM models in \cite{Munoz2020}. Future observations, particularly of the $21$-cm  line, could potentially  distinguish between these two possibilities as shown for example in \cite{Munoz2020}.
Also, the approximate view of SI-WDM linear theory presented here can be improved in many ways, some of the most obvious being extending this approach to more general models (such as light mediators or long-range interactions). Other improvements involve the implementation of the full collision terms shown in \cite{Yunis2020b}, or calculating precisely the relativistic to classical transition. 

In general, we believe this work can serve as an important stepping stone in the study of self interacting extensions to WDM. We have indeed shown here that the inclusion of self interactions to these general class of models can alter the resulting power spectra, thermal history and distribution today on a significant way, with important consequences for a wide range of DM models.

\acknowledgments

CRA has been supported by CONICET, Secretary of Science and Technology of FCAG and UNLP (grants G140 and G175), National Agency for the Promotion of Science and Technology (ANPCyT) of Argentina (grant PICT-2018-03743) and ICRANet. 
CGS is supported by ANPCyT grant PICT-2016-0081; and grants G140, G157 and G175 from UNLP.
DLN has been supported by CONICET, ANPCyT and UBA.

\clearpage
\bibliographystyle{jhep}
\bibliography{CLASS_Implementation,CLASS_Implementation_2}
%\bibliography{CLASS_Implementation,/home/rafael/Documents/DOCubuntu/bibtex/CLASS_Implementation}

\clearpage

\appendix

\renewcommand\thefigure{\thesection.\arabic{figure}}
\setcounter{figure}{0}

\section{Power Spectrum Oscillations and Tight Coupling}

\label{sec:Appx_Osc}

Here in this appendix we take a look at the phenomena of oscillations in the final linear power spectra we have obtained in section \cref{sec:CLASS_PS}. 
This oscillatory phenomenon only appears at a reasonably high value of $k$, and it only appears when the self interaction constant reaches values high enough as to have a significant effect in the matter dominated era. 
Surprisingly, the frequency of these oscillations seems unchanged with increasing coupling constant and indeed retains similar shape along different interaction models, only ever changing with different particle masses. The interaction constant only seems to increase or decrease the amplitude of such oscillations.

Indeed, phenomena like these appear to be linked to the phenomenon of tight coupling and fluid approximations themselves, as these oscillations in the power spectra appear not only for SI-WDM, as we show in \cref{fig:PowerSpectrum_Combined}, but also in interacting applications of CDM \cite{2016MNRAS.460.1399V} or even fluid approximations to WDM itself \cite{CLASSIV}. Here, we will try to shed some light on this phenomenon by reviewing initial insights in \cite{CLASSIV} and expanding upon them in this work, relating them with SI-WDM itself.

We start by recalling some results initially shown in \cite{CLASSIV}, where the authors developed the homogeneous part of the non-interacting Boltzmann hierarchy for WDM (see \cite{Ma1995}) into a more convenient form: 

\begin{equation}
\mathbf{\dot{\Psi}} = \frac{q k}{\epsilon} \mathbf{A} \mathbf{\Psi} \equiv \alpha(\tau) \mathbf{A} \mathbf{\Psi} \ ,
\label{eq:Appendix_Osc_BoltzHomogeneous}
\end{equation}

\noindent where $\mathbf{\Psi} = (\Psi_0, \Psi_1, ...)$ and $A$ is the following matrix:

\begin{equation}
A = \begin{bmatrix} 0 & -1 & & & & \\ 1/3 & 0 & -2/3 & & \\ \\ & \ddots & \ddots & \ddots & & \\ \\ & & l/(2l+1)& 0 & -(l+1)/(2l+1) & \\ \\ & & & \ddots & \ddots & \ddots \end{bmatrix}
\label{eq:Appendix_Osc_BoltzHomogeneous_Adefinition}
\end{equation}

There, they note that the formal solution to this homogeneous system is indeed the following matrix exponential:

\begin{equation}
\mathbf{\Psi} (\tau) = \mathbf{U} e^{\int_{\tau_i}^{\tau}d\tau ' \alpha(\tau ') \mathbf{D}} \mathbf{U}^{-1} \mathbf{\Psi}(\tau_i)
\label{eq:Appendix_Osc_FormalSolution}
\end{equation}
with $\mathbf{D}$ a diagonal matrix and $\mathbf{A}=\mathbf{U}\mathbf{D}\mathbf{U}^{-1}$. There, they show that the highest eigenvalue goes to $\pm i$ for $l_{max} \rightarrow \infty$ and the time dependent phase accompanying this oscillation is: 

\begin{equation}
\phi (\tau) = k \int_{\tau_i}^{\tau} d\tau ' \frac{q}{\epsilon(\tau ')}
\label{eq:Appendix_Osc_TimeDependentPhase}
\end{equation}

\begin{figure}[htb]
\centering
\includegraphics[width=0.8\textwidth]{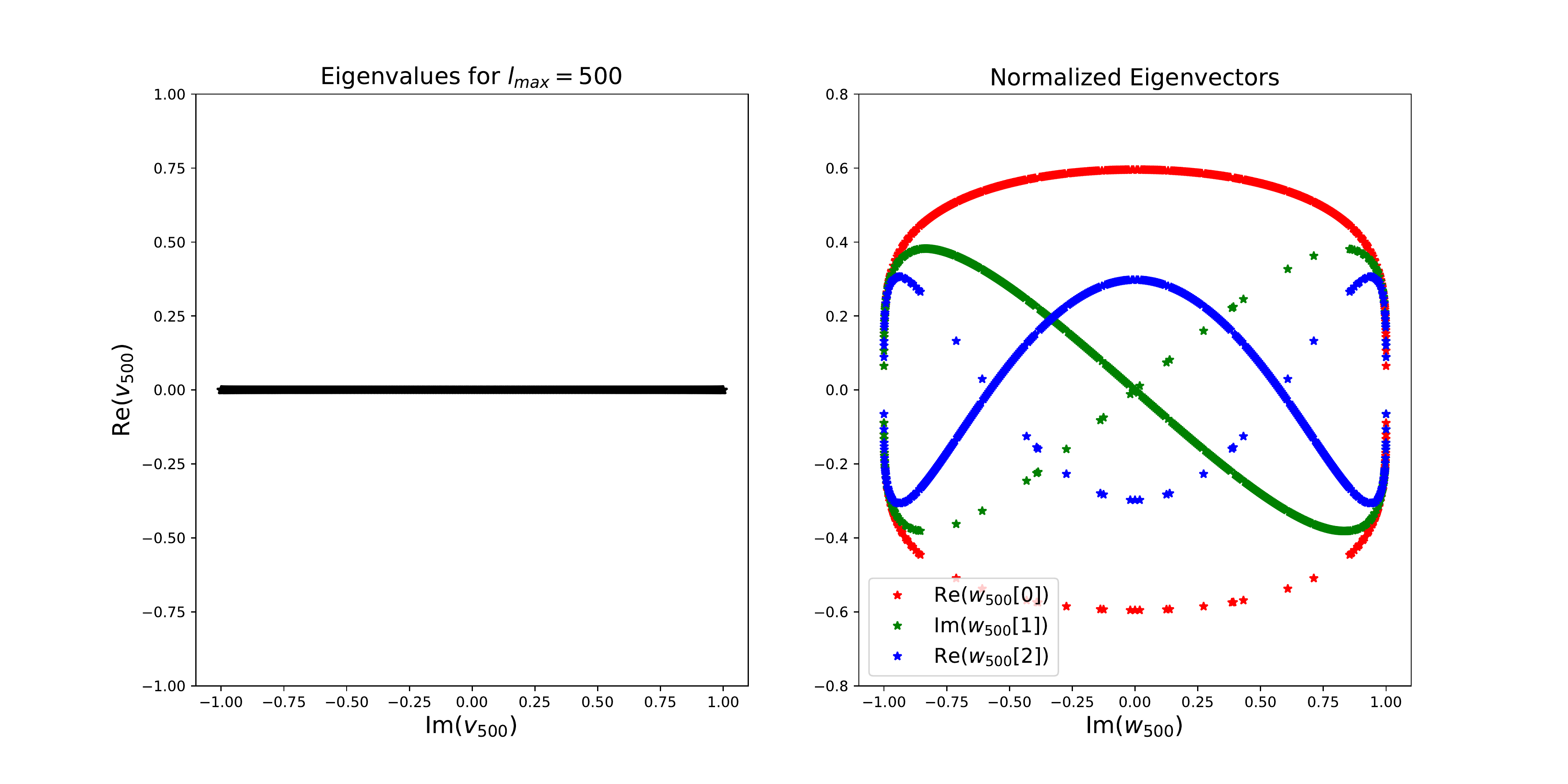}
\caption{\small 
Eigenvalue and eigenvector structure in the homogeneous Boltzmann Hierarchy without a damping term, for $l_{max}=500$. The left panel shows the eigenvalue distribution for this case in the complex plane, with the real part of most eigenvalues being proportional to the damping term in the case $\Gamma \gg 1$. The right panel shows the zero, first and second components of the normalized eigenvalues in the case $l_{max}=500$, as a function of the imaginary part of their eigenvalue. The real/imaginary parts of the eigenvalues not shown in the right hand plot are negligible.
}
\label{fig:Appendix_Osc_EigenvStructure_noInt}
\end{figure}

In this appendix, we will start  exploring this claim by analyzing the eigenvalue structure of the matrix $A$. 
Indeed, we can check in \cref{fig:Appendix_Osc_EigenvStructure_noInt} that the eigenvalues of this matrix are purely complex, with its highest eigenvalues in modulo approaching $\pm i$ as $l_{max} \rightarrow \infty$ (we use $l_{max} = 500$, for higher values the shown structure simply becomes more densely populated). 
The eigenvectors also show a rich structure, alternating between real and complex and the $l$-th component possessing $l$ zeros along the range of eigenvalues. This implies that an initial condition in $l=0$ (such as the initial condition for CPT we consider) would populate densely all the normal modes of oscillation in the system.

\begin{figure}[htb]
\centering
\includegraphics[width=0.8\textwidth]{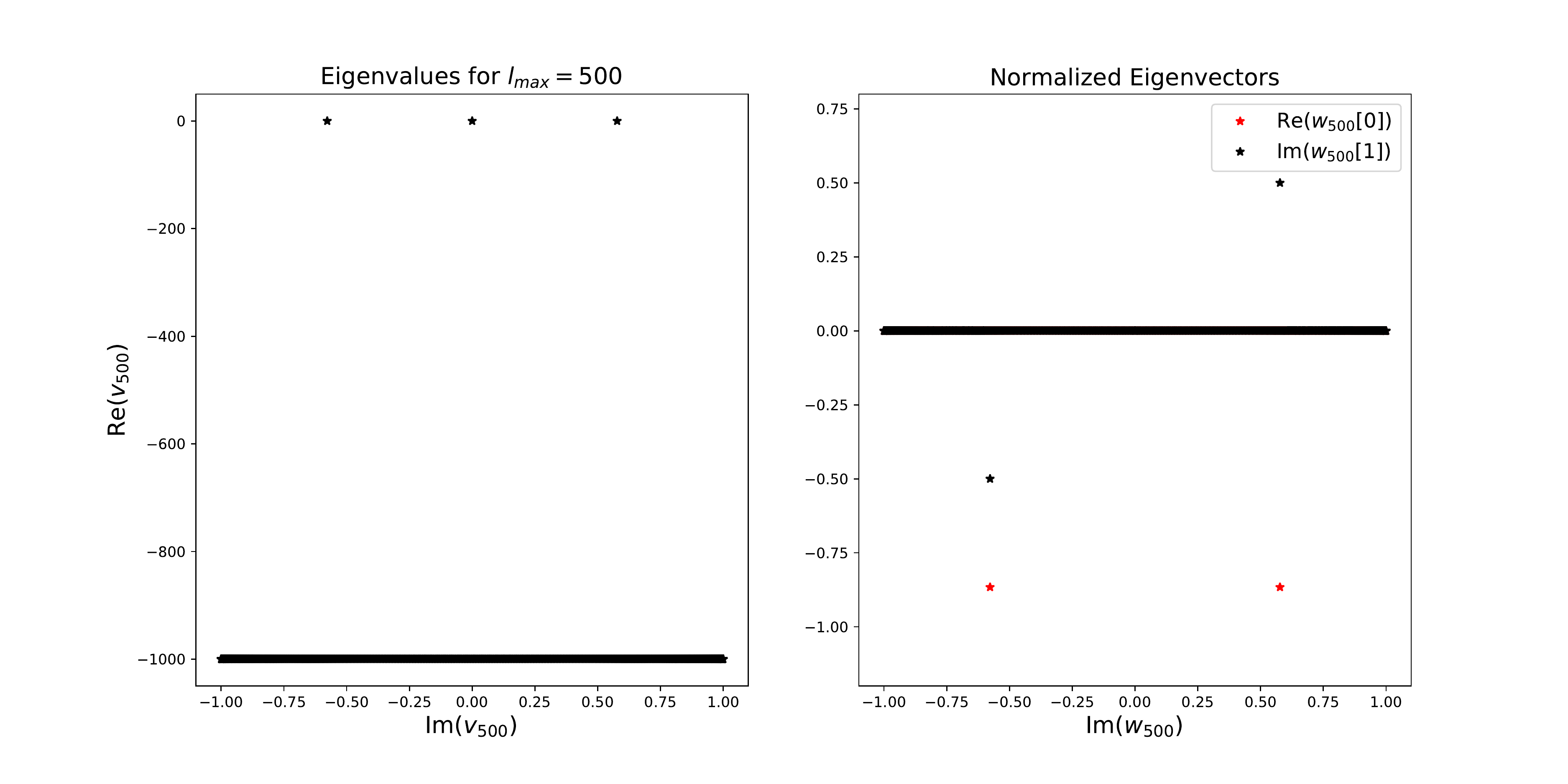}
\caption{\small 
Eigenvalue and eigenvector structure in the homogeneous Boltzmann Hierarchy with a damping term $\Gamma \sim 1\times 10^3$, for $l_{max}=500$. The left panel shows the eigenvalue distribution for this case in the complex plane, with the real part of most eigenvalues being proportional to the damping term in the case $\Gamma \gg 1$. The right panel shows the zero and first of the normalized eigenvalues in this case, as a function of the imaginary part of their eigenvalue. The real/imaginary parts of the eigenvalues not shown in the right hand plot are negligible.
}
\label{fig:Appendix_Osc_EigenvStructure_Int}
\end{figure}

But, what happens in the case of a damping term? In this case, the homogeneous system to resolve would be: 

\begin{equation}
\mathbf{\dot{\Psi}} = \alpha(\tau) \left[ \mathbf{A} \mathbf{\Psi} - \mathbf{\Gamma} \right]
\label{eq:Appendix_Osc_BoltzHomogeneous_plusIntTerm}
\end{equation}

\noindent with the damping term $\mathbf{\Gamma} = \mathrm{diag}(0,0,\Gamma,\Gamma,...)$. In the case $\Gamma \gg 1$ the eigenvalue structure, shown in \cref{fig:Appendix_Osc_EigenvStructure_Int}, becomes much different. 
Indeed, the only eigenvalues with a real part that is not equal to $-\Gamma \ll 0$ (thus, with modes that become highly suppressed during evolution) are the ones corresponding to the eigenvalues $0$ and $\pm i/\sqrt{3}$. 
We can also see in the eigenvector structure that only these last two modes are the ones initially populated by a perturbation in the $0$-th component (i.e. an initial density perturbation).

Thus, in this regime where a strong damping term is dominant over the homogeneous structure of the system, it will oscillate with a frequency of $\pm i/\sqrt{3}$ and a time dependent phase according to \eqref{eq:Appendix_Osc_TimeDependentPhase}. 
This damped system in particular represents the tightly-coupled regime in the case of SI-WDM, but indeed as this term would have the overall effect of suppressing all Boltmann hierarchy modes with $l \geq 2$ we can think of it as a representation of Boltzmann systems in general where higher modes are suppressed, such as what can happen in the various fluid or quasi fluid approximations for either WDM or other species. Obviously, according to the specific dynamics of the species the $0$-th and first equations in the hierarchy may involve different terms but the oscillatory nature is a common characteristic.

\begin{figure}[htb]
\centering
\includegraphics[width=0.8\textwidth]{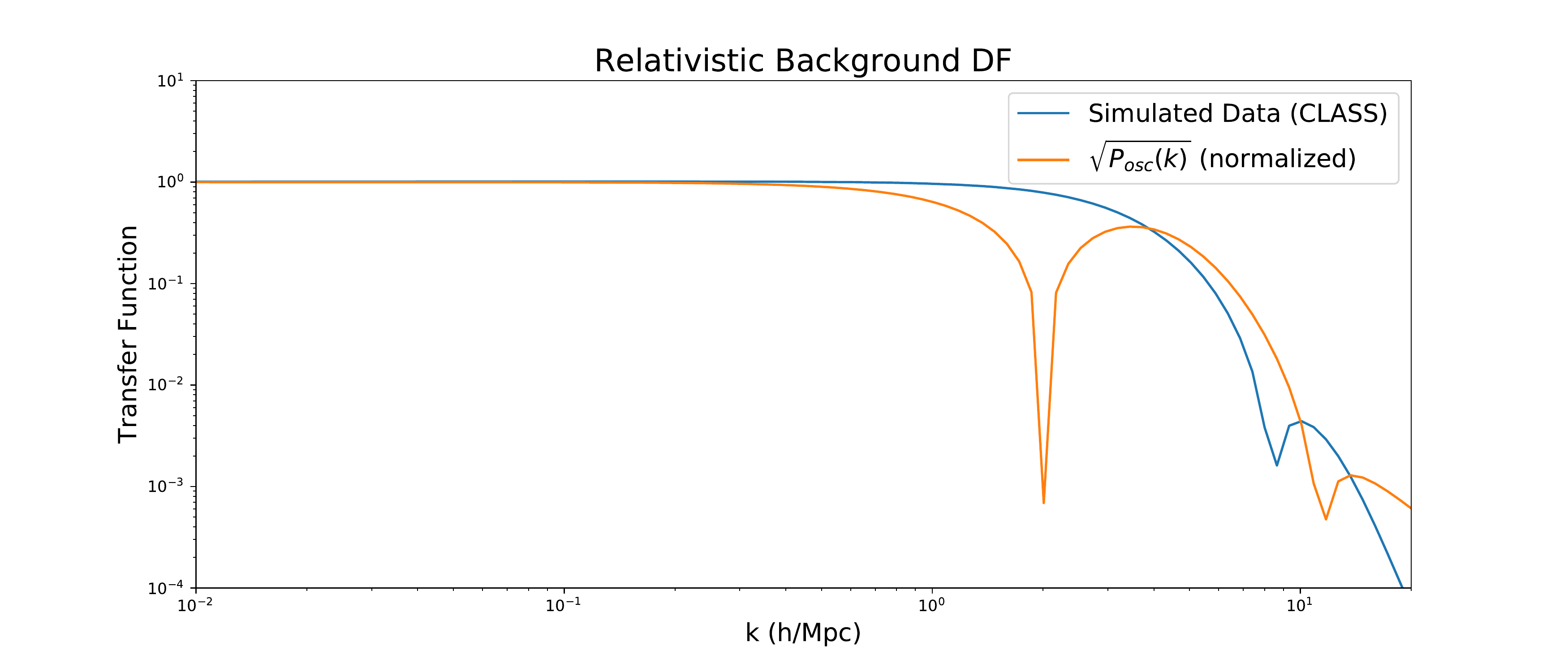}
\caption{\small 
Transfer function obtained with the approximate form \eqref{eq:Appendix_Osc_Posc} compared to simulated data using CLASS. We estimate the approximate transfer function using the value of $\sqrt{P_{osc}(k)}$ normalized to $1$ for $k \ll 1$. In this case, we take a self interaction model such as in \cref{fig:PowerSpectrum_Combined} for the highest interaction constant, but ignore the effects of non-relativistic self decoupling, and use a relativistic background distribution function assuming $T_{SI-WDM} = T_{\nu_a}$ with $T_{\nu_a}$ the temperature of active neutrinos.
}
\label{fig:Appendix_Osc_PSOscillations_noNRSD}
\end{figure}

\begin{figure}[htb]
\centering
\includegraphics[width=0.8\textwidth]{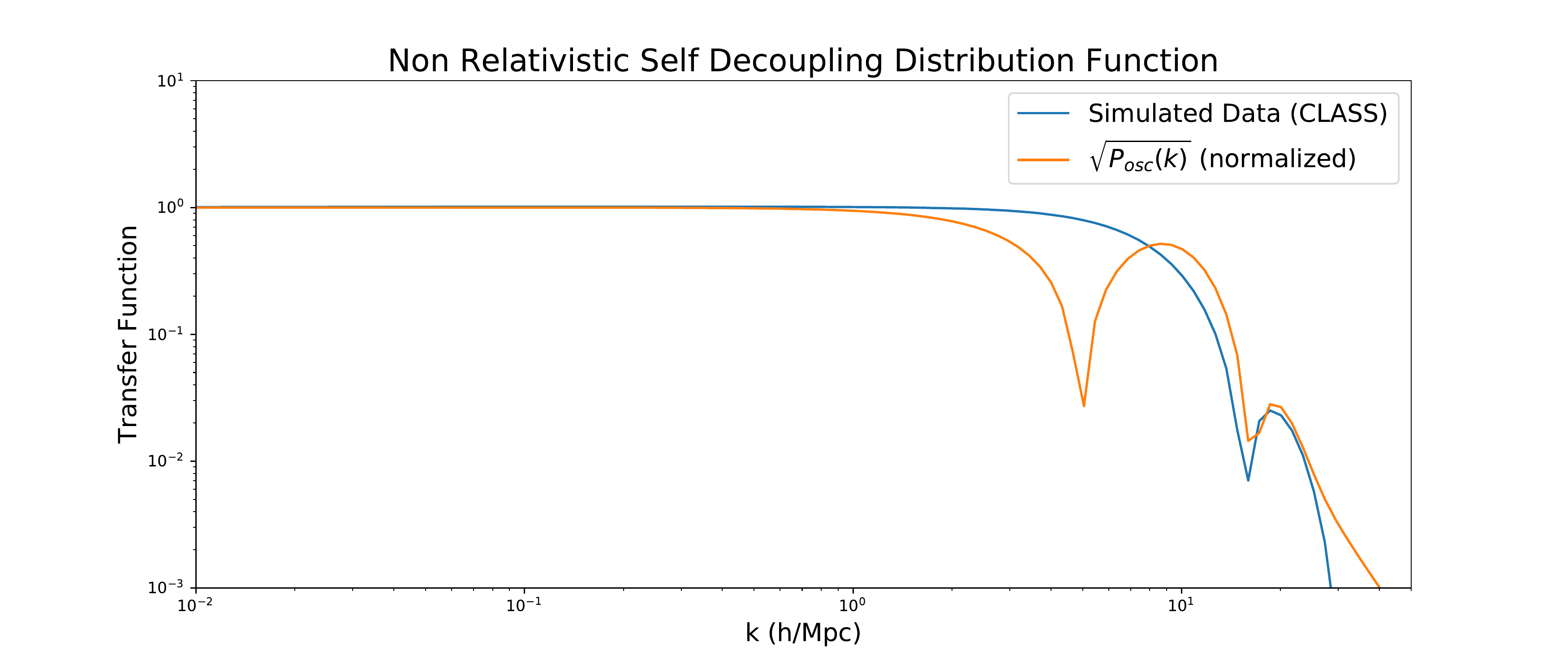}
\caption{\small 
Transfer function obtained with the approximate form \eqref{eq:Appendix_Osc_Posc} compared to simulated data using CLASS. We estimate the approximate transfer function using the value of $\sqrt{P_{osc}(k)}$ normalized to $1$ for $k \ll 1$. In this case, we take a self interaction model such as in \cref{fig:PowerSpectrum_Combined} for the highest interaction constant, with a background distribution function from a non-relativistic self decoupling scenario, according to the derivations in \ref{sec:SID} and \ref{sec:CLASS_NR}.
}
\label{fig:Appendix_Osc_PSOscillations_NRSD}
\end{figure}

This oscillatory nature of the system naturally has some impact in the power spectrum today. If this oscillation would continue until today, the linear power spectrum would be modulated by the oscillation in the density perturbation, itself governed by the oscillations in the $0$-th component of the Boltzmann hierarchy. 
If we focus on only the matter power spectrum, and ignore other effects we obtain the following approximate expression

\begin{equation}
P_{osc}(k) \propto \delta\rho^2(k) = \left[ 4 \pi T_{ncdm,0}^4 \int dq q^2 \epsilon \left( e^{ - i \phi / \sqrt{3}} + e^{i \phi / \sqrt{3}} \right) f_0(q) \right]^2
\label{eq:Appendix_Osc_Posc}
\end{equation}

\noindent where we see the both frequencies selected, $\pm 1/\sqrt{3}$, and this is just assuming that the WDM remains self coupled until today: the evolution after the self-decoupling is not considered here and is expected to modify the spectra. 
We can, however, compare this approximate power spectrum with the simulation results and we can see them side by side in \cref{fig:Appendix_Osc_PSOscillations_noNRSD} for the case without non-relativistic self decoupling and in \cref{fig:Appendix_Osc_PSOscillations_NRSD} for the full scenario, using a background DF as in \eqref{eq:SID_SelfCoupledEquilibrium_NRel}.

We can see in these figures that the frequency of the oscillation in the approximate form \eqref{eq:Appendix_Osc_Posc} is somewhat similar to the simulated data, even considering the surprising amount of approximations considered in the formula. 
However, the amplitude of the peaks and valleys are considerably different, especially in the first valley which has disappeared completely, which seems to point to more complex evolution governing the actual amplitude of the oscillations themselves. Nevertheless, we have gained some intuition about the origin of this behavior in the power spectrum. So, in summary:

\begin{itemize}
\item While the system is in tight-coupling (or potentially any other fluid approximation with negligible power in $l \geq 2$), the species shows an oscillatory nature that is imprinted in the power spectrum.

\item This oscillatory nature is not present in the full hierarchy and, by extension, in non-interacting WDM as the system shows a dense distribution of eigenvalues and eigenvectors that are uniformly populated for $l_{max} \gg 1$.

\item It is reasonable to assume that, after exiting the tight-coupling phase, free evolution erases some of the oscillations in the power spectrum, and that the longer the system remains tightly coupled the more evident these oscillations are.

\item Questions still remain about the mechanisms that govern the amplitude of the oscillations themselves, and the differences in amplitude between different valleys in the power spectrum/transfer function.
\end{itemize}

\clearpage

\section{Sterile Neutrino WDM Model Subsets}
\setcounter{figure}{0}
\vfill
\begin{figure}[h!]
\centering
\includegraphics[width=0.75\textwidth]{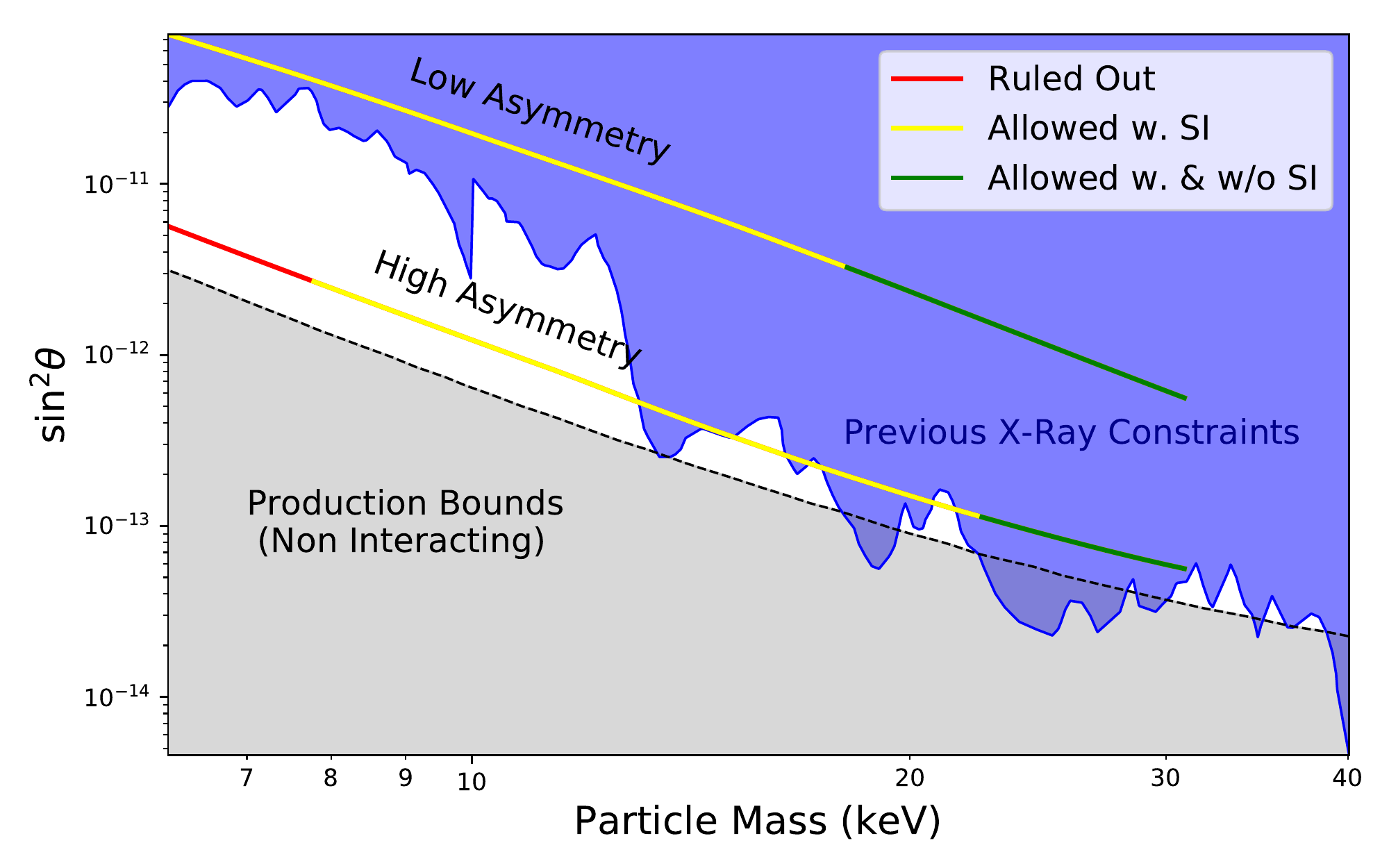}
\caption{\small 
Model subsets used for the resonant $\nu$MSM production model runs plotted over the $\nu$MSM model parameter space $\{\sin^2 \theta, m\}$, where $\theta$ is the sterile-active neutrino mixing angle and $m$ is the sterile neutrino mass. 
The models used here are sampled from the line labeled as ``high asymmetry'' for the high lepton asymmetry, low mixing angle models and from the line labeled as ``low asymmetry'' for the low lepton asymmetry, high mixing angle models. 
Over these lines, we distinguish the qualitative results of the analysis in \cref{sec:Obs_Nsub} and \cref{sec:Obs_deltaA}. In red, we represent models that are excluded by this analysis, for any values of the SI coupling constant and any SI model. In yellow we represent models that are allowed for some SI models but become forbidden as $C_{i} \rightarrow 0$, i.e. models that were previously forbidden but become allowed by the inclusion of self interactions. Finally, in green we represent models that are allowed for any SI coupling constant and any model in the ranges tested. 
We also plot other bounds to the $\nu$MSM parameter space for informative purposes, namely X-Ray indirect detection bounds \cite{Schneider2016,Cherry2017,Ng2019a} and sterile neutrino production bounds \cite{Boyarsky2018,Venumadhav2015}, again reminding the reader that these bounds do not include the effects of self interactions in sterile neutrino production.
}
\label{fig:Obs_Plots_sterileDMParameterSpace}
\end{figure}
\vfill
\clearpage

\begin{landscape}

\section{Fitting Formulae for the Transfer Function Coefficients}
\label{sec:Appendix_Table}
\setcounter{figure}{0}

\begin{table}[h!]
\tiny
\begin{tabular}{c|l|c|c|}
\cline{2-4}
                                                      & \multicolumn{1}{c|}{Production Mechanism} & \begin{tabular}[c]{@{}c@{}}Relativistic Decoupling \end{tabular}               & \begin{tabular}[c]{@{}c@{}}non-relativistic Decoupling \end{tabular}                                                            \\ \hline
\multicolumn{1}{|c|}{\multirow{4}{*}{\begin{tabular}[c]{@{}c@{}}Massive \\ Scalar\\ Mediator\end{tabular}}} & Non-Resonant                              & \begin{tabular}[c]{@{}c@{}}$i \equiv a_i \left( m_{DM} / \mathrm{MeV}\right)^{b_i}$ \\ $a_\alpha = 4.20 \times 10^{-4}$ , $b_\alpha = -0.894$\\ $a_\beta = 2.03$ , $b_\beta = -0.0181$\\ $a_\gamma = -3.42$, $b_\gamma = 0.0101$\end{tabular} & \begin{tabular}[c]{@{}c@{}} $i \equiv a_i \left( m_{DM} / \mathrm{MeV} \right)^{b_i} \cos^2 \left[ (C / \mathrm{eV}^{-4})/d_i + c_i \right] $ \\ $a_\alpha = 1.14 \times 10^{-4} $ , $b_\alpha = -9.45 \times 10^{-1}$ , $c_\alpha = 0.583$ , $d_\alpha = 1.002193 \times 10^{-32}$\\ $a_\beta = 2.17 $ , $b_\beta = -7.89 \times 10^{-3}$ , $c_\beta = -6.17 \times 10^{-2}$ , $d_\beta = 9.953530 \times 10^{-33}$\\ $a_\gamma = -1.79 \times 10^{-4}$, $b_\gamma = 0.134$ , $c_\gamma = -14.1$ , $d_\gamma = -2.692007 \times 10^{-31}$\end{tabular} \\ \cline{2-4} 
\multicolumn{1}{|c|}{}                                & Resonant (High Asymmetry)                 & \begin{tabular}[c]{@{}c@{}}$i \equiv a_i \left( m_{DM} / \mathrm{MeV}\right)^{b_i}$ \\ $a_\alpha = 1.59 \times 10^{-4}$ , $b_\alpha = -0.981$\\ $a_\beta = 2.13$ , $b_\beta = 0.00663$\\ $a_\gamma = 3.02$, $b_\gamma = -0.173$\end{tabular} & \begin{tabular}[c]{@{}c@{}}$i \equiv a_i \left( m_{DM} / \mathrm{MeV}\right)^{b_i} \left\{\mathrm{erfc} \left[ \log_{10}(C /\mathrm{eV}^{-4} / d_i) - c_i  \right] +e_i \right\}$ \\ $a_\alpha = 8.48 \times 10^{-5}$ , $b_\alpha = -0.694$ , $c_\alpha = 0.207$ , $d_\alpha = 1.871225 \times 10^{-33}$ , $e_\alpha = 0.217$ \\ $a_\beta = -0.128$ , $b_\beta = -0.00326$ , $c_\beta = 0.653$ , $d_\beta = 8.591565 \times 10^{-32}$ , $e_\beta = -16.5$ \\ $a_\gamma = 18.3$, $b_\gamma = -0.967$ , $c_\gamma = 1.262$ , $d_\gamma = 1.192073 \times 10^{-32}$ , $e_\gamma = -2.01$\end{tabular} \\ \cline{2-4} 
\multicolumn{1}{|c|}{}                                & Resonant (Low Asymmetry)                  & \begin{tabular}[c]{@{}c@{}}$i \equiv a_i \left( m_{DM} / \mathrm{MeV}\right)^{b_i}$ \\ $a_\alpha = 9.62 \times 10^{-5}$ , $b_\alpha = -1.01$\\ $a_\beta = 1.84$ , $b_\beta = -0.0239$\\ $a_\gamma = -16.7$, $b_\gamma = 0.189$\end{tabular} & \begin{tabular}[c]{@{}c@{}} $i \equiv a_i \left( m_{DM} / \mathrm{MeV}\right)^{b_i} \left\{\mathrm{erfc} \left[ \log_{10}(C /\mathrm{eV}^{-4} / d_i) - c_i  \right] +e_i \right\}$ \\ $a_\alpha = 2.43 \times 10^{-5}$ , $b_\alpha = -0.945$ , $c_\alpha = 0.485$ , $d_\alpha = 1.458468 \times 10^{-33}$ , $e_\alpha = 0.411$ \\ $a_\beta = -0.0567$ , $b_\beta = -0.00130$ , $c_\beta = 0.940$ , $d_\beta = 1.583597 \times 10^{-31}$ , $e_\beta = -37.9$ \\ $a_\gamma = 178$, $b_\gamma = 0.00807$ , $c_\gamma = 1.09$ , $d_\gamma = 1.918661 \times 10^{-32}$ , $e_\gamma = -2.10$\end{tabular} \\ \cline{2-4} 
\multicolumn{1}{|c|}{}                                & Thermal                                   & \begin{tabular}[c]{@{}c@{}}$i \equiv a_i \left( m_{DM} / \mathrm{MeV}\right)^{b_i}$ \\ $a_\alpha = 1.631 \times 10^{-5}$ , $b_\alpha = -1.173$\\ $a_\beta = 2.023$ , $b_\beta = -0.01332$\\ $a_\gamma = -2.919$, $b_\gamma = -0.01952$ \end{tabular} & \begin{tabular}[c]{@{}c@{}} $i \equiv a_i \left( m_{DM} / \mathrm{MeV}\right)^{b_i} \left\{\mathrm{erfc} \left[ \log_{10}(C /\mathrm{eV}^{-4} / d_i) - c_i  \right] +e_i \right\}$ \\ $a_\alpha = 1.0956 \times 10^{-5}$ , $b_\alpha = -1.0093$ , $c_\alpha = -1.0257$ , $d_\alpha = 6.50224 \times 10^{-32}$, $e_\alpha = 8.260 \times 10^{-2}$ \\ $a_\beta = -1.1856 \times 10^{-1}$ , $b_\beta = -4.141 \times 10^{-2}$ , $c_\beta = 9.391 \times 10^{-1} $ , $d_\beta = 1.244112 \times 10^{-31}$, $e_\beta = -1.65 \times 10^1$ \\ $a_\gamma = 3.711 \times 10^1$, $b_\gamma = -6.82 \times 10^{-1}$ , $c_\gamma = 1.077$ , $d_\gamma = 4.77270 \times 10^{-33}$, $e_\gamma = -1.948$\end{tabular} \\ \hline
\multicolumn{1}{|l|}{\multirow{4}{*}{\begin{tabular}[c]{@{}c@{}}Vector \\ Field\\ Mediator\end{tabular}}}   & Non-Resonant                              & \begin{tabular}[c]{@{}c@{}}$i \equiv a_i \left( m_{DM} / \mathrm{MeV}\right)^{b_i}$ \\ $a_\alpha = 4.20 \times 10^{-4}$ , $b_\alpha = -0.894$\\ $a_\beta = 2.03$ , $b_\beta = -0.0181$\\ $a_\gamma = -3.42$, $b_\gamma = 0.0101$\end{tabular} & \begin{tabular}[c]{@{}c@{}}$i \equiv a_i \left( m_{DM} / \mathrm{MeV} \right)^{b_i} \cos^2 \left[ (C / \mathrm{eV}^{-4})/d_i + c_i \right] $ \\ $a_\alpha = 3.14 \times 10^{-4} $ , $b_\alpha = -9.16 \times 10^{-1}$ , $c_\alpha = 1.02$ , $d_\alpha = 9.98 \times 10^{-33}$\\ $a_\beta = 2.27 $ , $b_\beta = -6.16 \times 10^{-3}$ , $c_\beta = -0.201$ , $d_\beta = 9.96 \times 10^{-33}$\\ $a_\gamma = -4.17 \times 10^{-4}$, $b_\gamma = 5.11 \times 10^{-2}$ , $c_\gamma = -14.0$ , $d_\gamma = -2.79 \times 10^{-32}$\end{tabular} \\ \cline{2-4} 
\multicolumn{1}{|l|}{}                                & Resonant (High Asymmetry)                 & \begin{tabular}[c]{@{}c@{}}$i \equiv a_i \left( m_{DM} / \mathrm{MeV}\right)^{b_i}$ \\ $a_\alpha = 1.59 \times 10^{-4}$ , $b_\alpha = -0.981$\\ $a_\beta = 2.13$ , $b_\beta = 0.00663$\\ $a_\gamma = 3.02$, $b_\gamma = -0.173$\end{tabular} & \begin{tabular}[c]{@{}c@{}}$i \equiv a_i \left( m_{DM} / \mathrm{MeV}\right)^{b_i} \left\{\mathrm{erfc} \left[ \log_{10}(C /\mathrm{eV}^{-4} / d_i) - c_i  \right] +e_i \right\}$ \\ $a_\alpha = 6.45 \times 10^{-5}$ , $b_\alpha = -0.766$ , $c_\alpha = 0.125$ , $d_\alpha = 6.506785 \times 10^{-34}$ , $e_\alpha = 0.211$ \\ $a_\beta = -0.126$ , $b_\beta = -0.00282$ , $c_\beta = 0.454$ , $d_\beta = 1.777554 \times 10^{-32}$ , $e_\beta = -17.1$ \\ $a_\gamma = 24.2$, $b_\gamma = -0.898$ , $c_\gamma = 0.635$ , $d_\gamma = 1.213926 \times 10^{-32}$ , $e_\gamma = -2.01$\end{tabular} \\ \cline{2-4} 
\multicolumn{1}{|l|}{}                                & Resonant (Low Asymmetry)                  & \begin{tabular}[c]{@{}c@{}}$i \equiv a_i \left( m_{DM} / \mathrm{MeV}\right)^{b_i}$ \\ $a_\alpha = 9.58 \times 10^{-5}$ , $b_\alpha = -1.01$\\ $a_\beta = 1.83$ , $b_\beta = -0.0241$\\ $a_\gamma = -16.8$, $b_\gamma = 0.192$\end{tabular} & \begin{tabular}[c]{@{}c@{}} $i \equiv a_i \left( m_{DM} / \mathrm{MeV}\right)^{b_i} \left\{\mathrm{erfc} \left[ \log_{10}(C /\mathrm{eV}^{-4} / d_i) - c_i  \right] +e_i \right\}$ \\ $a_\alpha = 3.02 \times 10^{-5}$ , $b_\alpha = -0.909$ , $c_\alpha = 0.00988$ , $d_\alpha = 3.831799 \times 10^{-34}$ , $e_\alpha = 0.390$ \\ $a_\beta = -0.0598$ , $b_\beta = -1.021 \times 10^{-2}$ , $c_\beta = 0.543$ , $d_\beta = 5.533459 \times 10^{-32}$ , $e_\beta = -36.3$ \\ $a_\gamma = 166$, $b_\gamma = 0.00665$ , $c_\gamma = 1.22$ , $d_\gamma = 1.442606 \times 10^{-33}$ , $e_\gamma = -2.08$\end{tabular} \\ \cline{2-4} 
\multicolumn{1}{|l|}{}                                & Thermal                                   & \begin{tabular}[c]{@{}c@{}}$i \equiv a_i \left( m_{DM} / \mathrm{MeV}\right)^{b_i}$ \\ $a_\alpha = 1.63 \times 10^{-5}$ , $b_\alpha = -1.173$\\ $a_\beta = 2.022$ , $b_\beta = -0.0134$\\ $a_\gamma = -2.92$, $b_\gamma = -0.0190$\end{tabular}  & \begin{tabular}[c]{@{}c@{}} $i \equiv a_i \left( m_{DM} / \mathrm{MeV}\right)^{b_i} \left\{\mathrm{erfc} \left[ \log_{10}(C /\mathrm{eV}^{-4} / d_i) - c_i  \right] +e_i \right\}$ \\ $a_\alpha = 5.99 \times 10^{-6}$ , $b_\alpha = -1.12$ , $c_\alpha = -1.88$ , $d_\alpha = 1.098403 \times 10^{-31}$, $e_\alpha = 8.32 \times 10^{-2}$\\ $a_\beta = -1.19 \times 10^{-1}$ , $b_\beta = -4.6 \times 10^{-2}$ , $c_\beta = 0.598$ , $d_\beta = 6.87690 \times 10^{-32}$, $e_\beta = -1.60 \times 10^1$ \\ $a_\gamma = 5.73$, $b_\gamma = -0.968$ , $c_\gamma = 0.772$ , $d_\gamma = 7.93625 \times 10^{-33}$, $e_\gamma = -1.95$ \end{tabular} \\ \hline
\end{tabular}
\label{tab:PS_Fits}
\caption{%
\scriptsize
Approximate fitting formula for the transfer function parameters in SI-WDM models, with respect to CDM cosmology, for all combinations of mediator model and production scenario analyzed in \cref{sec:Obs}. This table provides fitting functions for the parameters $\{\alpha,\beta,\gamma\}$ used in the parametrization \eqref{eq:Obs_TkFittingFormula} of the suppressed transfer function for ``non cold'' DM models. 
For each model/production mechanism combination, we provide the functional form that all parameters follow as a function of $\{m_{\rm DM},C_i\}$ with a number of free parameters, and the appropriate best fit values for each parameters (for example, for a given cell, a fitting formula for $\alpha$ is obtained by replacing the values of $\{a_\alpha,b_\alpha,...\}$ in the functional form at the top of the cell). 
}
\end{table}

\end{landscape}

\section{Power Spectra and Parameter Space Constraints for Other Mediator and Background Models}
\label{sec:Appendix_AllPlots}
\setcounter{figure}{0}
\vfill
\begin{figure}[htb]
\centering
\includegraphics[width=0.95\textwidth]{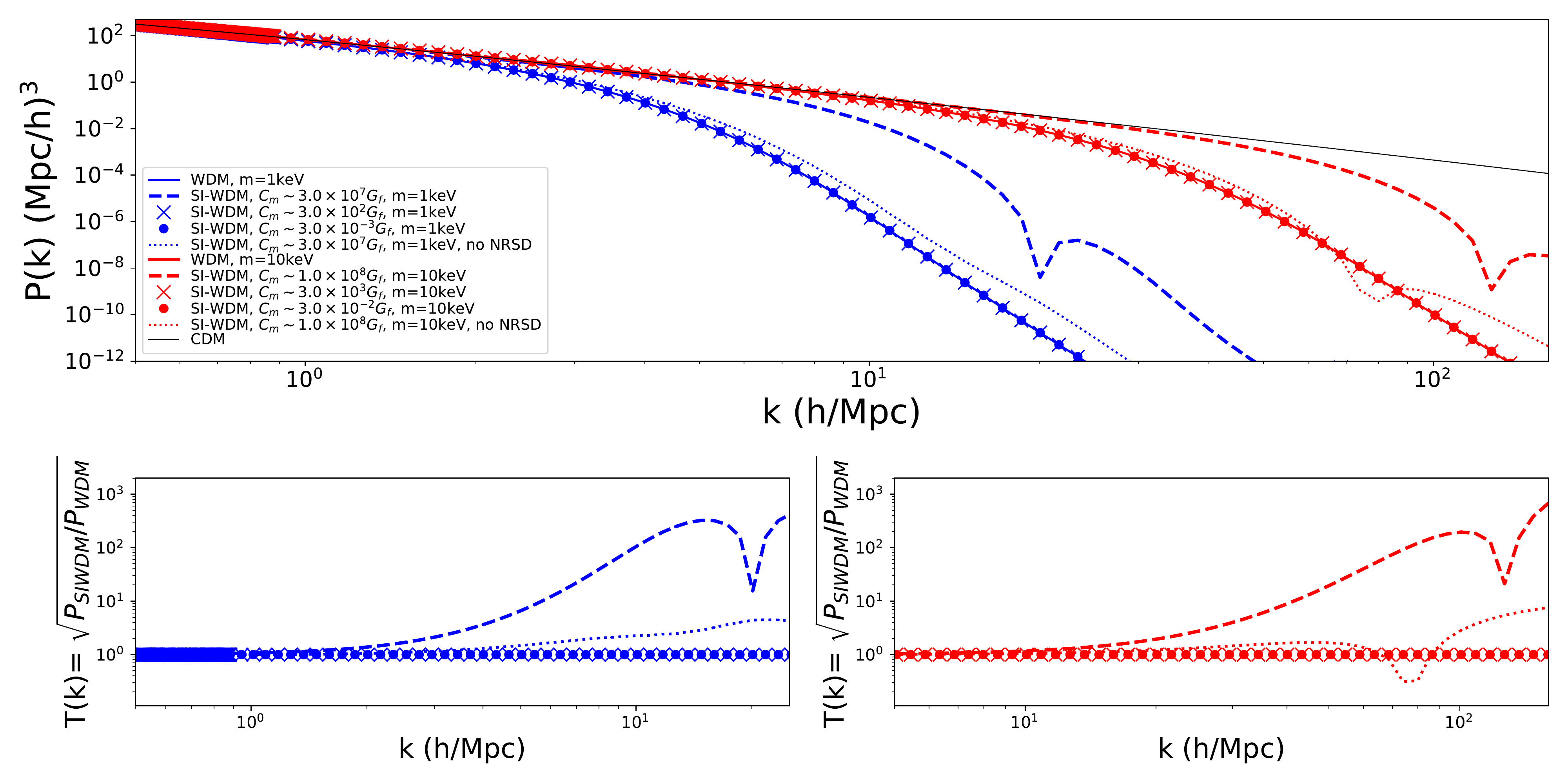}
\caption{\small 
Power Spectrum and Transfer Functions with respect to standard WDM for several SI-WDM models for the case of a Massive Scalar Mediator. See the caption in \cref{fig:PowerSpectrum_Combined} for more information. 
}
\label{fig:PowerSpectrum_MS_Combined_Appx}
\end{figure}
\vfill
%\begin{figure}[htb]
%\centering
%\includegraphics[width=0.95\textwidth]{plots/power-spectra/Plots_No_NR_SI/SIWDM_MS_Transfer_1and10keV_separateTransfers.pdf}
%\caption{\small 
%Power Spectrum and Transfer Functions with respect to %standard WDM for several SI-WDM models, ignoring the effects %of non-relativistic self decoupling, for the case of a %Massive Scalar Mediator. See the caption in %\cref{fig:PowerSpectrum_noNRSD} for more information. 
%}
%\label{fig:PowerSpectrum_noNRSD_MS_Appx}
%\end{figure}

%\begin{figure}[htb]
%\centering
%\includegraphics[width=0.95\textwidth]{plots/power-spectra/SI%WDM_MS_Transfer_1and10keV_separateTransfers.pdf}
%\caption{\small 
%Simulated Power Spectrum and Transfer Functions with respect %to standard WDM for several SI-WDM models, considering the %effects of non-relativistic self decoupling when appropriate, %for the case of a Massive Scalar Mediator. See the caption in %\cref{fig:PowerSpectrum_noNRSD} for more information. 
%}
%\label{fig:PowerSpectrum_NRSD_MS_Appx}
%\end{figure}

\clearpage

\begin{figure*}[h!]
\begin{adjustwidth}{-2cm}{-2cm}
\centering
    \begin{subfigure}{0.49\linewidth}
        \centering
        \includegraphics[width=\textwidth]{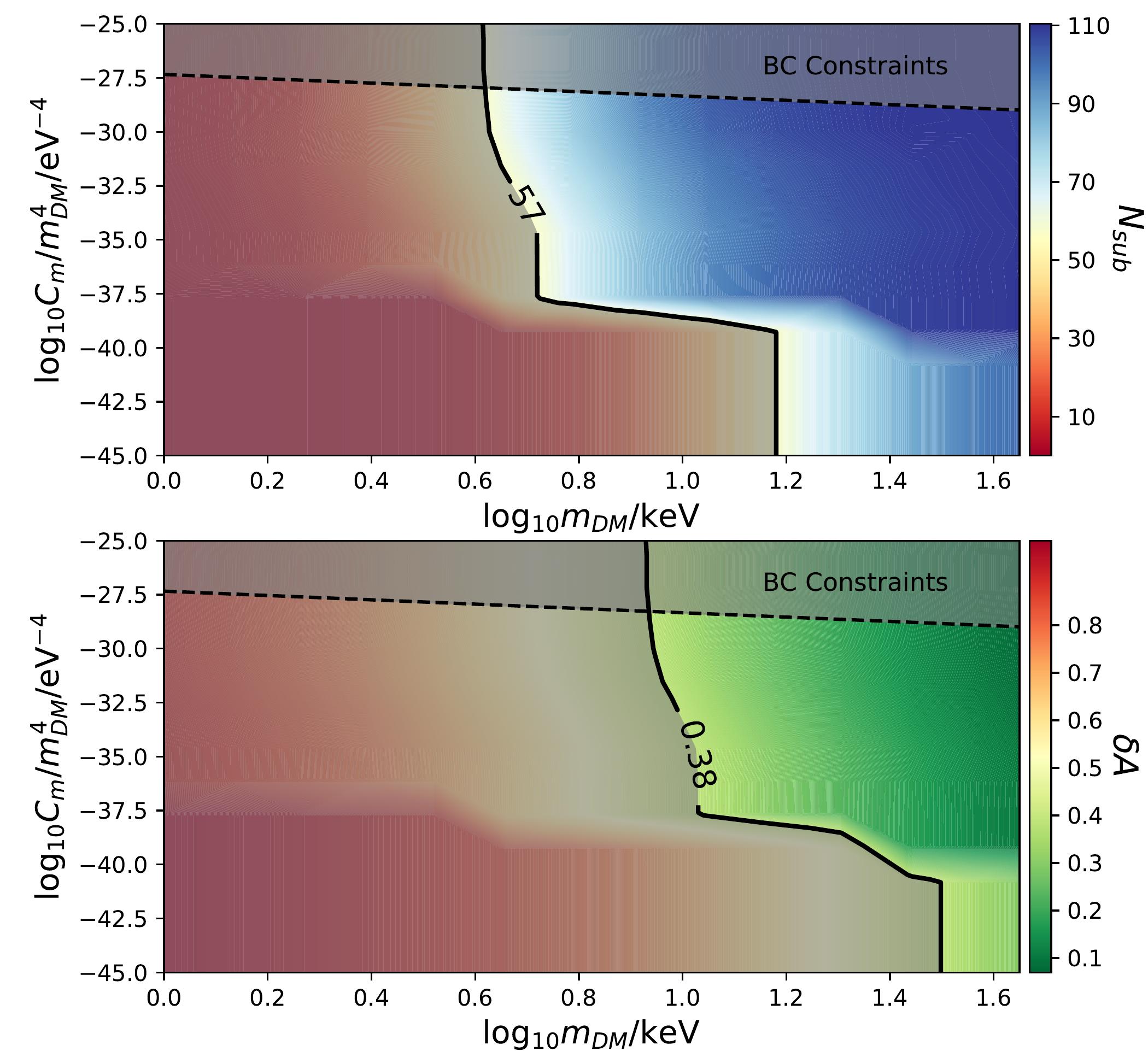}
        \label{fig:Obs_Results_Grid_MS_NR}
        \caption{Massive scalar mediator, non-resonant production}
    \end{subfigure}
    \begin{subfigure}{0.49\linewidth}
        \centering
        \includegraphics[width=\textwidth]{plots/MassiveScalar-SterileLowTheta-Simple.jpg}
        \label{fig:Obs_Results_Grid_MS_Res_l}
        \caption{Massive scalar mediator, resonant production (high asymmetry)}
    \end{subfigure}
    \par\bigskip
    \begin{subfigure}{0.49\linewidth}
        \centering
        \includegraphics[width=\textwidth]{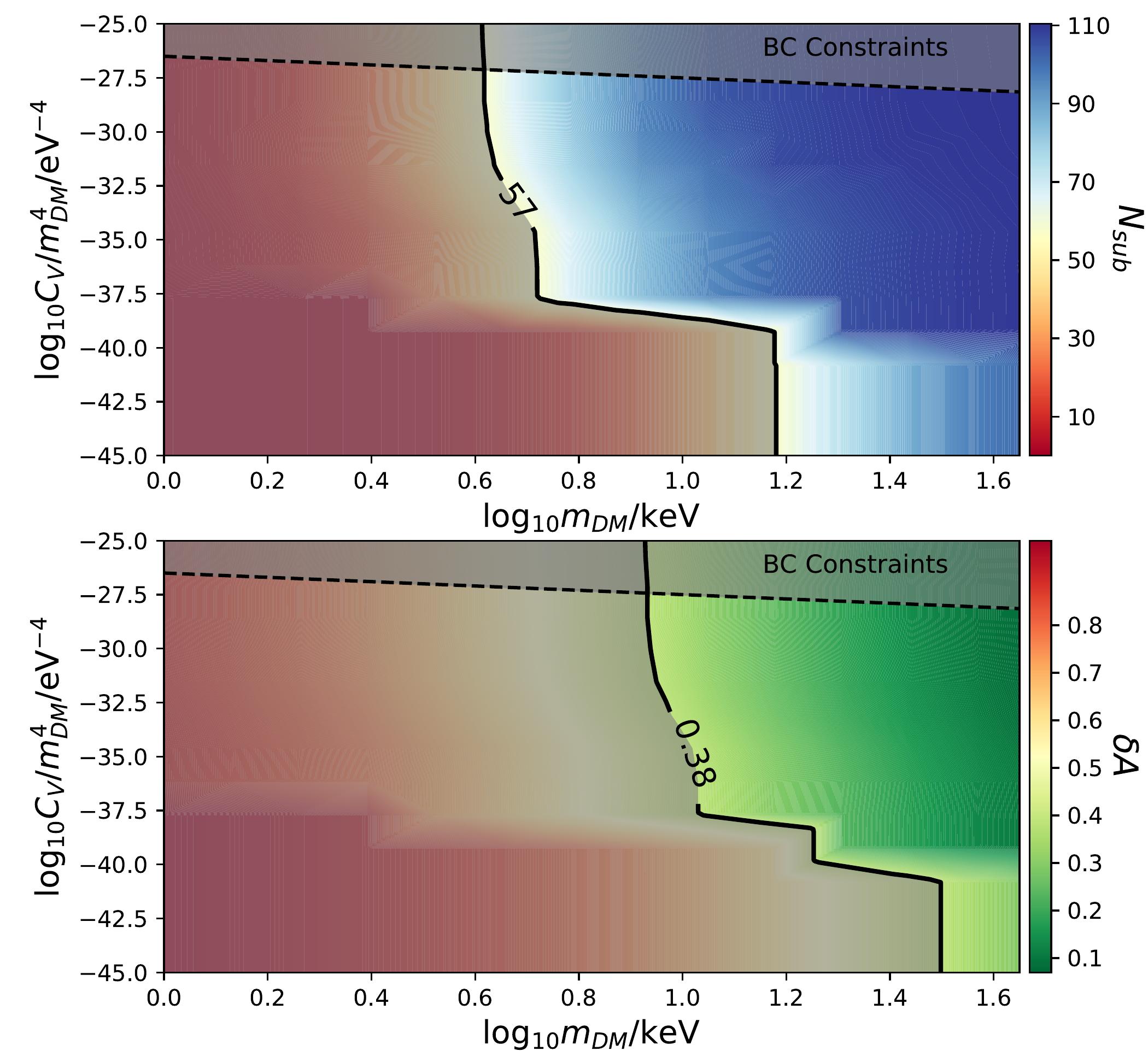}
        \label{fig:Obs_Results_Grid_VF_NR}
        \caption{Vector field mediator, non-resonant production}
    \end{subfigure}
    \begin{subfigure}{0.49\linewidth}
        \centering
        \includegraphics[width=\textwidth]{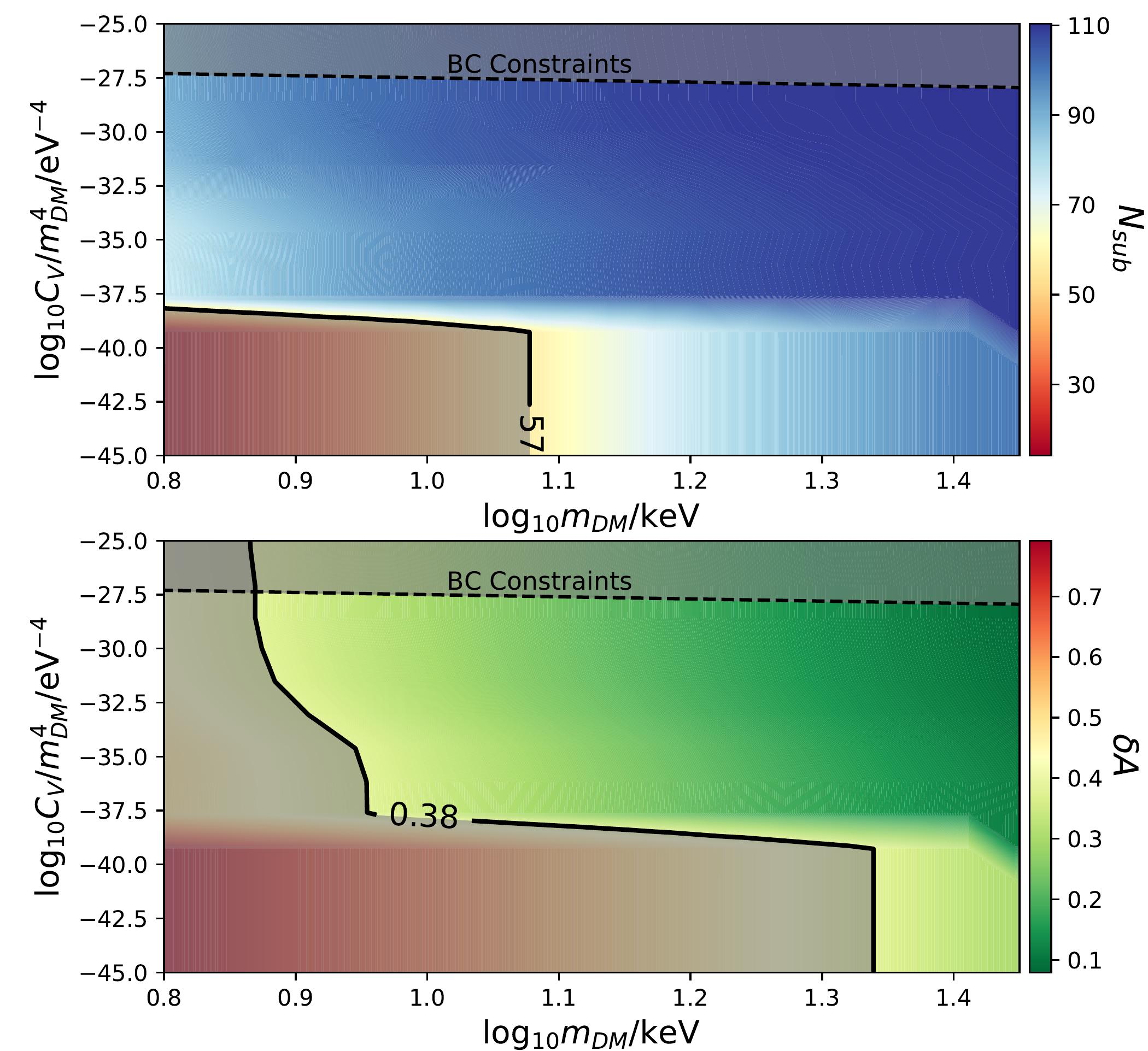}
        \label{fig:Obs_Results_Grid_VF_Res_l}
        \caption{Vector field mediator, resonant production (high asymmetry)}
    \end{subfigure}
%\caption{\small 
%AAJFGKASJHFBA,SNDB 
%}
%\label{fig:Obs_Results_Grid_a}
\end{adjustwidth}
\end{figure*}

\clearpage

\setcounter{figure}{2}

\begin{figure}[t!]
\begin{adjustwidth}{-2cm}{-2cm}
\centering
	\begin{subfigure}{0.49\linewidth}
    	\addtocounter{subfigure}{4}
        \centering
        \includegraphics[width=\textwidth]{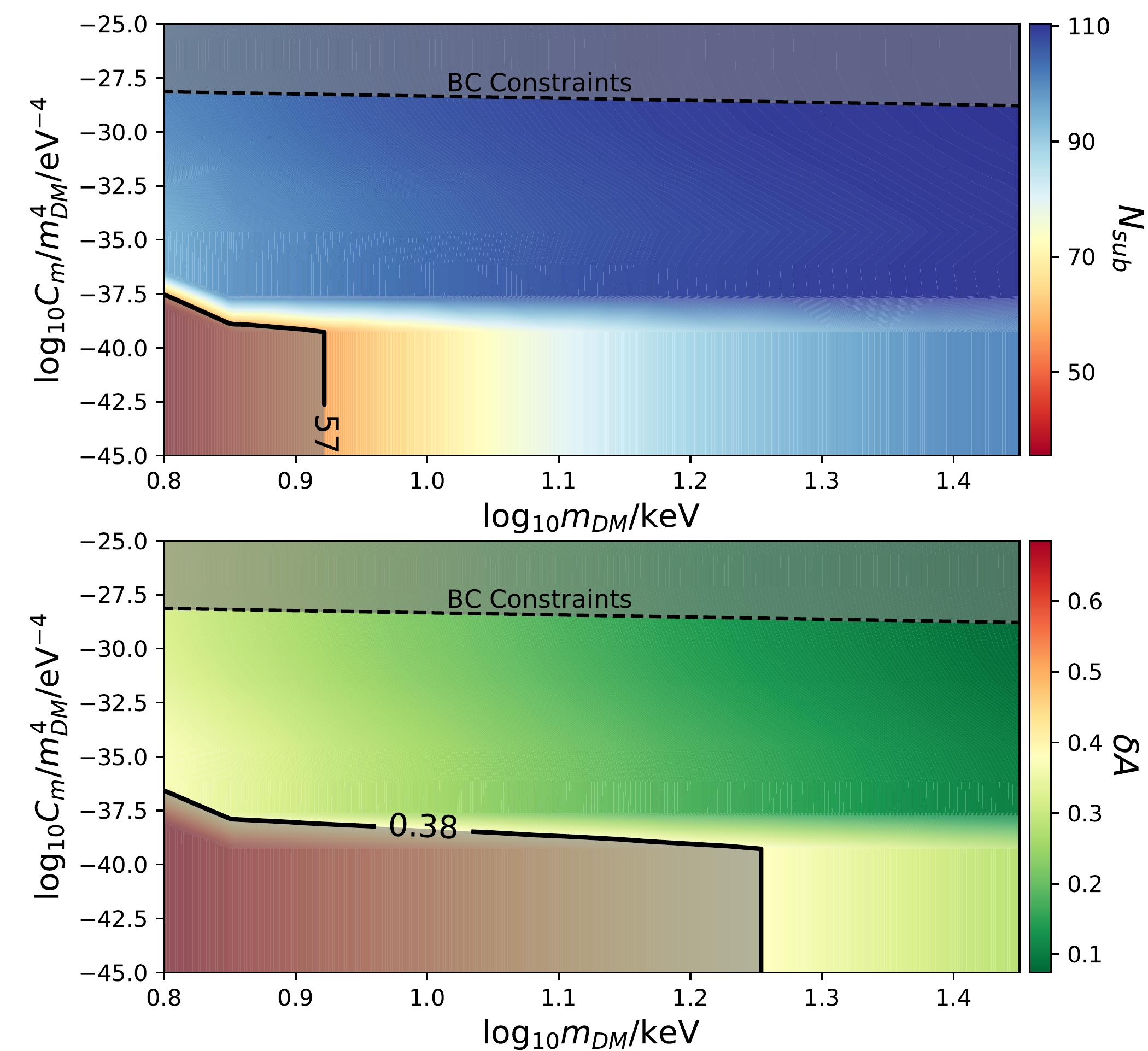}
        \label{fig:Obs_Results_Grid_MS_Res_h}
        \caption{Massive scalar mediator, resonant production (low asymmetry)}
    \end{subfigure}
    \begin{subfigure}{0.49\linewidth}
        \centering
        \includegraphics[width=\textwidth]{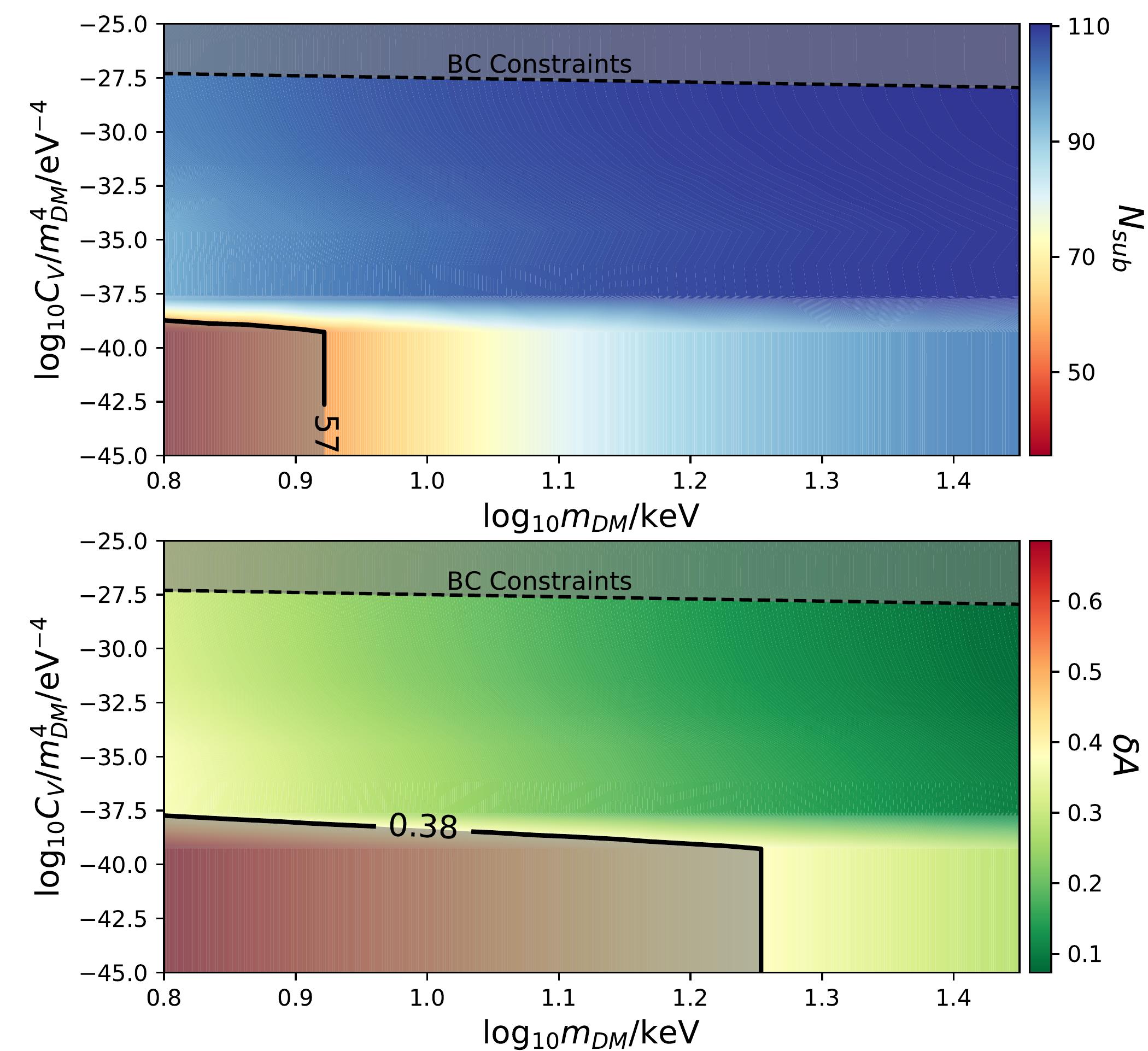}
        \label{fig:Obs_Results_Grid_VF_Res_h}
        \caption{Vector field mediator, resonant production (low asymmetry)}
    \end{subfigure}
    \par\bigskip
    \begin{subfigure}{0.49\linewidth}
    	\addtocounter{subfigure}{4}
        \centering
        \includegraphics[width=\textwidth]{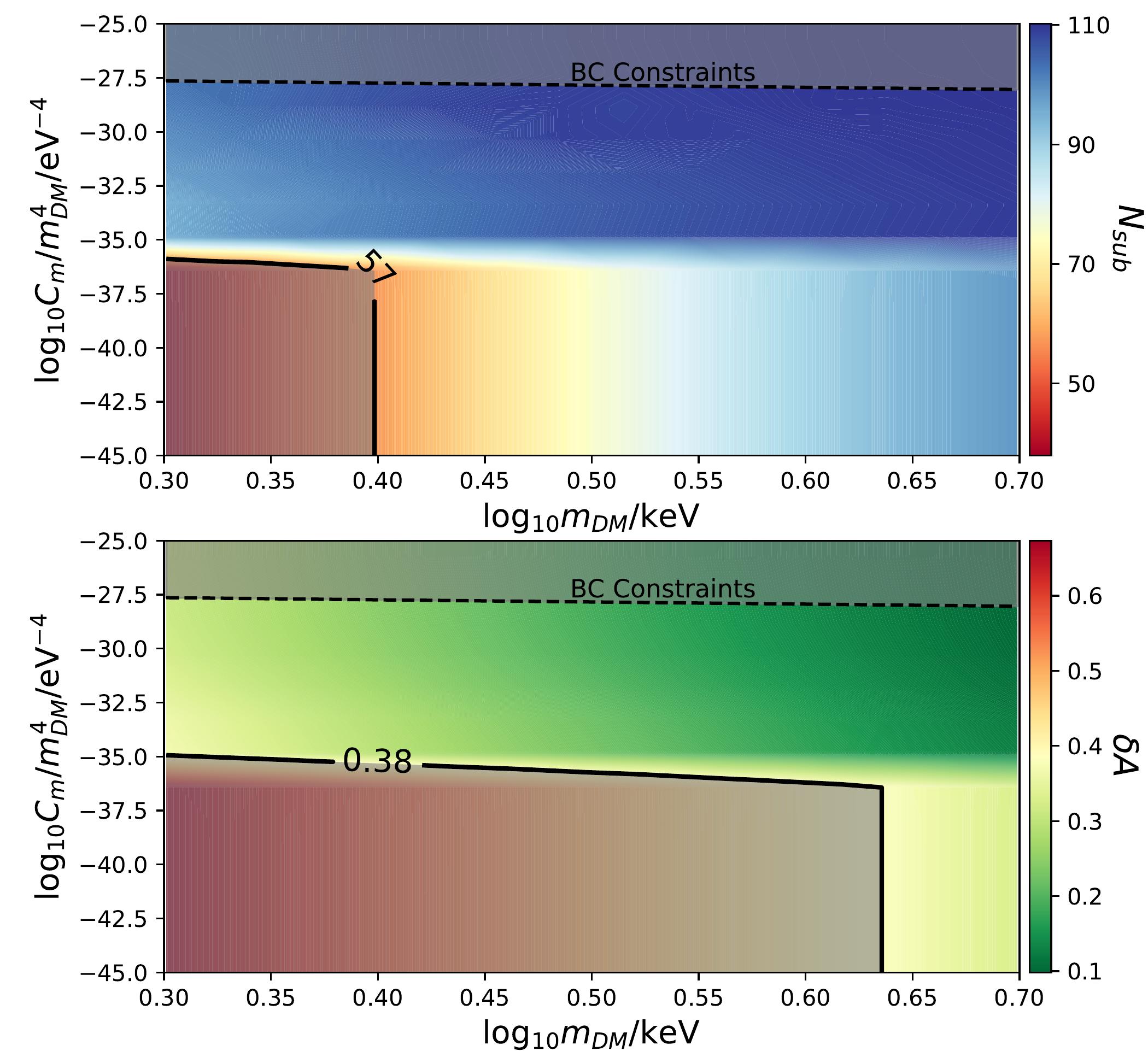}
        \label{fig:Obs_Results_Grid_MS_Th}
        \caption{Massive scalar mediator, thermal production}
    \end{subfigure}
    \begin{subfigure}{0.49\linewidth}
        \centering
        \includegraphics[width=\textwidth]{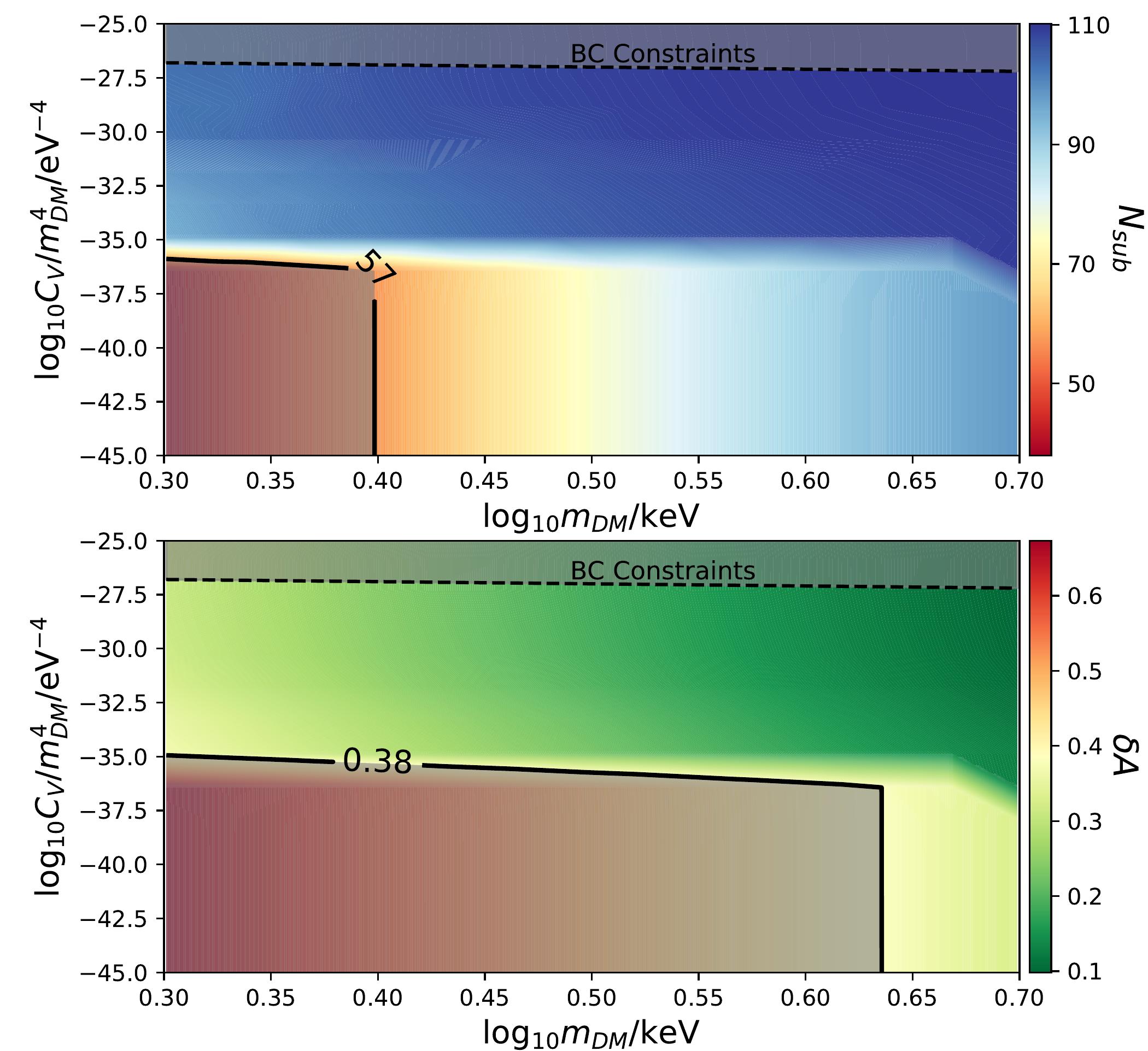}
        \label{fig:Obs_Results_Grid_VF_Th}
        \caption{Vector field mediator, thermal production}
    \end{subfigure}
\caption{\small 
Parameter space constraints for SI-WDM cosmologies, according to the analysis outlined in section \ref{sec:Obs}, for the combinations of Mediator models and background/production models not considered in \cref{fig:Obs_Results_Grid}. See the caption in \cref{fig:Obs_Results_Grid} for more information. 
}
\label{fig:Obs_Results_Grid_Appx}
\end{adjustwidth}
\end{figure}

%\null
%\vfill

\end{document}